 \documentclass[twocolumn, 10pt]{article}
\usepackage{float}
   \usepackage[round,authoryear]{natbib}

 \usepackage{soul}   %  for striking through text horizontally via  \st{...}

 \usepackage{hyperref}

  \usepackage[latin1]{inputenc}
 \usepackage{amsfonts}
 \usepackage{amsthm}
 \usepackage{amssymb, latexsym, amsmath}
 \usepackage{amsbsy}
\usepackage{amstext}
\usepackage{amsopn}
\usepackage{amsgen}
 \usepackage{graphicx}
 \usepackage[normalem]{ulem}
   \usepackage{wasysym}
   \usepackage{deluxetable}
  \usepackage{verbatim}
  \usepackage[T1]{fontenc}
  \usepackage[bottom]{footmisc}

  \usepackage{subfigure}

   \usepackage[dvips]{color}
   \usepackage{color}

\usepackage{txfonts}
\usepackage{xcolor}

\definecolor{mypink}{rgb}{0.858, 0.188, 0.478}
\definecolor{mygrey}{rgb}{0.55, 0.68, 0.55}

 \newcommand{\be}{\begin{equation}}
  \newcommand{\bs}{\begin{subequations}}
 \newcommand{\es}{\end{subequations}}
 \newcommand{\ba}{\begin{eqnarray}}
 \newcommand{\ea}{\end{eqnarray}}
 \newcommand{\s}{\nobreak\hspace{.11em}\nobreak}
   \newcommand{\m}{\nobreak\hspace{.05em}\nobreak}
  
\newcommand{\eb}{\begin{equation}}
\newcommand{\ee}{\end{equation}}

\definecolor{rkka}{RGB}{219,66,32}
\definecolor{nsgreen}{rgb}{0.1,0.5,0.1}
\definecolor{mypink}{rgb}{0.858, 0.188, 0.478}
\definecolor{mygrey}{rgb}{0.55, 0.68, 0.55}

\begin{document}
 \title{
 {\Large{\textbf{Evolution of a synchronous planet-moon pair due to solar tides.\\Demise of the synchronous moon that initiated Mars' triaxiality.\\ A possible link to
% rhythmites
tidalites
in Vastitas Borealis and Gale Crater
 \\}
            }}}
 \author{
  {\Large{Michael Efroimsky}}\\
 {\small{US Naval Observatory, Washington DC 20392 USA}}\\
 {\small{michael.efroimsky{\s}@{\s}gmail.com}}
  }
 \date{}

 \twocolumn[
  \begin{@twocolumnfalse}
 \maketitle
 \begin{abstract}
   Mars' asymmetric figure\,---\,with two opposing equatorial elevations\,---\,originated from a frozen tidal bulge raised by
a primordial synchronous moon Nerio. Nerio's emergence, through in situ accretion or by capture in the disk's remnants, and its
synchronisation with Mars' rotation preceded or was coeval with crust formation. The submoon and antimoon regions hypothetically
developed thinner crusts, intensifying the tectonics that further amplified Mars' triaxiality.

We investigate Nerio's orbit stability and demise, and its impact on Mars' rotation. Nerio may also have been the cause of the tidal
rhythmites discovered in Martian sediments.

We derive a relation governing the orbit evolution of a synchronous planet-moon pair under the influence of stellar tides in the planet.
These tides gradually shrink the orbit of the synchronous pair, so the synchronous orbit is stable only transiently. On reaching a critical
separation from the planet, the moon departs synchronism and spirals down, accelerating the planet's spin.

The application of this development to Mars and Nerio shows that Mars' rotation rate at the desynchronisation time matches the present-day rate to the first decimal place.  This coincidence should not be overinterpreted,  as post-desynchronisation evolution included Mars' continued spin-up during Nerio's descent until Nerio's destruction amid the LHB, followed by Mars' despinning by solar tides.
  Nerio's reaching the Roche limit intact is questionable. Beyond LHB hazards, it would imply Mars' larger spin-up, necessitating values up to $k_2/Q\simeq 18$ to allow subsequent despinning to the present-day rate. Such values may be high even for  shallow oceans.

  Absent future evidence supporting such elevated $k_2/Q$ values, Nerio likely perished during the LHB. This viewpoint
may be reconsidered should new data on Mars' palaeo-ocean show up. The existence and subsequent demise of Nerio are indirectly
confirmed by the presence of rhythmites in Vastitas Borealis and Gale Crater. In the course of its post-desynchronisation descent,
Nerio was capable of generating tidalites within the palaeo-ocean. After Nerio's demise, this generation may have been continued by
its massive remnant(s).   \\~\\
\end{abstract}
  \end{@twocolumnfalse}
  ]

  % Keywords :  {Planets and satellites: general --
  %             Planets and satellites: individual: Mars  --
  %             Planets and satellites: physical evolution --
  %             Planets and satellites: dynamical evolution and stability --
  %             Planets and satellites: formation --
  %             Planets and satellites: terrestrial planets  --
  %             Planets and satellites: oceans  --
  %             Celestial Mechanics
  %            }

   \maketitle

 \section{Introduction}\label{Introduction}\label{1.1}

 % \enlargethispage{\baselineskip}
 %  \enlargethispage{\baselineskip}

 About $9.3$ times less massive than the Earth, Mars has had a very different geological history.  It had been losing its internal heat much faster, and its volcanism eventually came to an end {or near-end \citep{Carr}}.  Due to the rapid cooling, Mars has preserved a manifestly nonhydrostatic shape.  Highly triaxial and unusually asymmetric for a terrestrial planet, this shape is marked with two large near-equatorial elevations located almost opposite to one another\,---\,a highly improbable configuration. One of these uplifts is, of course, the colossal Tharsis Rise ($0^{\rm o}\s$N $260^{\rm o}\s$E).
 {The other is a less prominent elevation comprising
  %  Elysium Planitia ($3^{\rm o}\s$N  $154.7^{\rm o}\s$E),
 Syrtis Major Planum (centered around $8.4^{\rm o}\s$N $69.5^{\rm o}\s$E) and an adjacent
  %  to Syrtis Major
 part of
 %  Terra Sabaea
 Tyrrhena Terra (centered around $14.8^{\rm o}\s$S $90^{\rm o}\s$E),
 see \citet{atlas}.}

 %  \subsection{Primordial synchronous moon as a source of Mars' initial asymmetric triaxiality}

 Positioned on opposite sides of the equator, but differing in mean height, these two great elevations are likely to have provened from a fossilised tidal bulge generated by a massive synchronous moon \citep{Mars}.
 In {\it Ibid.}, it was also suggested that in the tidally elevated submoon and antimoon areas the subsequent tectonic and volcanic activity became more intense, leading to further elevation of those provinces.  It was demonstrated there that to cause the initial asymmetric triaxiality of an early Mars, the moon should have had a mass
 \ba
 M_m \s\lesssim\s 0.032\s M\,,
 \label{}
 \ea
 $M$ being the mass of Mars. Following Beno\^it Noyelles'  advice, the moon was named {\it Nerio}, after a female companion of Mars in early Roman mythology.

 Despite the large chaotic excursions in Mars' obliquity over geological time \citep{Touma, Laskar},
 the moon\,---\,once captured near the equator\,---\,remained tightly coupled to the moving equator, exhibiting only
 small inclination oscillations.\,\footnote{~For uniform equinoctial precession, this was proved by \citet{Goldreich} who, as it turned out, carried out his development in nonosculating orbital elements. A later analysis in osculating elements confirmed his conclusion \citep{Efroimsky2005}. The case of nonuniform variations of the planet's obliquity required a further study\,---\,which established that also in this situation a near-equatorial satellite is never repelled considerably from the equator of date \citep{Gurfil}.}

{
 At the same time, the existence of such a massive moon ought to have reduced the amplitude of those chaotic excursions \citep{Laskar, Laskar2, Lissauer} and, consequently, to have mitigated the resulting variations in insolation. Whether this mitigation had implications for the Martian climate is beyond the scope of our project.}

 \section{Principal points of this paper}

 \subsection{Dynamics of a mutually synchronised planet-moon pair,
             and the implications for Mars and Nerio}

\begin{figure}
 	\hspace*{-4.91mm}\includegraphics[width=1.46\columnwidth]{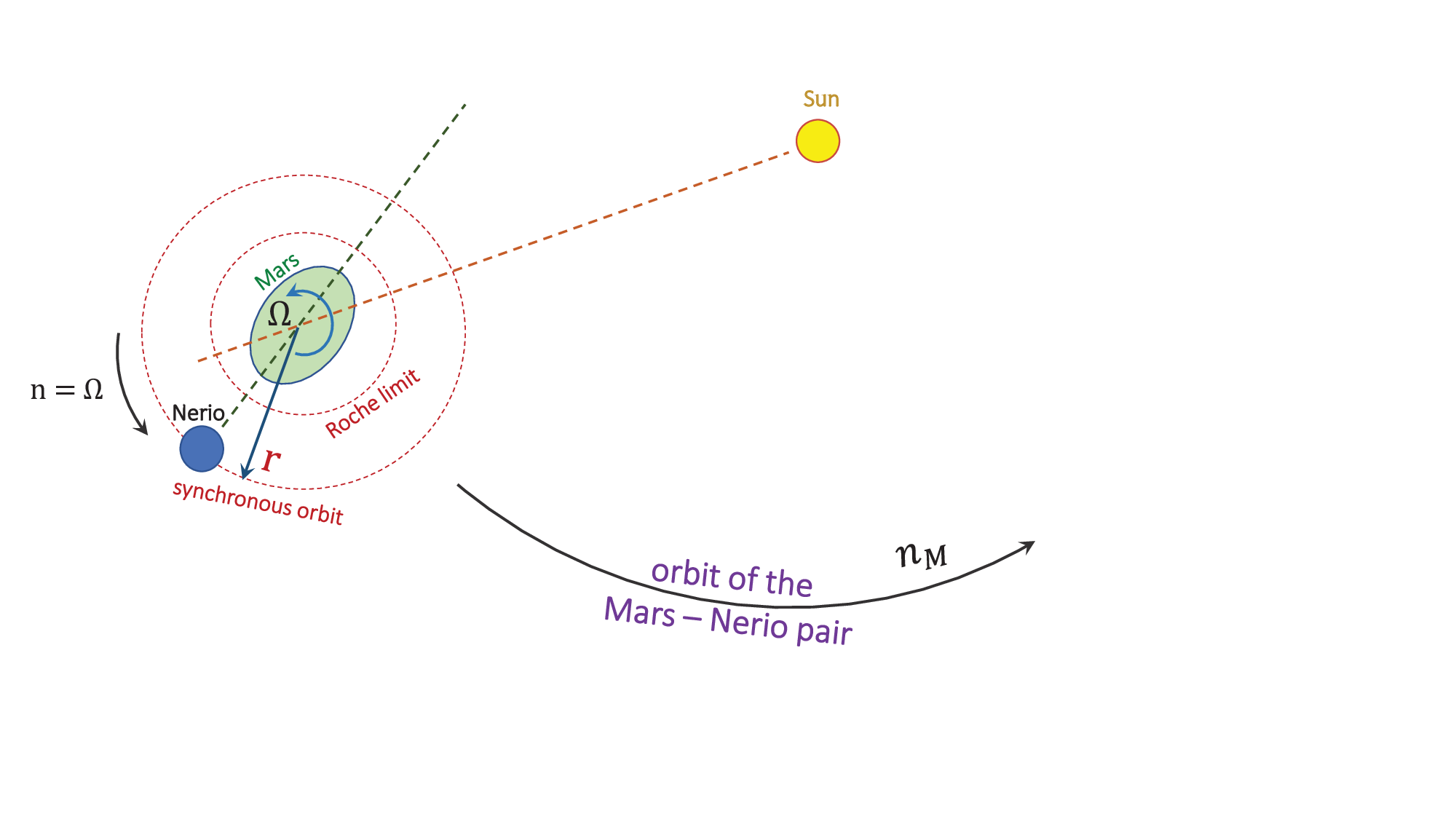}
   \caption{. \small{Schematic showing that the spin rate of Mars, $\Omega$, is equal to Nerio's mean motion $n$, but is much faster than the mean motion $n_M$ of the Mars-Nerio pair about the Sun.
   % Consequently, the semidiurnal solar tidal bulge on Mars is leading.
   }}
    \label{picture}
   \end{figure}

 A synchronous moon with mean motion $n$ maintains the planet's spin rate at $\Omega=n$,  neglecting external influences.
 {This rate is much faster than the mean motion of the planet-moon system as a whole. In application to Mars and Nerio, this situation is depicted in Figure \ref{picture}.}

 Solar tides acting on the planet, however, change this picture. In the absence of a synchronous moon, solar tides would ordinarily act to slow the planet's rotation. With a synchronous moon present, by contrast, solar tides cause the moon's orbit to contract, transferring angular momentum to the planet and thereby increasing its spin rate\,---\,a counterintuitive outcome. As we shall see, the system's slow evolution ultimately destabilises  the moon's orbit. The moon then departs from synchronism and spirals inward, further accelerating the planet's spin.

 For Mars and Nerio, a simple calculation demonstrates that Mars's rotation rate at the moment of Nerio's desynchronisation is remarkably close to the present-day rate.  This means that Nerio was destroyed {\it after} the desynchronisation, during its tidal descent towards Mars.\,\footnote{~We recall that after Nerio's demise, solar tides began working to slow Mars' spin. Therefore, at the moment of Nerio's destruction, Mars' rotation rate ought to be higher than it is at present.}

 We shall discuss whether Nerio disintegrated at the Roche radius or was destroyed by the LHB after desynchronisation but before reaching the Roche limit.

\subsection{A possible link to tidal rhythmites in Gale Crater and Vastitas Borealis}

{
While synchronous, Nerio was unable to generate tidalites within the Martian palaeo-ocean.  However, Nerio acquired this capability after it got desynchronised.
The post-desynchronisation spiralling-down phase became a window for tidalite generation in sediments.  This generation continued until Nerio either reached the Roche limit or was destroyed during the LHB.  After Nerio's demise, its large fragment(s) could have sustained this periodic deposition process. This mechanism is consistent with the hypothesis proposed by \citet{Das} and \citet{Sarkar}, who analysed the rhythmites in Gale Crater and deduced that they may have been generated by the tidal action of a massive lost moon.  Similar sedimentary structures were found by the Zhurong Rover in the Vastitas Borealis Formation \citep{Xiao}.
}

 \section{Chronology}

 \subsection{Origins}
 \label{birth}

 As explained above,  Nerio must have emerged during Mars' earliest history, when the newborn planet was sufficiently soft and deformable, about $4.56$ Gyr ago or shortly thereafter.
 Accretion {\it in situ} being a trivial option, a viable alternative to it could be capture in the remnants of the disk \citep{Hunten}. {
 A different option could be binary dissociation \citep{2006Natur.441..192A, Williams_and_Zugger}.
 }

 {
 A still different option could be accretion after an impact.
  Short of a giant impact, a moon comprising 3\% of the planet's mass could form in a
   kiss-and-capture event, a scenario wherein two bodies become temporarily connected, and upon detachment comprise a binary system \citep{Denton2025}.
 }

\begin{figure}[h]
 	\hspace*{-5.1mm}\includegraphics[width=1.1\columnwidth]{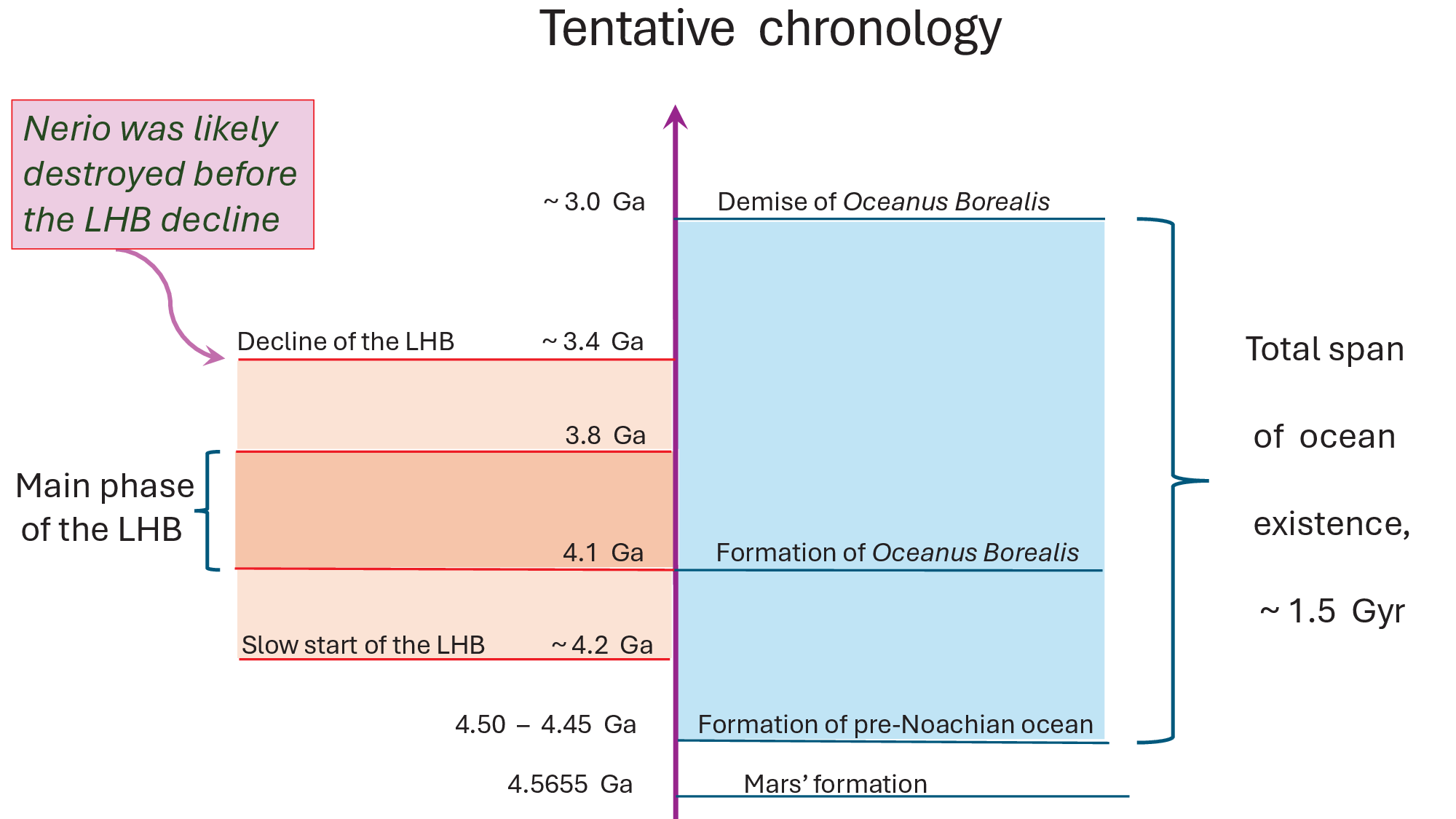}
   \caption{. \small{Tentative chronology of Mars.\vspace{0.7mm}\newline
   Based on meteorites' studies \citep{Borg, Zircon} and on the presence of extensive Fe/Mg deposits \citep{Andrews},  Mars had a liquid water ocean within its first 100 Myr\,---\,a conclusion confirmed by theoretical modelling of accretion \citep{Elkins}. Thus, a massive body
of water was likely present continuously from the pre-Noachian
through the Noachian, with additional exogenous water delivered by the LHB. \vspace{0.3mm}\newline
Initially constrained to between 4.1 and 3.8 Ga \citep{Tera, Gomes}, the LHB is now known to have  been spread over a much longer interval. Based on the impact record of the Moon, Zellner
(2017) demonstrated that the LHB began at 4.2 Ga and continued
till at least 3.4 Ga. Other studies, however, contend that the LHB tail experienced a more protracted decline, extending into much later history \citep{Hartmann,JohnsonMelosh,Bottke2012,Marchietal2014,LOWE201839}.
   }}
    \label{Chronology}
   \end{figure}

 \subsection{Environment}
 \label{life}

{
Over its lifetime, the Mars-Nerio pair was subject to two main influences that defined its dynamical evolution at large: solar tides on Mars and the Late Heavy Bombardment (LHB).  Hence, it is necessary to determine the intensity of Mars' tidal response and the LHB timing.
}

\subsubsection{Tides on Mars}

{
Tidal response of Mars comprises two components: that of the solid body and that of the ocean. Since each component exhibits a distinct frequency dependence, the planet's aggregate response can be highly complex. When an ocean is present, its contribution is dominant and may exceed that of the solid body by one to two orders of magnitude, depending on the depth and shape of the basin \citep{Auclair2018, Farhat, Auclair}.
}

{The concept of a pre-Noachian ocean dates back to the influential work by \citet{Borg}. Using the ages of secondary alteration minerals in Martian meteorites to obtain absolute ages when liquid water was at or near the surface of Mars, the authors concluded that Mars had a liquid water ocean within its first 100 Myr, i.e., 4.50 - 4.45 Gyr ago.}

{
Their conclusion was later supported by the modelling of \citet{Elkins} who demonstrated that terrestrial planets can retain, upon accretion, sufficient water to form oceans without addition from exogenous sources. Through collapse of their primordial atmosphere, these planets must produce water oceans within tens to hundreds of Myr after their last
major accretionary impact.
(See also \citeauthor{Russel} \citeyear{Russel}.)
}

{
The presence of liquid water in the pre-Noachian era
%  \footnote{~The pre-Noachian is the time period bounded by two defining events\,---\,the dichotomy-forming impact ($\sim 4.5$ Gyr ago) and the Hellas-forming impact ($\sim 4.1$ Gyr ago). }
was deduced also by  \citet{Andrews} based on the mineralogical record, specifically on the presence of extensive Fe/Mg deposits. Another confirmation of a wet pre-Noachian Martian crust was obtained by \citet{Zircon} who found that the impact-shocked zircon from the regolith breccia meteorite NWA7034 bears textural and chemical indicators of hydrothermal conditions on Mars during crystallisation 4.45 Gyr ago.  A similar conclusion was reached by \citet{Malarewicz}, who pointed out that quartz-bearing granitic clasts in the regolith breccia meteorite NWA7533 must have formed under the combined action of hydrothermal activity and impact melting.}

%  ~\\ REFERENCES:
% https://theconversation.com/a-4-45-billion-year-old-crystal-from-mars-reveals-the-planet-had-water-from-the-beginning-243172

% https://www.hou.usra.edu/meetings/earlymars2017/pdf/3078.pdf
% ~\\
{
Given the mineralogical and theoretical evidence for liquid water on early Mars, we adopt Borg and Drake's hypothesis of a pre-Noachian ocean.
A body of water was thus present continuously from the pre-Noachian through the Noachian, with additional exogenous water delivered by the Late Heavy Bombardment (LHB). Consequently, Oceanus Borealis can be viewed as a continuation of the primordial pre-Noachian ocean.}

{
Originally, it was believed that the main phase of {\it Oceanus Borealis} lasted over the Noachian period only, 4.1 - 3.7 Ga \citep{Clifford}. Recent studies, however,  provide evidence in favour of the ocean remaining stable until the end of the Hesperian period, around 3 Ga \citep{Schmidt}.}

{
 Taken together, these findings indicate that a palaeo-ocean existed on Mars from $\sim\!4.5$ Ga through $\sim\!3$ Ga, i.e., for about $1.5$~Gyr, see Figure \ref{Chronology}. Over this time span, Mars possessed a high value of the tidal ratio $k_2/Q\s$.}

 \subsubsection{Late Heavy Bombardment (LHB)}
 \label{LHB}

{
 The boundaries of the LHB\,---\,both its onset and its end\,---\,remain unresolved.   This interval was initially constrained to between 4.1 and 3.8 Ga. \citep{Tera, Gomes}.
 Later, however, it was demonstrated by \citet{Hartmann} that the LHB was declining gradually until at least $\sim 3.1$ Ga. Consistent with that point of view was the result by \citet{JohnsonMelosh} who used spherule layers on Earth as a source of impact chronology and found that at 3.5 Ga the impactor flux was significantly higher than now.  Analysing the age of craters on the Earth and Moon, and of the Earth spherule beds, \citet{Bottke2012} demonstrated that intense bombardment had continued for much longer. (Whether these authors effectively extended the duration of the LHB to 2.5 Ga or perhaps even 1.7~Ga is a matter of interpretation of the tail end.) Modelling the history of the Earth's Hadean crust,
    \citet{Marchietal2014} argued that massive bombardment of the Earth was quite early a process, a continuous flux spanning the entire Hadean {\AE}on, $\sim 4.56$ to $4.0$~Ga.
  \citet{LoweByerly2015, LOWE201839} pointed out that Earth experienced at least 17 major asteroid impacts between 3.5 and 2.5 Ga, implying that the LHB lasted until at least 2.5 Ga.}

{  Finally, based on the impact record of the Moon, \citet{Zellner2017} concluded that the LHB started at 4.2 Ga and continued till {\it at least} 3.4 Ga, see Figure \ref{Chronology}. Compared to the previously cited estimates, this is a moderately conservative value, which we shall adopt.
 }

  \subsection{Elimination}
 \label{elimination}

 For Nerio's presence to overlap with the era of crustal solidification on Mars\,---\,and thereby to allow the tidal bulge to become frozen into the planet's figure\,---\,its synchronous orbit would need to remain stable for a sufficient time.

  On the other hand, Nerio ought to have disintegrated  early enough, so the resulting  latitudinal distribution of craters could then be gardened out of recognition
  %  or, at least, smeared considerably
  by the LHB.
  In this way,
  % having accreted {\it in situ} simultaneously with Mars, or captured shortly after Mars' formation some 4.5 - 4.6 Gyr ago,
 the moon should have disappeared when the LHB was still intense.

{
As explained above, we adopt the LHB timeline of 4.2 - 3.4~Ga proposed by \citet{Zellner2017}.}

 While Nerio's  destruction can plausibly be attributed to the LHB, it is worthwhile to investigate {\it{inter alia}} if Nerio could have reached the Roche limit after its orbit was destabilised by solar tides acting to decelerate Mars' rotation.

 \section{The synchronous radius linked to the time of Nerio's orbital synchronisation\label{geoph}
 }

 {As shown in Figure \ref{picture}, Mars is rotating at a rate $\Omega$, which is equal to Nerio's mean motion $n$, but is much faster than the mean motion $n_M$ of the Mars-Nerio pair about the Sun.
 %  Therefore, the semidiurnal solar tidal bulge on Mars is leading.
 }

The synchronous radius $r$ relates to the planet's rotation rate $\Omega$ by an expression easily deducible from Kepler's Third Law:
\ba
 r^3\,=\,\frac{G\,(M + M_m)}{ \Omega^{\,2} }\,\;,
 \label{r}
 \ea
 $M$ and $M_m$ being the planet's and moon's masses, and $G$ being the gravitational constant.

 The centrifugal force can be expanded into components~---~one purely radial, another imitating the axially symmetrical part of a quadrupole tidal perturbation. The former component negligible for incompressible bodies, the latter produces a dynamical oblateness $J_2$. Hence a relation connecting $J_2$ with the rotation rate $\Omega$ and the quadrupole Love number (see, e.g. Appendix C to \citeauthor{Mars} \citeyear{Mars}):
 \ba
 \Omega^{\s 2}\s=\;\frac{3\, G\s M\s J_2}{R^{\s 3}\, k_2}\;\,,
 \label{bills}
 \ea
 $R$ being Mars' early radius  (different from the present radius by $\sim 0.6\s\%$ due to the Late Veneer and LHB, {see \citeauthor{Mars} \citeyear{Mars}}).

 In the former and latter formulae, $\Omega$ is the planet's spin rate established at the moment of synchronisation, i.e., equal to the mean motion of the moon at that time. In the absence of solar tides, the value $\Omega$ would then be sustained by the synchronous moon for as long as it existed. Consequently, $r$ would remain unchanged through the moon's lifetime. Solar tides, however, altered the dynamics by applying a torque to the planet. Since the torque was decelerating, one might expect it to have been mitigating Mars' spin.  This indeed is what happens to a planet in the absence of a synchronous moon. In its presence, however, the situation becomes more complex. Since Mars' rotation was locked to the Mars-Nerio mutual orbit, the solar torque's action on Mars was propagating into this orbit's shrinkage, i.e., into a slow decrease in $r$. Therefore, quite counterintuitively, the solar tides were forcing the synchronous state to adiabatically evolve in the direction of a slow {\bf{increase}} in $\Omega$. Here, {\it{slow}} means that the evolution of the synchronous state was {\it much slower} than the process of Mars' figure solidification.\,\footnote{~On the one hand, magma-ocean crystallisation and primordial-crust formation on Mars were extremely rapid, occurring in less than 20~Myr.
 %  Even if the crust were reworked by impacts $\sim\s100$ Myr later (as proposed by \citeauthor{Magma} \citeyear{Magma}), it remains safe to assume that the Martian figure fossilised within several hundred Myr after the planet's accretion~---~certainly well within 0.5 Gyr.
 \newline
   $\phantom{In}$ On the other hand, we shall see in Section \ref{questions} that prior to the formation of a palaeo-ocean on Mars, the tidal response of the Mars-Nerio system to solar tides was weak, rendering the increase in $\Omega$ and the ensuing possible change in $J_2$ negligible. The system's response to solar tides was boosted by the ocean, but this enhancement occurred when Mars' figure was already shaped.\label{Footnote}}
 This enables us to identify the value of $\Omega$ at the time of Mars' figure solidification with the value of $\Omega$ at the time of Nerio's synchronisation.

 Combining equations (\ref{r}) and (\ref{bills}), we express the synchronous radius' early value $r$ through the early values of the Love number and oblateness:
\bs
 \ba
 \frac{r}{R}\,=\,\left(\frac{k_2}{3\s J_2}\;\frac{M+M_m}{M}  \right)^{\,1/3}\;\,.
 \label{int}
 \ea

 Nerio's life time spanned from $\s 4.56 - 4.5\,$ through, at most, $\,3.4\s$ Gyr ago {(see Section \ref{1.1})}. Since the Martian figure solidified before Nerio's demise, it is reasonable to approximate the then $J_2$ of Mars with its present value {taken from the table in Appendix \ref{description}}: $J_2\approx\s J_2^{\rm (present)}
   = 1.9566\times 10^{-3}
$. Also be mindful that $J_2$ enters the above expression raised to the power of $1/3$, which makes $r$ less sensitive to a possible uncertainty in the value of $J_2\s$.
 %  For $M_m/M \ll 1$, the value of $r/R$ is not very sensitive to that of $M_m$. In Section \ref{limitations}, we shall find that $M_m$ was close to $3\times 10^{-2}\s M\s$.
 Formula (\ref{int}) therefore becomes:
 \ba
 \frac{r}{R}\;\approx\;5.54\,\left(1\,+\;\frac{M_m}{M}  \right)^{1/3}\!k_2^{\s 1/3}\,\;.
 \label{inta}
 \ea
 \label{inn}
 \es
 Through this expression, the synchronous radius at the time of the moon's synchronisation is linked to the quadrupole Love number value at that time.

 The title of this subsection indicates the necessity to express the planet's synchronous radius and rotation rate as functions of the time of the moon's orbital synchronisation. This could be accomplished by using equations (\ref{inta}) and (\ref{r}), were the time dependence $k_2(t)$ known.  A search for this time-dependence would, however, amount to a major project focused on modeling the evolution of a {\it nonspherical} nascent planet. Taking into account the asymmetric shape, the model should specifically account for the change of rheology during the cooling of the young Mars, a process involving in particular the solidification of an early magma ocean, and the formation and thickening of Mars' stagnant lid. To sidestep this major work~---~and still obtain physically meaningful estimates~---~the timing question can be tackled through a simpler parameterisation of the solidification process. Following \citet{Mars}, we are using, instead of time, the evolving mean rigidity $\mu$ as a parameter through which to express the Love number, $k_2(\mu)$. This approach is sound, because the values of $\mu$ can be reasonably related to the stages of Mars' early history. The mean rigidity grew from $\approx 0.20$ GPa, at the end of the magma-ocean stage, to $\approx 18$ GPa at the stage of solidification, see {\it Ibid.} and references therein:
 \ba
 0.20~\mbox{GPa}\,\lesssim\,\mu\,\lesssim\,18~\mbox{GPa}\,\;.
 \label{mu}
 \ea

 {\it Under static loading}, the Love number is defined by the mean shear rigidity $\mu$ via {the formula pioneered  by \citet[p.2]{Thomson}:}
  \footnote{~{According to the
  % (very self-referential)
  {\it Arnold principle}, if a notion bears a personal name, then this name is not the name of the discoverer. The principle
   %  is self-referential, i.e.,
  applies to itself~---~for in reality it was formulated by Michael Berry \citep{Arnold}.
  \newline
  $\phantom{In}$ In humanities, this principle is known as {\it Stigler's law of eponymy} and also is self-referential, because it was authored by Robert K. Merton \citep{Stigler}.}}
\ba
 k_2\s=\,\frac{3}{2}\;\frac{1}{1\s+\s{\cal B}_2\,\mu}\,\;,
 \label{kl}
 \ea
 where
\ba
 {\cal B}_2\,=\,\frac{57}{8\,\pi\, G\,(\rho\, R)^2}\,=\,0.193\times 10^{-9}\;\mbox{kg}^{-1}\,\mbox{m}~\mbox{s}^{2}\;\;,
 \label{Bl}
 \ea
 $\rho$ being the young Mars' mean density.

 Substituting the bounds (\ref{mu}) on $\mu$ into equation (\ref{kl}), and inserting the result into expression (\ref{inta}), shows that {for $M_m \simeq 0.03 M$} the values of the synchronous radius are constrained to the interval
  \ba
  3.89  \,\lesssim\,  \frac{r}{R}  \,\lesssim\,
  %  6.26  <-- for M_m = 0
  6.32
  \;\,.
 \label{double}
 \ea

 The upper bound on $ r/R$ is appropriate, via equation (\ref{int}), to the upper bound on $ k_2$ and therefore, through equation (\ref{kl}), to the lower value $\mu\approx 0.20\s$~GPa~---~which, in its turn, corresponds to a situation where Nerio became synchronous and shaped Mars' figure already at the magma-ocean stage.

 The lower bound on $r/R$ is, on the other hand, defined by the lower bound on $k_2$ or, equivalently, by the upper value $\mu\approx 18$ GPa~---~which is the case of Nerio synchronising its orbit and shaping the planet at the time when Mars' solidification was already going on.

 It should be noted that the lower bound on $r/R$ is well above the Roche limit, which in this case is $2.32\s R$ \citep[Section 7.1]{Mars}.

 In Section \ref{limitations}, we shall see that a different, purely dynamical reasoning imposes a lower bound on $r/R$ more stringent than in equation~(\ref{double}).
 % {The upper bound on $r/R$ also will be refined, but only minusculely.}\,\footnote{~{While estimate (\ref{double}) was obtained for $M_m \simeq 3\s\%\s\m M$, we shall derive in Section \ref{limitations}  a more accurate constraint $M_m < 3.23\s\%\s\m M$, which will provide via formula (\ref{inta}) a more accurate upper bound $r/R \lesssim 6.33$, the difference with estimate (\ref{double}) being in the second decimal place only.}\label{Footnote_on_constraint}}

   \begin{figure}[h]
 	\hspace*{-9.6mm}\includegraphics[width=1.12\columnwidth]{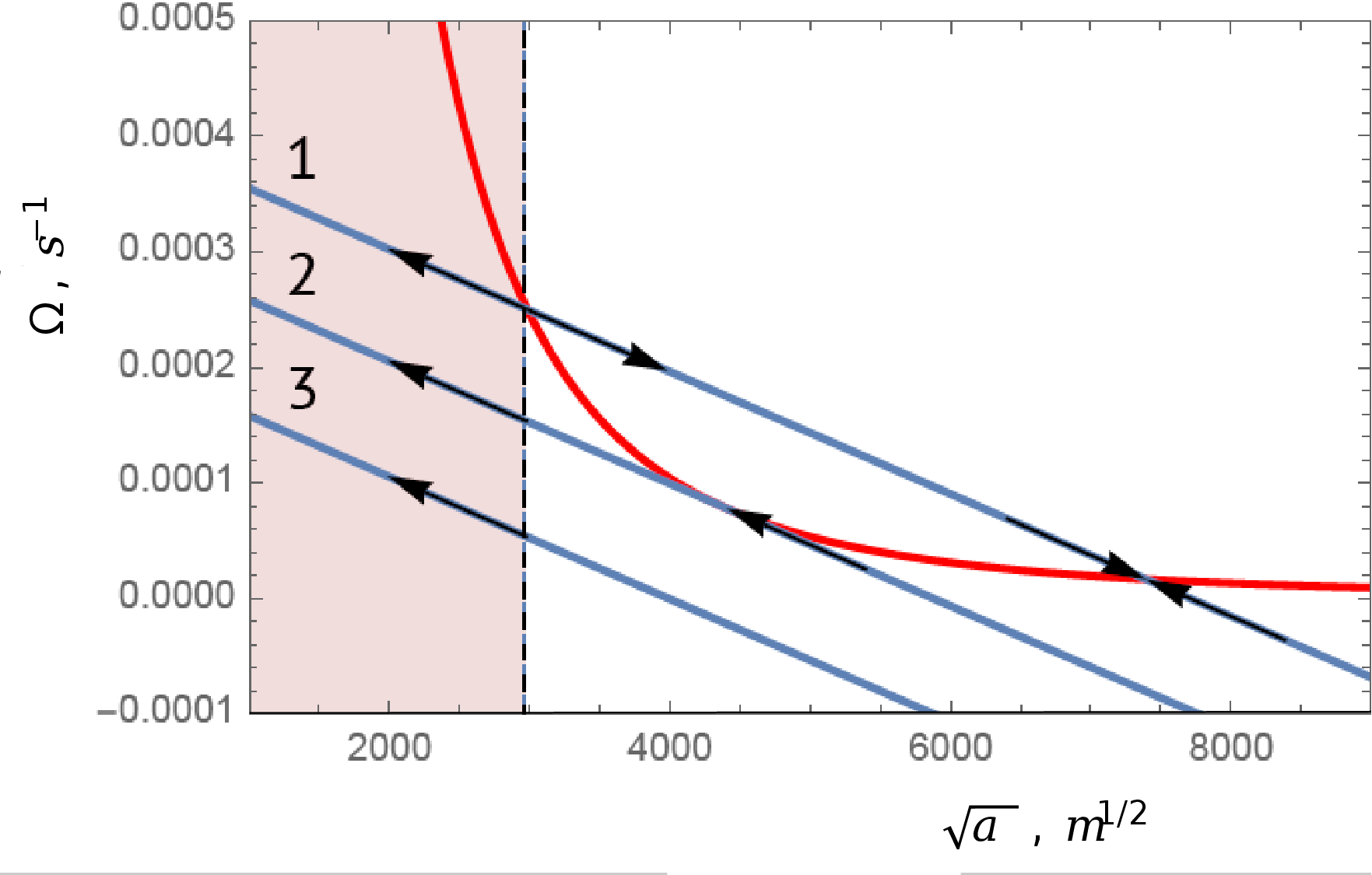}
   \caption{. ~\small The blue lines {\bf 1}, {\bf 2}, {\bf 3} depict different dynamical tracks given by equation (\ref{5}) for different sets of initial conditions $\{a(t_0)\s,\,\Omega_0\}$, with the mass values $M$ and $M_m$ fixed. The red cubic hyperbola comprises, for fixed $M$ and $M_m$,
   all synchronous states given by equation (\ref{6}). Track {\bf 1} crosses the red curve in two points, only one of which is a stable synchronous state, see equation (\ref{insynchro2}) and the paragraph after it. Track {\bf 2} is tangent to the red curve, and shares only one point with it. As demonstrated in Section \ref{solar} and Figure \ref{Figure_2}, this synchronous state is unstable under a perturbation generated by solar tides. Track {\bf 3} originates from initial conditions excluding the possibility of synchronism. The vertical dashed line and the pink shaded area indicate the Roche limit.
   If the moon is not destroyed by the LHB prior to crossing that limit, it gets disintegrated there by tidal forces.
   }
    \label{Figure_1}
   \end{figure}

 \section{(In){\,}stability of synchronous orbits}
\label{instabi}

 \subsection{Is a synchronous orbit always stable?}

 For the first time, this question was addressed yet by \citet{Darwin1879}. Later, it was tackled by \citet{Counselman} and \citet{Hut}, whose analysis was probably more mathematically complex than necessary.

 To gain intuition, we first approach the problem in elementary terms.

 The orbital angular momentum \citep{book},
 \ba
 H_{\rm orb} = \, \frac{M\s M_m}{M +\s M_m}\,\sqrt{G\s (M +\s M_m )\,a\,(1\,-\,e^2)\,}\;\,,
 \ea
 is proportional to a positive power of ${a}$.  Therefore, an infinitesimal increase $\delta a > 0$ of the semimajor axis entails an increase $\delta H_{\rm orb} > 0$.  Owing to the conservation law, this yields a decrease in the spin angular momentum of the planet, and therefore a negative change $\delta\Omega < 0$ of its rotation rate.
 The synchronous radius
  \ba
  r\s=\s\left[\frac{\textstyle G\s (M+M_m)}{\textstyle \Omega^{\s 2}}\right]^{\s 1/3}
  \label{sy}
  \ea
  of the planet scales as $\Omega$ to a negative power. Hence $\delta\Omega < 0$ produces an increase in the synchronous radius, $\delta r > 0$. We  observe that the infinitesimal variations of $a$ and $r$ have the same sign:
 \ba
 \delta a > 0\;~\Longrightarrow\;~
 \delta H_{\rm orb} > 0\;~\Longrightarrow\;~
 \delta \Omega < 0\;~\Longrightarrow\;~
 \delta r > 0\;\,.
 \label{}
 \ea
 The configuration is stable if $\delta r > \delta a$, because this inequality ensures that an infinitesimal increase in $a$ leads to a subsequent infinitesimal tidal descent.

 Alternatively, if we reverse all signs,
 \ba
 \delta a < 0\;~\Longrightarrow\;~
 \delta H_{\rm orb} < 0\;~\Longrightarrow\;~
 \delta \Omega > 0\;~\Longrightarrow\;~
 \delta r < 0\;\,,
 \label{}
 \ea
 the stability condition implies that a negative $\delta a$ must produce a negative $\delta r$ of a larger absolute value, so the moon finds itself above synchronism and performs an infinitesimal tidal ascent.

 It can be shown that for fixed values of the primary and secondary masses, a tidal binary can have up to two synchronous spin-orbit configurations. When these configurations are two, one is stable, the other is not. This was the conclusion to which  \citet{Darwin1879}, \citet{Counselman} and \citet{Hut} arrived by various means.  We here prefer to link these two possible outcomes to the initial conditions, as was done in \citet{synchronisation}.

 \subsection{A stability criterion\label{Formalism}}

 For a planar configuration with a small eccentricity, smaall obliquities, and a negligible spin angular momentum of the secondary, the angular momentum conservation law yields \citep[Appendix D]{Pathways}:
 \ba
 \frac{\stackrel{\bf\centerdot}{\Omega}}{\stackrel{\bf\centerdot}{n}}
 \,=\,\frac{1}{3\,\xi}\,\frac{M_m}{M+ M_m}
 \,
 \left(\frac{a}{R}\right)^2+\,O(e^2)\;\,,
 \label{3ksi.eq}
 \ea
 $\xi$ being the moment of inertia (MOI) factor of the primary mass,
 and $n$ being the mean motion.

 Integrating this equation in neglect of $O(e^2)$, performed in \citet[Section 2]{synchronisation}, produces
 \ba
 \Omega\s=\s-\s X \sqrt{ a}\,+\,C\,\;,
 \label{5}
 \label{Omega}
 \ea
 with the integration constant given by
 \ba
 C\,=\,X\sqrt{a(t_0)\s}+\s\Omega(t_0)
 \label{C}
 \ea
 and the slope equal to
 \ba
 X\s=\s\frac{M_m}{\xi}\,\sqrt{\frac{G}{\s M\s +\s M_m\s}\s}\s R^{\s -2}
 \;\,.
 \label{X}
 \label{4}
 \ea
 The value of the slope $X$ parameterises the angular-momentum exchange rate between the planet's spin and moon's orbit.

 We observe that for a small eccentricity and small obliquities, and in neglect of the spin angular momentum of the moon, each dynamical history makes a linear function in the $(\!\sqrt{a}\s,\,\Omega)$ coordinates.

The synchronicity condition $\Omega = n$ can be written, {using equation (\ref{r}),} as
 \ba
 \Omega=\!\sqrt{G\,(M\s +\s M_m)\,}\,(\!\sqrt{a})^{-3}\;,
 \label{6}
 \label{OmegaS}
 \ea
 thus producing a cubic hyperbola in the $(\!\sqrt{a}\s,\,\Omega)$ coordinates.

 These observations are illustrated in Figure~\ref{Figure_1} where each inclined blue line implements a dynamical history (\ref{5}) defined by some initial condition $\s(\!\sqrt{a(t_0)}\s,\,\Omega(t_0)\,)\s$. The red curve is the cubic hyperbola (\ref{6}).  The figure explains why for a fixed pair of mass values, $M$ and $M_m$, the number of available synchronous orbits can be zero, or one, or two\,---\,depending on the initial conditions.

As we mentioned above,
%  \citet{Darwin1879}, \citet{Counselman} and \citet{Hut}
several authors had
established, by different methods, that whenever two synchronous states are available, one is stable, another not. We, however, favour a still different approach, which is not only straightforward but also more general. It comprises two steps. First, approximate the tidal rate $da/dt$ of the semimajor axis with its quadrupole semidiurnal part $\s\left({da}/{dt}\right)^{\rm (pqs)}$ caused by the tides in the planet only.\,\footnote{The superscript \s``$\rm(pqs)$''\s means: {\it planet, quadrupole, semidiurnal}.}  Second, take the full (not partial) derivative of $\s\left({da}/{dt}\right)^{\rm (pqs)}$  over $a$, in the synchronous state, i.e., for  $n=\Omega$. This yields \citep[Appendix A]{synchronisation}:
 \bs
 \ba
 \frac{d}{da}
 \left(\frac{da}{dt}\right)^{\rm (pqs)}
   _{\,n={\Omega\s}}=
 9\frac{\;M_m}{M\,}\left(\frac{R}{a}\right)^5 n^2
 \left(1-\frac{d{\Omega}}{dn}  \right)\frac{dK_2(\omega)}{d\omega}\s
 \Bigg{|}_{\,\omega=0}
 \;\,\;,
 \label{insynchro}
 \ea
 where $K_2$ is the quality function, and the shortened notation $\omega\equiv\omega_{2200}=2(n-{\Omega}_p)\s$ is introduced. For realistic rheologies, ${\textstyle dK_2(\omega)}/{\textstyle d\omega}$ at the point $\omega=0$ is positive, as was illustrated by the case of a Maxwell planet in {\it Ibid}. The derivative ${d{\Omega}}/{dn} $ is given by equation (\ref{3ksi.eq}) above.

  A synchronous spin-orbit state is stable (unstable) when the derivative $ \frac{\textstyle d}{\textstyle da}
 \left(\frac{\textstyle da}{\textstyle dt}\right)^{\rm (pqs)}
   _{\,n={\Omega\s}} $
 is negative (positive). Hence the stability criterion deduced in
{\it{Ibid.}}:
 % \citet{synchronisation}:
 \begin{itemize}
 \vspace{1mm}
 \item[] The synchronism is stable if $\s{d\Omega}/{dn}>1\s$;\\ unstable otherwise.
 \vspace{1mm}
 \end{itemize}

\noindent
 The criterion indicates that for a straight line intersecting the cubic curve in Figure
  %   1,
   \ref{Figure_1},
 the left crossing point is unstable, the right is~stable. To understand the reason for this, rewrite expression (\ref{insynchro}) as
 \ba
 \frac{d}{da}
 \left(\frac{da}{dt}\right)^{\rm (pqs)}
   _{\,n={\Omega}}=
 9\frac{\,M_m}{M}\left(\frac{R}{a}\right)^5 \!\!n^2
 \left(1-\frac{d{\Omega}/d\sqrt{a}}{dn/d\sqrt{a}}  \right)\frac{dK_2(\omega)}{d\omega}
 \Bigg{|}_{\omega=0}  \s
 \label{insynchro2}
 \ea
 \es
 and recall that ${d{\Omega}/d\sqrt{a}}$ and ${dn/d\sqrt{a}}$ are the slopes of the straight line and the cubic curve, correspondingly (both slopes being negative). At the left crossing point in Figure \ref{Figure_1}, the absolute value of the slope of the linear function, $|\s {d{\Omega}/d\! \sqrt{a}}\s|$, is larger than that of the cubic curve, $|\s{dn/d\!\sqrt{a}}\s|$, so the state is unstable. At the right crossing point, the situation is opposite; hence the stability of that state
%
% Thus, for a dynamical track permitting two synchronous states, i.e., intersecting the cubic hyperbola in two points,  the state on the left is always unstable, the one on the right always stable
(assuming that the orbit is located between the Roche and reduced Hill radii).
%  (, and that the tides in the moon may be neglected).

 This criterion from {\it Ibid.} is more generic than the techniques proposed
 %  different forms by \citet{Darwin1879}, \citet{Counselman} and \citet{Hut},
 in the literature until then, because the stability analysis based on examining the sign of the derivative $ \frac{\textstyle d}{\textstyle da} \left(\frac{\textstyle da}{\textstyle dt}\right)
  % ^{\rm (pqs)}
 _{\,n={\Omega\s}} $
 can be applied to settings incorporating finite eccentricity and obliquities, higher-than-quadrupole terms, tides in the moon, as well as the planet's nonsphericity, the Sun's gravitational pull, and other perturbations. Within this approach, one simply begin with an expression for $\,\left(\frac{\textstyle da}{\textstyle dt}\right)
  %  ^{\rm (pqs)}
   _{\,n={\Omega\s}}$ written down to any required precision in the powers or $e$ and $i$, and with any additional physical effects included. A subsequent turn-of-the-crank differentiation over $a$ provides the result. Furthermore, this method is equally effective for analysing the stability of higher-order spin-orbit resonances, as was demonstrated in Section 4 of
   %  {\it Ibid.}
     \citet{synchronisation}
   where the stability of the 3:2 spin-orbit state was addressed.

   \begin{figure}[h]
 	\hspace*{-1mm}\includegraphics[width=1.12\columnwidth]{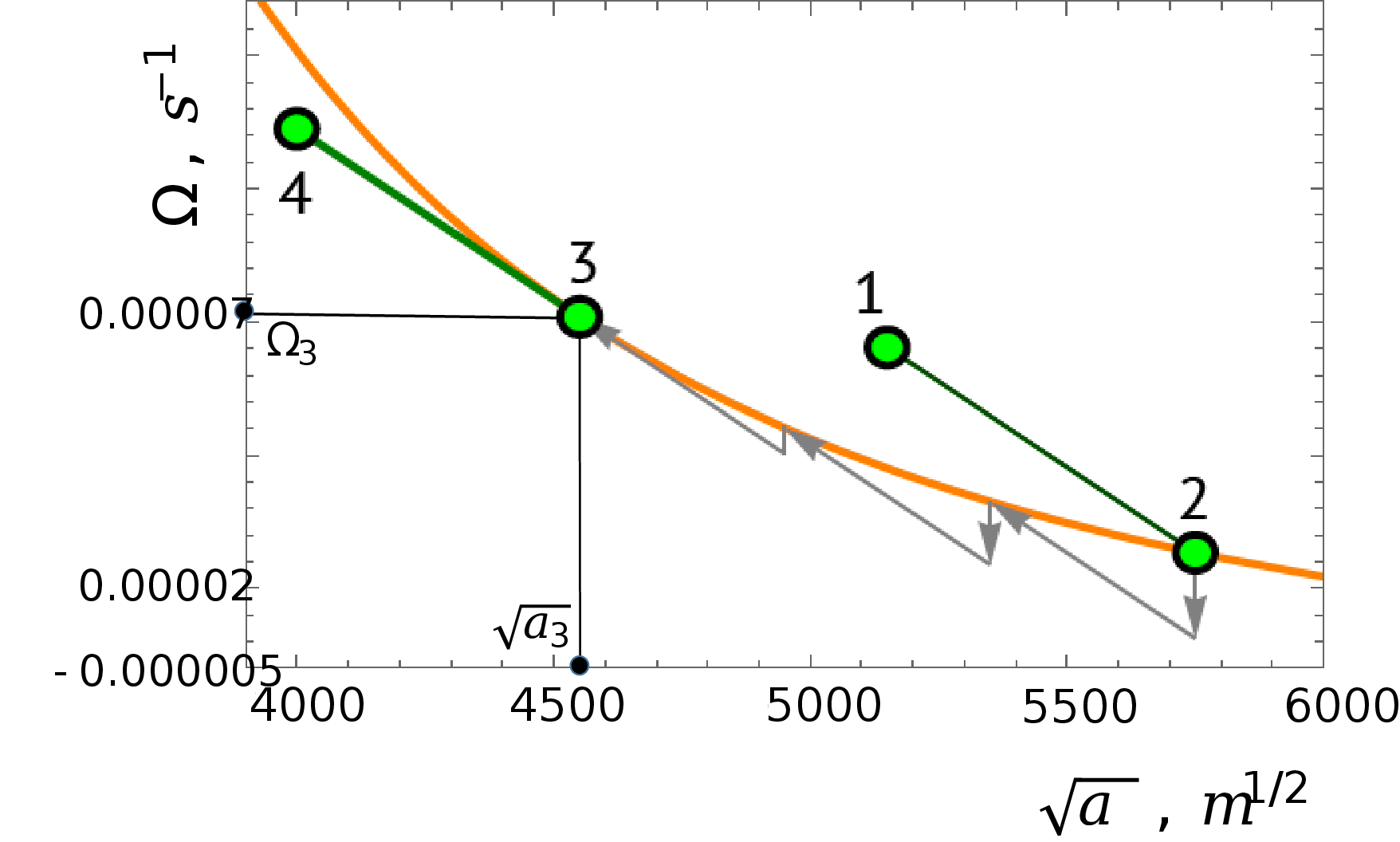}
   \caption{. \small
    An instantaneous action of solar tides entails an infinitesimal reduction of the planet's rotation rate, as shown by a short vertical arrow pointed downwards from the stable point of synchronisation 2. This action pushes the system to a different dynamical history terminating at a synchronous state with a slightly smaller $a$ and a slightly higher $\Omega=n$.  After numerous pushes, the system arrives at the marginal state 3. That state is unstable, and an infinitesimal action by the solar tides in that state finally sets the system on its final trajectory along which it starts moving towards the Roche limit, point 4. The green disks and connecting lines show a plausible evolution scenario: (1) Initial state where Mars rotates faster than the orbital motion. (2) Capture of Mars into a long-term synchronous equilibrium. (3) Critical point attained due to the action of solar tides, and Mars is released from the resonance. (4) Satellite migrates towards the Roche limit, and is likely destroyed by the LHB {\it en route}.\newline
    As an aside (not relevant to Mars, but of general interest),  $\Omega$ may in principle have a negative sign~---~corresponding to rotation retrograde w.r.t. the mean motion of the moon. In that setting, a moon undergoing inward tidal migration can reverse the planet's rotation, depending on the initial conditions and the mass ratio    \citep{2023Univ...10...15M}.
   }
   \label{Figure_2}
   \end{figure}

~\\

 \section{Solar tides on the planet make\\ the marginal  state unstable} \label{solar}\label{marg}

 Weak yet relentless, solar tides are adiabatically driving an initially stable spin state toward instability \citep{Ward, synchronisation}.
  In Figure \ref{Figure_2}, each short downward-pointing vertical arrow shows how solar tides slightly slow down the planet's rotation rate over a brief period. This infinitesimal action pushes the system towards a different dynamical history, a straight line parallel to the initial dynamical track and located slightly to the left of it. Now residing on this new trajectory, the system moves along it towards the new intersection with the cubic hyperbola. This new equilibrium is located on the hyperbola slightly to the left of the preceding equilibrium state. This shifted equilibrium corresponds to a slightly higher value of $\Omega$ and a slightly smaller synchronous radius $r$.

   Slowly but steadily, the state of the system is migrating leftwards along the cubic hyperbola, over a set of transiently stable synchronous states. Were it not for the synchronous moon, solar tides would be working to slow down Mars' rotation. The presence of a synchronous partner, however, results in a slow shrinking of the orbit and therefore in a slow spin-up of Mars. The system eventually arrives at the marginal state depicted by point 3 in Figure \ref{Figure_2}. In Figure \ref{Figure_1}, this borderline state is a point where the blue line 2 is tangent to the cubic curve.

 %  \section{The marginal point}
 %  As we saw in Figure~\ref{Figure_1},  up to two synchronous states are possible. The twain meet in the marginal point, one  where line 2 is tangent to the cubic hyperbola.
 The borderline spin state is transiently stable under an infinitesimal increase in the planet's spin rate $\Omega\,$---\,but is unstable under a decrease, because a decrease in $\Omega$ shifts the system to an adjacent trajectory (line 3 in Figure \ref{Figure_1}) lacking intersection with the cubic curve. Along that new trajectory, the system starts moving leftwards to the Roche radius.
 Solar tides thus are rendering the marginal state unstable.

 In the marginal state, the straight line depicting a physical history is tangent to the cubic curve. Consequently, their slopes are equal:
 \ba
 \frac{dn}{d\sqrt{a}}  \,=\, \frac{d\Omega}{d\sqrt{a}}\qquad\Longleftrightarrow\qquad  \frac{ d\Omega }{ dn }\;  =  \;1\;\,.
 \label{}
 \ea
 Combining this with equation (\ref{3ksi.eq}), we find that the marginal point corresponds to a synchronous orbit of the radius $r$ satisfying
  \begin{eqnarray}
 \frac{1}{3\,\xi}\,\frac{M_m}{M+ M_m}\,\left(\frac{r}{R}\right)^2\s=\,1\quad\Longleftrightarrow\quad
 \frac{r}{R}\,=\,\sqrt{ 3\,\xi\,\frac{M+ M_m}{M_m} }
 \;\,.
 \label{marginal1}
  \label{marginal}
 \end{eqnarray}
 {For $M_m\simeq 0.03\s M$ and $\xi\simeq 0.36$, the resulting values of $r$ come out to be $\simeq 6 R$, which is well above the Roche limit $r_{\rm R} = 2.32\s R\s$.}

 % \textcolor{red}{REMOVE}
 % ~~~{For $3\s\xi=1$ and $\s M_m=\s 0.0323\s M\s$, this becomes}
 %  {
 %  \begin{eqnarray}
 %  \frac{r}{R}\,=\,5.56 \;\,.
 %  \label{marginal2}
 %  \end{eqnarray}
 %  \label{marginal}
 %  }
 %  {This value safely falls within the interval (\ref{double}), being closer to its upper boundary, one corresponding to a higher $k_2$ and, consequently, a lower $\mu$. Indeed, from equation (\ref{inta}) we have
 %  } {
 %  \ba
 %  k_2\s=\,\left( \frac{\;r/R\;}{5.56}  \right)^3=\,1.01\;\,,
 %  \label{}
 %  \ea
 %  }  {
 %  whereafter formulae (\ref{kl}) and (\ref{Bl}) give us the value
 %  \ba
 %  \mu = \frac{\;\frac{\textstyle 3}{\textstyle 2\,k_2} - 1\;}{{\cal B}_2}\,=\,2.49~\mbox{GPa}
  %  \label{}
 %  \ea
 %  } {
 %  corresponding to a transition from magma ocean to solidification.
 %  }

\section{Constraints on the parameters of the moon}
\label{limitations}

 From equation (\ref{marginal1}) and the stability criterion $d\Omega/dn > 1$ obtained in Section \ref{Formalism}, we deduce that the initial synchronous state of the moon must satisfy the condition
 \ba
 \frac{r}{R} > \sqrt{3\s\xi\;\frac{M+M_m}{M_m}}
 %  \,\approx\,\sqrt{3\s\xi\;\frac{M}{M_m}}
 \;\,.
 \label{23}
 \ea
 {The fulfilment of this condition} ensures that the synchronisation initially takes place at a point residing on the stable branch of the cubic curve in Figure \ref{Figure_2} (point 2 in that plot).

 Combined with the expression (\ref{int}) for $r/R$, inequality (\ref{23}) produces
 %   \ba
 %   \left(\frac{k_2}{3\s J_2}\;\frac{M+M_m}{M}  \right)^{\,1/3} > \sqrt{3\s\xi\;\frac{M+M_m}{M_m}}\;\,.
 %   \label{}
 %   \ea
 %  or, equivalently,
   \ba
   3\s\xi\s\left(1\,+\,\frac{M_m}{M}  \right)^{1/3}\s\left(\frac{3\s J_2}{k_2}\right)^{2/3}<\;\frac{M_m}{M}\;\,.
   \label{inequality}
   \ea

 In \citet[Eqn 21]{Mars}, we demonstrated that
\ba
\frac{M_m}{M}\,\approx\,\frac{J_{22}}{J_2}\;\,,
\label{expr}
\ea
where $J_{2}$ and $J_{22}$ are Mars' oblateness and triaxiality at the synchronisation time.

Above, in Section \ref{geoph}, we noted  that the then value of $J_2$ was not very different from its present value,
 %  $\s J_2\,\approx\,J_2^{\rm(present)}\s$,
 \ba
 J_2\,\approx\,J_2^{\rm(present)}\;,
 \label{Hegel}
 \label{J}
 \ea
because the synchronous moon was, during its existence, sustaining the planet's spin rate and therefore the planet's shape\,---\,until the shape fossilised.
 %  If the early fossilisation was not complete, and slight residual adjustment of shape to the tidally decelerating spin continued for some time, the said approximation should be given a more conservative form:
 %  \ba
 %  J_2\,\gtrsim \,J_2^{\s\rm(present)} %  =\s 1.9566\times 10^{-3}\;.
 %  \label{Hegel}
 %  \label{J}
 %  \ea

By distinction, the value of $J_{22}$ may have slightly increased since then, because the Nerio-generated seed triaxiality could have produced a thinner crust in the submoon and antimoon zones, and caused higher tectonic activity in those two zones, leading to greater elevations and, as a result of that, to a further increase in $J_{22}$. Therefore,
\ba
J_{22}
\lesssim
J_{22}^{\s\rm(present)}
%  =\s 6.3106 \times 10^{-6}
\;.
\label{29}
\label{Plato}
\ea
Owing to equations (\ref{Hegel}) and (\ref{Plato}), expression (\ref{expr}) becomes:
\ba
\frac{M_m}{M}\,\lesssim\,\left(\frac{J_{22}}{J_2}\right)^{\rm (present)}
\s_{\textstyle{_{\textstyle _.}}}
\label{up}
\ea
In combination with inequality (\ref{23}), it defines the interval
 \ba
  3\s\xi\s\left(1\,+\,\frac{M_m}{M} \right)^{1/3}\left(\frac{3\s J_2}{k_2}\right)^{2/3}<\,
 \frac{M_m}{M}\,<\,\left(\frac{J_{22}}{J_2}\right)^{\rm (present)}\;\,.
  \label{doubleineq}
 \ea
  %  Thus the above constraint becomes
  %  \ba
  %  1.0914\s\left(1\,+\,\frac{M_m}{M} \right)^{1/3}\left(\frac{3\s J_2}{k_2}\right)^{2/3}<\,
  %  \frac{M_m}{M}\,<\,\left(\frac{J_{22}}{J_2}\right)^{\rm (present)}\;\,.
  %  \label{}
  %  \ea
Together with formulae (\ref{inn}) and (\ref{kl} - \ref{Bl}), this double inequality  enables us to constrain the values of $\mu$, $k_2$, and $r/R$ at the time of synchronisation, and then to write down a slightly more accurate constraint on $M_m/M$ itself. Derived in Appendix \ref{Appendix1}, these constraints read:
 \ba
 1.17\,< \,k_2\,<\,1.44\,\;,
 \label{refinedk2}
 \ea
 \ba
 1.45~\mbox{GPa}\,>\,\mu\,>\,0.20~\mbox{GPa}\,\;,
 \label{refinedmu}
 \ea
 \ba
 5.91  \,<\,  \frac{r}{R}  \,<\, 6.32 \;\,.
 \label{refinedr}
 \ea

 The right-hand bounds in these inequalities are dictated by geophysical considerations provided in Section~\ref{geoph},  equations (\ref{mu}), (\ref{kl}), and (\ref{double}).
 %\,\footnote{~{The upper bound on $r/R$ in equation (\ref{refinedr}) differs in the second decimal place from that in estimate (\ref{double}), because the upper bound in equation (\ref{refinedr}) is derived not from a crude estimate $M_m\lesssim 3\m\%\s M$, but from a more accurate limitation $M_m < 3.23\m\%\s M$, see equation (\ref{refinedM}).}}
 Given the low value of $\mu$, these bounds relate to a situation where the moon synchronised Mars at an early time (possibly, before the end of the magma ocean stage).

 The left-hand bounds in the above inequalities stem from the synchronous-state stability condition, equation (\ref{23}). Incorporating, specifically, the higher value of $\mu$ in equation (\ref{refinedmu}), these limits correspond to a scenario where the moon synchronised Mars at the beginning of solidification when the crust was already present or forming.  This timing, however, is  hypothetical, and its confirmation requires detailed modelling.

 The resulting constraint on Nerio's mass is:
  \ba
  3.23\times 10^{-2}
   % \gtrsim 3.225\times 10^{-2}
   \gtrsim
  \frac{M_m}{M}\,\gtrsim\, 2.81\times 10^{-2}\;\,.
 \label{refinedM}
 \ea
 Here, like in inequalities (\ref{refinedk2} - \ref{refinedr}), the right-hand bound
 %  , $2.81\times~10^{-2}$,
 corresponds
 (via equation \ref{inequality}) to the maximum value of $k_2$ and (through equations \ref{mu} and \ref{kl}) to the minimum $\mu$.
 The left-hand bound
 %  $3.225\times 10^{-2}$ corresponds in a similar way to the minimal value of $k_2$ and the maximal $\mu$.
 %  We have deliberately abstained from rounding it, in order to keep it separate from an almost identical bound
 % , $3.23\times 10^{-2}$,
 originates from %  very different considerations,
 the dynamical equation (\ref{up}).  %  With rounding, the two upper-bound values coincide, despite their different provenance.

 \section{Did Nerio reach the marginal point?\\
 Did it reach the Roche limit?}
 \label{questions}

 In agreement with equation (\ref{refinedM}), we set the mass of Nerio to be
 $$
 M_m= 0.03\s M =  1.89\times 10^{22}\,\operatorname{kg}\,\;.
 $$
 % Assuming differentiation to have  finished or almost finished by the end of magma-ocean stage, the MOI coefficient of the young Mars after that stage ought to be close to or only very slightly higher than its present value provided in Table~\ref{description}. In this calculation, we adopt a slightly higher value of $\xi=0.37\s$.
 The semimajor axis $a_3$ at the marginal point 3 in Figure \ref{Figure_2} is the radius $r$ of the synchronous orbit at that point, and is given by formula (\ref{marginal1}).  For  $M_m= 0.03\s M\s$, it is
 \ba
 a_3 =  2.0629\times 10^7~ \mbox{m}\;.
 \label{a3}
 \ea
 Its insertion into expression (\ref{OmegaS}) gives us the rotation rate of Mars at that moment of time:
 \ba
 \Omega_3\,=\,7.0294\times 10^{-5}~\mbox{rad}~\mbox{s}^{-1}\,\;.
 \label{Omega3}
 \ea
 The current rate of Mars' rotation is
 \ba
 \Omega^{\rm{(present)}}=\,
 7.088
 %  0.6556
 \times 10^{-5}~\mbox{rad}~\mbox{s}^{-1}\,\;.
 \label{Omegacurrent}
 \ea
 %  only a tiny bit faster than $\Omega_3$.
 Given the uncertainties in Nerio's mass and in Mars' radius prior to the LHB, the fourth, third, and possibly even the second decimal places in the values of $a_3$ and $\Omega_3$ should be interpreted with caution, see  Appendix \ref{E1}.  So, {\it to the precision available}, we obtain:
 \ba
 \Omega^{\rm{(present)}}\simeq\Omega_3\;\,.
 \label{}
 \ea
 This coincidence is so exact as to be somewhat fortuitous, though certainly not displeasing.  It would be too risky to conclude from it that the moon disintegrated exactly at the marginal point, and that Mars' rotation rate has survived unchanged since that event.
  Assuming that Mars' differentiation had long been completed by that time, and recalling that since then the solar tides have been slowing down Mars' spin, we deduce that Mars' rotation rate at the instant of Nerio's destruction should have been slightly higher than $\Omega_3$~---~to be later slowed down by solar tides back to $\Omega^{\rm (present)}\simeq\Omega_3\s$.  This means that Nerio had transcended  the marginal point and started moving up the tangent track in Figure~\ref{Figure_2}, reducing its separation from Mars and spinning Mars up.

  As we explain below, Nerio's creep from point 2 to point 3 in Figure \ref{Figure_2} had been slow before the emergence of the palaeo-ocean (i.e., while Mars' $|K_2|=k_2/Q$ was small),
  but accelerated greatly after the ocean's emergence, because $|K_2|$ increased by an order or two of magnitude. Therefore, Nerio did reach the marginal point right after the ocean formation,
  %  i.e., shortly after the beginning of LHB, \,\footnote{~A large amount of water was delivered to Mars by the comets and water-rich carbonaceous asteroids arriving in abundance during the LHB. This is why the birth of the Martian ocean coincided with the beginning of the LHB.}
  see Section \ref{palaeo} below.

   Nerio's destruction during the LHB did not necessarily take place right at the marginal point, but could as well have happened slightly later. This would give Nerio some time to crawl a bit up the tangent track in Figure \ref{Figure_2}, and to accelerate Mars' rotation slightly above its present value.

Now, we would like to know whether it is possible that Nerio somehow survived the LHB and migrated all the way inward to the Roche limit, disintegrating only there?
Had it done so, it would have endowed Mars with a faster rotation rate than today. For solar tides to subsequently slow down Mars' spin back to the present value $  \Omega^{\rm (present)} \simeq \Omega_{3}  $, the early Martian palaeo-ocean would then have needed to be much more dissipative and/or to have persisted far longer than current models suggest.
This scenario therefore appears questionable, based on our present understanding of tidal dissipation in oceans.
   %  On the other hand, if after the departure from the marginal point Nerio was losing mass sufficiently quickly through the sesquinary mechanism, the slope $X$ of its evolutionary track in Figure \ref{Figure_2} was rapidly decreasing  ($X$ is proportional to Nerio's mass $M_m$, see equation \ref{X}). In this case, the track of a leaner Nerio in Figure \ref{Figure_2} would cross the Roche radius at a value of $\Omega_R$ low enough for the solar tides in Mars to later decelerate it to $\Omega^{\rm (present)}$.

  To discuss these issues quantitatively, we need to know how Mars' rotation rate evolved as a synchronous Nerio migrated from point 2 toward the marginal point 3, and how it evolved later, after Nerio was released from synchronism at point 3 and began travelling inward along the tangent straight track in Figure~\ref{Figure_2}.

  \section{Action of solar tides on a planet synchronised by its moon. General formalism}
 \label{action}

We begin by deriving a general analytic expression for the tidal evolution of a planet's spin under the influence of tides raised by its host star, in a situation where the planet's rotation is staying synchronous with its moon.
 To our knowledge, such an expression has never appeared in the literature heretofore.

 \subsection{Master equation}

 The planet's angular acceleration is produced by the two counter-directed actions from the tidal torques generated by the star and the moon:
 \ba
 \dot{\Omega}\;=\;\dot{\Omega}\s\big{|}_{\rm\s solar}\s+\;\dot{\Omega}\s\big{|}_{\rm\s {``lunar"}}\,\;,
 \label{equality}
 \ea
 where, for brevity, the star's and moon's actions are termed as ``solar'' and ``lunar'', in analogy with the Earth's case.

 The semidiurnal part of the solar torque acting on the planet is given by the third line of equation (116) in \citet{Efroimsky2012}:
 \ba
 \nonumber
 {\cal T}\s\big{|}_{\rm\s solar} &=& \frac{
 3}{2}\,{M_\odot}\,\frac{GM_\odot}{a_{\rm_M}^3}\,\left( \frac{R}{a_{\rm_M}}  \right)^3 \! R^{\s 2} K_2(\omega)\\
 \label{torque.eq}\\
&=&-\;\frac{3}{2}\,{M_\odot}\,\frac{GM_\odot}{a_{\rm_M}^3}\,\left( \frac{R}{a_{\rm_M}}  \right)^3 \! R^{\s 2}\,{\cal I}\!{\it m}\left\{ \bar{k}_2(\omega) \right\}\;\,,
 \nonumber
 \ea
 with $M_\odot$ being the solar mass, and $M$, $R$,  $K_2(\omega)$, and $\bar{k}_2(\omega)$ respectively representing the planet's mass, radius, quality function, and complex Love number.\,\footnote{~We follow the convention $K_2\equiv\s-\s {\cal I}\!{\it m}\left\{ \bar{k}_2 \right\}\s$, with a ``minus'' sign, in order to ensure that the negative argument of the complex number $\bar{k}_2$ has the meaning of a phase lag, not advance.}
 The notation $a_{\rm _M}$ stands for the planet's semimajor axis.
 The letter $\omega$ denotes the semidiurnal tidal Fourier mode of the solar tides exerted on the planet: $\omega\s\equiv\s\omega_{2200}\s=\s2\s(n_{\rm_M}\!-\Omega)\s$, where $n_{\rm_M}$ is the mean motion of the planet-moon pair about the Sun, while
 $\Omega$ is the planet's rotation rate coinciding with the moon's mean motion $n$.
   For $n_{\rm_M}\ll \Omega$, we have
 $\s\omega\s\approx\,-\s2\s\Omega\s$ and therefore
 \ba
 K_2(\omega)\,\approx\;-\;K_2(2\s\Omega)\;=\;-\;\frac{k_2(2\s\Omega)}{Q(2\s\Omega)}\;\,.
 \label{minus}
 \ea
 This yields:
  \ba
 %  \nonumber
 \dot{\Omega}\s\big{|}_{\rm\s solar}
 %  &=& -\,\frac{1}{\xi M\s R^{\,2} }\;\frac{3}{2}\,\frac{G\s M_\odot^2}{a}\left( \frac{R}{a}  \right)^5 K_2   \\  \label{sol}\\
    = \frac{ {\cal T}\s\big{|}_{\rm\s solar}}{\xi\s M\s R^{\s 2}}\;=
 %  &=&
 \;-\;\frac{3}{2\s\xi}\,\frac{M_\odot}{M\,}\,\frac{GM_\odot}{a_{\rm_M}^3}\,\left( \frac{R}{a_{\rm_M}}  \right)^3 K_2(2\s\Omega)
 \,\;.
 %  \nonumber
 \label{sol}
 \ea

 The synchronous moon is acting on the planet with a torque which, according to equation (\ref{Omega}), contributes the following input into the planet' angular acceleration:
 \ba
 \dot{\Omega}\s\big{|}_{\rm\s {``lunar"}}=\;-\;\frac{1}{2}\;X\;a^{-1/2}\stackrel{\bf\centerdot}{a\s}\;.
 \label{lun}
 \ea
 This input is positive, because $\stackrel{\bf\centerdot}{a\s}$ is negative when the moon is approaching the planet. This inward evolution is taking place when the system, while remaining synchronous, is slowly moving from point 2 to point 3 along the hyperbola in Figure \ref{Figure_2}. The slowly evolving synchronism implies that the separation between the planet and the moon is always equal to the evolving synchronous radius $r$ given by equation (\ref{sy}):
  \ba
  a\;=\;r\;=\;\left(\frac{G\,(M + M_m)}{ \Omega^{\,2} }\right)^{1/3}\;,
  \label{a}
  \ea
  which leads to
  \ba
  \dot{\Omega}\;=\;-\frac{3}{2}\;\sqrt{G\s(M+M_m)\s}\,a^{-5/2}\stackrel{\bf\centerdot}{a\s}\,\;.
  \label{tot}
  \ea

 Together, equations (\ref{equality}), (\ref{sol}), (\ref{lun}), and (\ref{tot}) produce the following {{\it{Master Equation}}}:
 \ba
 \nonumber
 \frac{1}{2}\,\left(X\;-\;3\,\sqrt{G\s(M+M_m)\s}\,a^{-2}\right) a^{-1/2}\stackrel{\bf\centerdot}{a\s}\;=\\
 \label{hyper.eq}
 \label{Master}\\
 \qquad\qquad\qquad\qquad-\,
 \frac{3}{2\s\xi}\,\frac{M_\odot}{M\,}\,\frac{GM_\odot}{a_{\rm_M}^3}\,\left( \frac{R}{a_{\rm_M}}  \right)^3 K_2(2\s\Omega)
 \;\,,
 \nonumber
 \ea
 {
 a general result applicable to any planar system comprising a host star and a synchronised planet-moon pair with a negligible eccentricity.
 To avoid confusion, we reiterate that $a$ and $a_{\rm _M}$ denote the semimajor axes of the moon (Nerio) and the planet (Mars), respectively.}

 Below we shall employ equation (\ref{Master}) to estimate the duration of Nerio's stay in the transient synchronism with Mars.

 \subsection{The case of a Maxwell planet}

  As demonstrated in Appendix \ref{TidalMaxwell}, equation (\ref{C.16}), a Maxwell planet's quality function is of the form
 \ba
 K_2(2\Omega)\,=\;\frac{D}{\Omega}\,\;,
 \label{qualitet}
 \ea
 where
 \ba
 D\,=\,\frac{3}{4}
 \left( \frac{{\cal B}_2\;\mu}{ 1\s +\s{\cal{B}}_2\mu } \right)^2\!\frac{1}{\s{\cal B}_2\s\eta\s}\;\,.
 \label{}
 \ea
 {
 Approximation (\ref{qualitet}) becomes valid after the short magma-ocean stage,
 i.e., after the Maxwell time becomes longer than several hours, see Footnote \ref{estimate} in Appendix \ref{SolarMaxwell} below. This approximation is inapplicable to planets going through the magma-ocean stage or experiencing a thermal runaway.}

 Combined with equation (\ref{a}), expression (\ref{qualitet}) becomes
 \ba
 K_2(2\Omega)\,=\;D\;\frac{a^{3/2}}{\sqrt{G(M+M_m)}}\,\;.
 \label{K}
 \ea  {
 The insertion of this expression into formula (\ref{hyper.eq}) yields the following differential equation for spin-orbit evolution of a synchronised planet-moon pair, where the planet is Maxwell and subject to star-generated tides:
 \ba
 % \nonumber
 \left(\frac{M_m}{M+M_m} x^{-2}-{3}\s\xi\s   x^{-4}\right)dx =
 %  ~\\  \label{capt.eq}\\
 -\,\frac{3\, D\; M}{M+M_m}  \left(\frac{M_\odot}{M}\right)^2\left(\frac{R\s}{a_{\rm_M}}\right)^6dt\,,
 \label{vavila}
 \ea
 $x$ being the dimensionless semimajor axis:
 \ba
 x\s\equiv\s\frac{a}{R}\,\;.
 \label{}
 \ea
 }
% Inserting this expression into equation (\ref{hyper.eq}) furnishes us with
% \ba
% \left(X\s a^{-2}\s-\s Y\s a^{-4}\right)\s da\s=\,Z\s dt\;\,,
% \label{equation}
% \ea
% where
%  \ba
% X\s=\s\frac{G\s M_m}{\xi}\,\frac{1}{\sqrt{\s G(M\s +\s M_m)\s}\s}\s R^{\s -2}
% \;\,,
% \ea
% \ba
% Y\,=\,3\s\sqrt{G\s(M\s+\s M_m)}
% \label{}
% \ea
% and
% \ba
% Z\,=\,\frac{3}{\xi}\,
% \frac{M_\odot}{M\,}\,\frac{GM_\odot}{a_{\rm_M}^3}\,\left( \frac{R}{a_{\rm_M}}  \right)^3\frac{D}{\sqrt{G(M+M_m)}}
% \,\;.
% \label{}
% \ea
%This equation describes the slow evolution of the transient synchronous state of the system.

 \subsection{The case of an Andrade planet}

  {
  For reasons explained in Appendix \ref{TidalAndrade}, we set the Andrade exponential $\alpha$ equal to $1/4$, wherefore
 \ba
K_2(2\Omega)\s=\s\frac{H\;}{\Omega^{1/4}}\,\;.
\label{ho}
\ea
 The expression for the constant $H$ through the rigidity, viscosity and the Andrade time is given in  Appendix \ref{SolarAndrade}, equation (\ref{HH}).
}

 {
 Approximation (\ref{ho}) becomes valid after the magma-ocean stage only.}

{Derived in Appendix \ref{SolarAndrade}, the resulting equation for time evolution of the dimensionless semimajor axis $\s x\equiv a/R\,$ is:
 \ba
 \left(\frac{M_m}{M+M_m} x^{-7/8}\s-\s{3}\s\xi \s x^{-23/8}\right)\s dx\, =\,-\,F\,dt\;\,,
 \label{danila}
 \ea
$F$ being a constant given by expression (\ref{F}).
 %  \ba
 %  F\,=\,3\,\frac{M_\odot}{M}\,\left(\frac{n_{\rm{_M}}}{b}\right)^2\s\left(\frac{R}{a_{\rm {_M}}} \right)^3 b^{3/4}
 %  H\,=\,5.9\times 10^{-22\pm 1}\,\mbox{s}^{-1}\;\,,
 %  \label{}
 %  \ea
 %  where $b\equiv\sqrt{G(M+M_m)/R^3\,}\s$.
}

 \subsection{A terrestrial planet with an ocean}

 {
 For an oceanless terrestrial planet, both dependencies of $K_2$ on the tidal frequency look simple\,---\,equations  (\ref{qualitet}) and (\ref{ho}), for a Maxwell and Andrade planets, correspondingly.
 Consequently, the evolution law of the semimajor axis $a$ of a synchronous planet-moon pair, equations (\ref{vavila}) and (\ref{danila}), can be integrated analytically, see Appendices \ref{SolarMaxwell} and \ref{SolarAndrade}.
 }

{
 In the presence of an ocean, the frequency dependence of $K_2$ is complex and contains sharp peaks corresponding to tides' resonances with the normal modes of the ocean's basin\,---\,see, e.g.  \citet{Auclair2018} or \citet{Auclair2019}.\,\footnote{~{ Equation (55) in \citet{Auclair2019} illustrates the frequency dependence of the purely oceanic part of $K_2$ in the vicinity of each peak. The function first increases linearly with the tidal frequency, then demonstrates a complex behaviour, and finally, in the infinite-frequency limit, falls off as the inverse cube of the frequency. However, the overall (solid plus ocean) shape of the overall $K_2$ remains complex, especially at intermediate frequencies. This leaves one no chance for analytical modelling of tidal evolution unless the peaks are trimmed.}
} With such a dependence inserted, integration of the Master Equation (\ref{Master}) can be performed only numerically.
A crude approximation, however, may be achieved by ignoring the peaks  and approximating the resulting average frequency dependence of $K_2$ with a smooth scaling law.
 Since the peaks add a lot to dissipation, the approximation ignoring these peaks will be a lower bound for the actual evolution rate.
}

{
  It should be added that in the presence of an ocean, the quality function $K_2$ in the Master Equation (\ref{Master}) should be substituted with $K_{22}$, because in that case not only the degree but also order matters, and the term with $\left\{lm \right\} =\left\{22 \right\} $ renders the leading input when the inclination of the planet's orbit on the equator is small, see Appendix \ref{ocean} for details and references.
  }

 \section{Evolution of Mars' rotation before Nerio's demise}
\label{Section 6.2}

  The reasoning provided in Section \ref{marg} explains that  synchronisation took place, in Figure \ref{Figure_1}, on a dynamical track 1 residing above track 2, which is tangent to the cubic curve. To understand how far above track 2 track 1 lies,  we first recall that the action of solar tides is an adiabatic process, in that a dynamical track (straight line 1 in Figure \ref{Figure_1}) is very slowly evolving leftwards (or downwards), approaching the tangent straight line 2 and always staying parallel to it. Owing to this slowness, the evolving synchronous state (the right-hand crossing point of the slowly moving straight line and the cubic curve in Figure \ref{Figure_1}) reaches the marginal point when the Martian crust is already formed, see Footnote \ref{Footnote} in Section \ref{geoph}.

   In Figure \ref{Figure_2}, this evolution of the transient synchronous state is depicted by a slow motion of point 2 to the marginal point 3.

 \subsection{Motivation}

 On the one hand, Nerio ought to have reached the marginal point 3 in Figure~\ref{Figure_2}. For had Nerio been eliminated by a collision somewhere between points 2 and 3, Mars' rotation rate $\Omega$ corresponding to that moment of time would be lower than $\s \Omega^{\rm{(present)}}\simeq\Omega_3\s$. This would be unsatisfactory, given that the subsequent action by solar tides has been working since then to further decelerate Mars' rotation.

 On the other hand, Nerio's elimination ought to have predated the termination of the LHB
 { at $\simeq 3.4$ Ga.}\,\footnote{~{In Section \ref{LHB}, we reviewed the available estimates for the LHB termination time. While some authors extend the LHB to much later epochs, we chose a conservative estimate of 3.4 Ga \citep{Zellner2017}.}} Given the current absence of any significant latitudinal distribution in the early Martian cratering record, this time constraint is required to leave the LHB enough time to smear the distribution of craters left by the falling remnants of Nerio after its destruction. {With the adopted termination time, the duration of Nerio's migration from point 2 to point 3 would not have exceeded $\sim 1.2$~Gyr. This estimate may be refined as more constraints on the history of the LHB tail become available.}

 Our goal is to estimate for how long Nerio stayed in the transient synchronism\,---\,i.e., to estimate
the duration $t_3-t_2$ of the creep from point 2 to point 3  in Figure \ref{Figure_2}, for realistic rheologies and parameters' values.

Owing to the extreme shortness of the magma-ocean stage, all or most of the evolution from point 2 to point 3 was taking place when the mantle was already viscoelastic, not liquid, and the viscosity was high enough to ensure that the forcing frequency was located to the right of the peak in Figure \ref{figure1}, see Footnote \ref{estimate} in Appendix \ref{SolarMaxwell}. For the Maxwell and Andrade rheologies, the quality function therefore assumed the forms given by equations (\ref{qualitet}) and (\ref{ho}), correspondingly.

\subsection{Before the ocean formation.\\ A Maxwell Mars}\label{max}

 {
Presented in Appendix \ref{SolarMaxwell},
 integration of equation (\ref{vavila}) demonstrates that for realistic values of the parameters involved, even a short tidal descent of $1000$~km would requires prohibitively long times, leaving Nerio no chance to reach the marginal point 3 before the formation of a palaeo-ocean on Mars. The evolution is so slow because the Maxwell rheology produces unphysically low values of $K_2$ at realistic frequencies\,---\,a fundamental limitation of the Maxwell model.}\,\footnote{~Applicable to description of mantles' response at extremely long timescales, the Maxwell model is insufficient at ordinary timescales, because it ignores transient processes and therefore greatly underestimates the dissipation rate.  Long known in seismology, in planetary science this fact was pointed out for the first time by \citet{Bills} who determined that the rate of Phobos's tidal descent is consistent with that of a homogeneous Maxwell  Mars with effective viscosity of
  $ 8.7\times 10^{14}~\operatorname{Pa~s}\,$---\,which is six to eight orders of magnitude lower that the actual mean viscosity of Mars' mantle
 %  , known to be $(2 - 6)\times 10^{22}$~Pa~s
\citep{Guinard, Broquet}.
}

\subsection{Before the ocean formation.\\ An Andrade Mars}

{
To adequately describe the crawl of a synchronous Nerio towards the marginal point 3 in Figure \ref{Figure_2}, while Mars is still oceanless, we must invoke the Andrade model which comprises both the viscoelastic reaction and transient processes and therefore describes well the tidal reaction of mantles, see
%  \citet{Efroimsky2012} and
\citet{2022AdGeo..63..231B} and references therein.
}

{Provided in Appendix \ref{SolarAndrade}, integration of equation (\ref{danila})  renders the time interval $t_3 - t_2$ required for traversing the difference $a_2 - a_3$ in semimajor axis' values, see equation (\ref{prohor}) in Appendix \ref{SolarAndrade}.
Calculation demonstrates that even for the Andrade time much shorter than the Maxwell time (i.e., for a stronger input of transient phenomena into dissipation), the time interval $t_3-t_2$ is in the billions of years, unless we squeeze the altitude change $a_2-a_3$ to less than a thousand kilometres (which would be a fine tuning of sorts). Therefore, while for an Andrade Mars the evolution of the synchronous state was faster than for a Maxwell Mars, Nerio still could not reach the marginal point 3 before the
ocean formation.}

\subsection{After the ocean formation}\label{palaeo}

 Governed by equations (\ref{vavila}) or (\ref{danila}) before the formation of an ocean, the system's evolution accelerated greatly after the ocean emerged.  Depending on its shape and depth, the ocean  boosted the value of Mars' $K_2$  and thereby evicted Nerio by shortening the time of its stay in the synchronous orbit.
 For an ocean-bearing planet, the estimate {$K_2\simeq 0.7$} is  conservative, see Appendix~\ref{ocean} for details.

 {
 To integrate the Master Equation (\ref{hyper.eq}), we need not only a quantitative estimate of $K_2$, but also some understanding of its frequency dependence.
 Removal of peaks and smoothening of the frequency dependence  of $K_2$ would reduce the efficacy of tidal interaction and provide a low bound for the resulting evolution rate.}

 {
 Figure 3 in \citet{Auclair} indicates that the {\it average} shape of a smoothened frequency dependence of an ocean-bearing planet is not very different from that of a viscoelastic planet: a near-linear increase at very low frequencies, followed by a gradual fall-off at higher frequencies.}
 {Gwena\"{e}l Bou\'{e} has kindly generated for us a version of that figure, in which the quadrupole tidal torque is collated with a curve falling off as the inverse frequency, see Figure \ref{Boue} in Appendix \ref{Appendix_Boue}. The curve turns out to be a good fit for the torque.  Except in the low-frequency limit, the average tidal reaction of an ocean planet is similar to that of a Maxwell sphere. At the frequencies where difference is appreciable, the inverse-frequency curve is serving as a safe lower bound.  This motivates us to rely on equation (\ref{vavila}).}

{
 Through equation (\ref{qualitet}), the value $K_2 = 0.7$  corresponds to $D=4.2\times 10^{-5}$~s$^{-1}$, whose substitution into expression (\ref{equation_c}) renders reasonably short times. For example, for a
 $\, a_2\s -\s a_3\,=\,10,000\;\mbox{km}\s$ descent, the resulting time is $\s t_3-t_2= 1.0$~Gyr, which means that Nerio reached point 3 well before the end of the LHB.}\,\footnote{~{For a shorter fall of $\, a_2\s -\s a_3\,=\,1000\;\mbox{km}\s$, the evolution of Nerio's synchronous state takes only $t_3 - t_2 =  0.023$ Gyr.
 Such a short fall, however, is less realistic, because it would imply the formation of Nerio in too close a vicinity of the marginal state\,---\,which would be fine tuning.}
 }

 {
 We would reiterate that
  % emphasise that without a detailed model of an ocean-bearing Mars at our disposal, our employment of equation (\ref{qualitet}) remains an approximation.
  % Therefore our estimates, like $0.29$ Gyr for a $10,000$ km descent, or $0.073$ Gyr for a $1000$ km descent should not be taken too literally.  Our goal was simply to illustrate that the emergence of an ocean greatly accelerates the process.
  % Finally,
  the possible presence of peaks in the frequency dependence only adds to dissipation and shortens the resulting interval $\s t_3-t_2\s$.
  }

\subsection{Conclusions to Section \ref{Section 6.2}}

Before the formation of an ocean, Nerio was crawling from its synchronisation point 2 towards the marginal point 3 in Figure \ref{Figure_2} extremely slowly.

 {
After the ocean came to existence, Nerio sped up and arrived at the marginal state\,---\,and departed therefrom onto the track leading to the Roche limit. In Figure \ref{Figure_2}, that track is the straight line which is tangent to the cubic curve at point 3. Shortly after its departure from that point, Nerio was eliminated by the LHB.
 %  , likely through the sesquinary mechanism rather than by a single hit.
 %   In the single-hit case,
The angular velocity of Mars was then gradually slowed by solar tides.
}
 %  In the sesquinary-mechanism case, the rate of Mar's rotation rate increase is getting slower with the reduction of Nerio's mass $M_m\s,\s$ as prescribed by the expression (\ref{X}) for the slope of the track. The track is getting less steep and more horizontal as the moon is becoming slimmer.

{
The timescale of evolution of Nerio's transient synchronous state was much longer than the time of fossilisation of Mars' figure. By the moment of Nerio's departure from the marginal point,   Mars' tidal bulge had long been frozen, giving the planet its distinct asymmetrical triaxial shape.
}

 \section{Evolution of Mars' rotation after Nerio's demise}

After Nerio disintegrates, Mars becomes subject to solar tides alone. Its angular velocity $\Omega$ is thus gradually decelerated.

\subsection{Despinning of an oceanless Mars}

 As we saw in Section \ref{questions}, Mars' rotation rate at the moment of Nerio's demise virtually coincided with the current rotation rate.

Mars' secular despinning rate is
 \ba
 \stackrel{\bf\centerdot}{\Omega}\;=\;
 \frac{
 1
 }{\xi\s M\s R^2}\;
 {\cal T}\s\big{|}_{\rm\s solar}
 \,\varpropto\,\frac{
 1
 }{\xi\s M\s R^2}\;{{\cal T}_{22}}\s\big{|}_{\rm\s solar}\;\,,
 \label{evalua}
 \ea
where ${{\cal T}_{22}}$
 %  \ba
 %  {{\cal T}_{22}}\s\big{|}_{\rm\s solar}\,=\;\frac{3}{2}\,{M_\odot}\,\frac{GM_\odot}{a_{\rm_M}^3}\,\left( \frac{R}{a_{\rm_M}}  \right)^3 \! R^{\s 2} K_{22}(\omega)+\,O(e^2)
 %  \label{}
 %  \ea
is the semidiurnal component of the solar tidal torque $\cal T$, and is given by expression (\ref{torque.eq}).

 Mars is now cold, its quality function assuming as low a value as $\s K_2\s\equiv\s k_2/Q\s =\s 0.174/93.0\s=\s 1.87\times 10^{-3}$ \citep{Konopliv2020, Pou}. Inserting it into the expression (\ref{torque.eq}) for ${\cal T}_{22}$, we find that at present this despinning torque is extremely weak. By equation (\ref{evalua}), its effect on Mars' spin evolution is only
 \ba
 \nonumber
 \stackrel{\bf\centerdot}{\Omega}\s\big{|}_{\rm\s solar}^{\rm\s (present)}
  &=&
 -\,8.801\times 10^{-25}\,\;\mbox{rad~\,s}^{-2}
   ~\\    \label{ti}\\   &=&
  -\,1.808\times 10^{-1} \;\,\mbox{mas~\,yr}^{-2}\;\,.
 \nonumber
 \ea
 Superimposed on this secular effect are  rotation variations induced by geophysical and atmospheric processes (like the redistribution of carbon dioxide),\,\footnote{~Basing their analysis on {\it InSight} radio tracking, \citet{LeMaistre} concluded that Mars' rotation is currently accelerating at the rate of
 \ba
 \dot{\Omega}\s\big{|}_{\rm\s total}^{\rm\s (present)}
 \s=\s 19.0\times 10^{-24}\,\;\mbox{rad~s}^{-2}
 \s=\s 3.9\,\;\mbox{mas~yr}^{-2} \,\;,
 \nonumber
 \ea
 exceeding by a factor of $\simeq 22$ the solar decelerating input~(\ref{ti}). We presume that this acceleration, if confirmed by future missions, is not a secular effect, by distinction from the permanent input (\ref{ti}).} but at this point we are interested in the secular effect solely.

 If we assume that the mean tidal deceleration rate over the past {\ae}ons has been the same as today (equation \ref{ti}), and if we also accept that Nerio was gone by 3.4 Ga,
 this will leave us with the following estimate for the ensuing change in the angular velocity:
 \ba
 \Delta\Omega^{\rm(body~tide)}=\,
 \stackrel{\bf\centerdot}{\Omega}\s\big{|}_{\rm\s solar}^{\rm (present)}\times 3.4\;\mbox{Gyr}
 % =\;-\;8.801\times 10^{-25}\,\mbox{rad}~\mbox{s}^{-2}\times\s 4\s\times 3.1536\times 10^{16}~\mbox{s}
 =-\s 0.94
 \times 10^{-7}\s\mbox{rad/s}\,\;.
 \label{estime}
 \ea
  This is three orders of magnitude less than the range of spin rates we are dealing with in Figure \ref{Figure_2}. We see that for solar tides to despin an oceanless Mars to the present rotation rate value, the timing of Nerio's demise ought to have been very precise, a kind of fine tuning.

 In reality, the $K_2$ of the solid Mars was certainly higher in the past than today.  Still, even an order of magnitude increase of our estimate (\ref{estime}) would be insufficient to evade fine tuning.
 The presence of an ocean, however, changes the situation.

 \subsection{Despinning due to a shallow ocean}

Sufficiently long-lived in geological terms, the ocean should have had significant implications for the planet's rotational history.
 Our goal is to estimate how efficient Mars' despinning by solar tides was
 after Nerio's destruction. To this end, we write:
 \ba
 \Delta\Omega=\int \frac{{\cal T}\s\big{|}_{\rm\s solar}}{\xi\s M\s R^2}  \,dt\varpropto
 \frac{\Delta t}{\xi\s M\s R^2}\,{\cal T}\s\big{|}_{\rm\s solar}\varpropto\,\frac{
 \Delta t
 }{\xi\s M\s R^2}\,{\cal T}_{22}\s\big{|}_{\rm\s solar}\;\,,
 \label{evaluation}
 \ea
 the torque being given by expression (\ref{torque.total}).

 While the above integral covers the entire interval from Nerio's demise to the present time, the overwhelming contribution to the spin change $\Delta\Omega$ occurred over the period $\Delta t$ spanning from Nerio's demise to the ocean's disappearance.

 As we agreed in Section \ref{palaeo},
 a palaeo-ocean existed on Mars from $\sim\!4.5$ Ga through $\sim\!3$ Ga, i.e., for about $1.5$~Gyr. If  Nerio perished  $\sim 3.4$~Gyr ago, thereafter the ocean still existed for
 \ba
 \Delta t\,\simeq\,0.4~\mbox{Gyr}\;\,.
 \label{deltate}
 \ea

Equation (\ref{evaluation}) should not be taken too literally. It is not simply a crude estimate, but certainly an underestimate. As explained in Appendix \ref{ocean}, an ocean basin can sustain a spectrum of eigenfrequencies, one or several of which may resonate with some tidal mode.
Over the time span $\Delta t$, Mars' rotation rate and therefore the tidal frequencies were changing. Consequently, as the system evolved over $\Delta t$, the frequency-dependent torque might exhibit sharp, high-amplitude peaks caused by the basin shape's resonances  with the semidiurnal mode present in expression (\ref{evaluation})\,---\,and with other modes (e.g., with $\left\{lm\right\}$=$\left\{21\right\}\s$), which are neglected in that expression.
These peaks were working to amplify tidal interaction and increase $|\Delta\Omega |$. With this caveat in mind,
equations (\ref{evaluation} - \ref{deltate}) and (\ref{torque.total} - \ref{estim}) produce:
\ba
-\,\Delta\Omega \gtrsim \frac{3}{2\s\xi}\frac{\s M_\odot}{M}\frac{G\s M_\odot}{a_{\rm_M}^{3}}\left(\frac{R}{a_{\rm_M}}\right)^3 K_2\,\Delta t\approx\s 0.46\times 10^{-5}\s\mbox{rad~s}^{-1}.\,
\label{deceleration}
\ea
Consequently, Mars' angular velocity at the moment of Nerio's demise was
\ba  \Omega\;\simeq\;\Omega^{\rm\s(present)}-\,\Delta\Omega\;\gtrsim \;7.55\times 10^{-5}\s\mbox{rad~s}^{-1}\,,
\label{}
\ea
which is only slightly larger than Mars' rotation rate $\Omega_3\approx 7.03\times 10^{-5}$ rad/s at the marginal point.  {This is an argument in favour of Nerio's demise by a cataclysmic collision (or  collisions) not far from the marginal point, well above the Roche limit.}

We reiterate that owing to the possible peaks in its frequency dependence, the mean value of $K_2$ over the interval $\Delta t$ may have been higher. On the other hand, had Nerio's demise happened earlier than 3.4 Ga, the residual time $\Delta t$ would be longer. Despite these caveats, estimate (\ref{deceleration}) indicates that Mars' tidal despinning following Nerio's destruction was minimal.

 %  \section{The fate of the remnants}
 %
 %  After Nerio's demise in the LHB, the share of its mass (and angular momentum) deposited on Mars depended on the speed and direction of the collision. While the topic deserves a separate consideration, we would mention that the Martian surface contains numerous craters resulting from grazing collisions.
 %    %  The oldest of them are located on Lunae Planum, a Hesperian-aged (3.0 - 3.7 Gyr old) province.
 %  Their distribution demonstrates a pronounced latutudinal dependence  \citet[Figure 5c]{Schultz}.
 %  While this dependence certainly bears an impact from erosional and depositional processed in polar regions, it was demonstrated in {\it Ibid.} that the near-equatorial reagios are overall favoured, and that many of the impactors had been areocentric, not heliocentric.

\section{Determining whether Nerio crossed the Roche radius}

Could Nerio have somehow survived the LHB hazards and made it all the way to the Roche limit, disintegrating only there?

As demonstrated in Appendix \ref{E1} (see also Figure \ref{Figure_3}), Mars' rotation rate at the time of Nerio's hypothetical entering the Roche limit was (for $M_m\s=\s 0.03\s M$) as high as
   $\Omega_{\rm R} = 1.4823\times 10^{-4}$~rad/s. For this rate to be subsequently reduced by solar tides down to $\Omega^{\rm (present)} = 0.7088 \times 10^{-4}$ rad/s, Mars'  quality function would have to be up to $K_2\simeq 12$.

   {
   Moreover, if the entire mass and angular momentum of Nerio were deposited on Mars after Nerio disintegrated at the Roche radius, then the resulting Mars' rotation rate would be even higher.  To slow it down by solar tides to its current value, Mars'  quality function would have to be up to $K_2\simeq 18$,   see Appendix \ref{E2}.
   }
   %  This by more than an order of magnitude exceeds our conservative  estimate $K_2\simeq 0.2$ acceptable within the presently available models.

 Presently available models provide no indication that planets with oceans are so highly dissipative. Unless a similar indication appears in future models of Mars' ocean, we should deem this scenario of Nerio's demise questionable.

  \begin{figure}
 	\hspace*{-5.05mm}{
    \includegraphics[width=
    102mm
    % \columnwidth
    ]{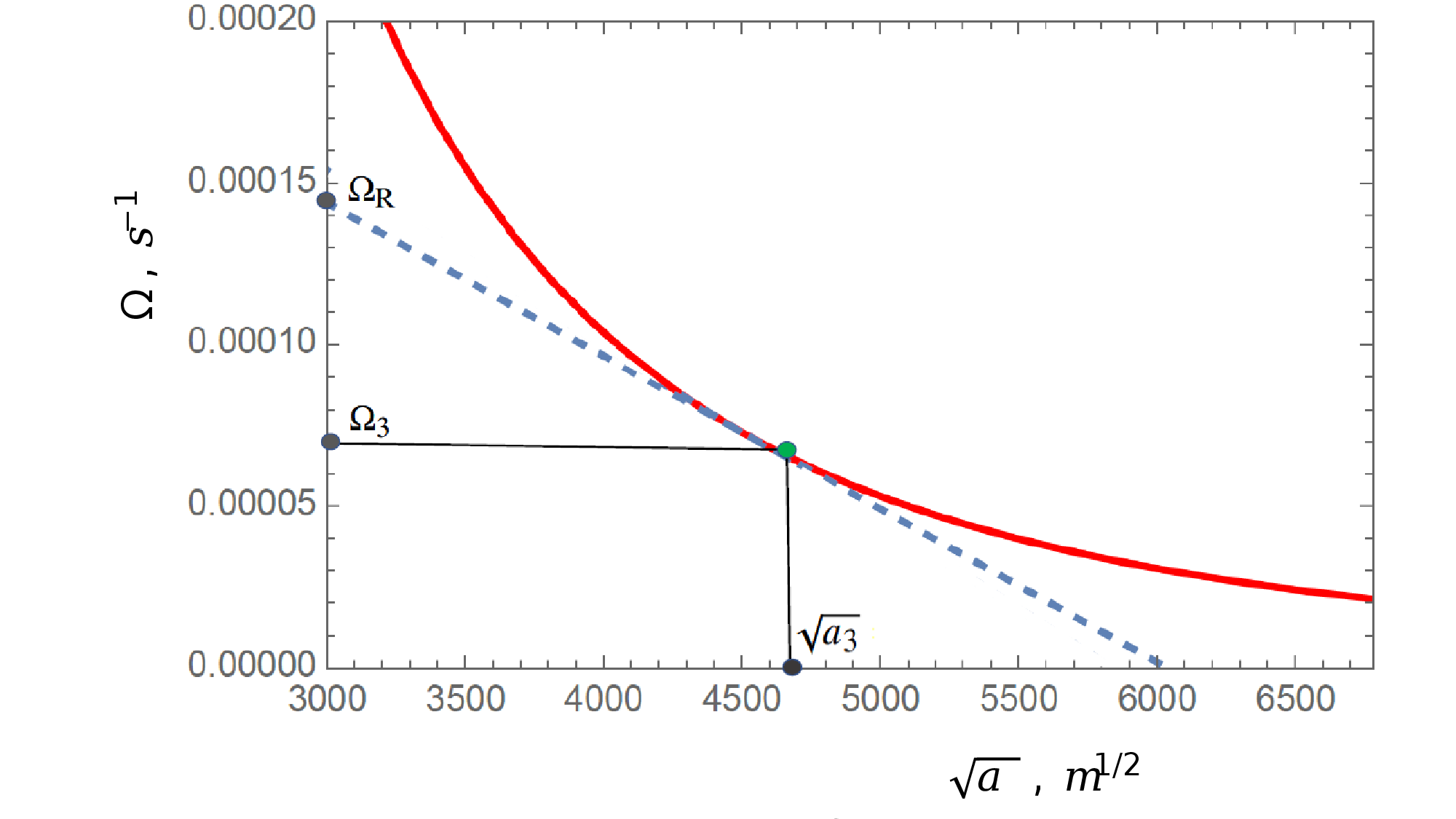}
    }
   \caption{. \small If Nerio survived the LHB and made it all the way to the Roche limit, then, for $M_m = 0.03 M$,
   the spin rate of Mars at the time of Nerio's crossing the Roche radius would be
     $\s\Omega_{\rm R}=1.4823\times 10^{-4}$ rad/s$\s$, equation~(\ref{r_R}).
    }
   \label{Figure_3}
   \end{figure}

\section{Conclusions and outlook}

\subsection{General result}

{Our study builds on the observation that stellar tides acting on a planet synchronised by its moon render the planet-moon synchronism transient (see, e.g., \citet{synchronisation} and references therein).}

{Motivated by that observation, we have derived the Master Equation (\ref{hyper.eq}), a generic relation governing the orbit evolution of a planet-moon pair under the influence of stellar tides in the planet.
 In the absence of a synchronous moon, stellar tides work to slow down the planet's rotation; instead,
 %  (and to either synchronise the planet's spin with its mean motion about the star or, in some cases (like Mercury), to bring the planet to a higher spin-orbit resonance. In
in the presence of a synchronous moon, they work to shrink the orbit of the planet-moon pair, thus accelerating the planet's spin.  This process adiabatically continues until the moon either disintegrates at the planet's Roche radius or gets obliterated before that by a collision or a series of collisions.}

{
This result was used by us to investigate the orbital stability and the demise of the hypothetical moon Nerio, which had endowed Mars with its initial triaxiality.
}

\subsection{Application to Mars and Nerio}

Under the action of solar tides, Nerio's synchronous state was adiabatically evolving towards a marginal orbit (marginal point 3  in Figure \ref{Figure_2}), the one where Nerio's orbit became unstable. Except for an extremely brief magma-ocean stage at the start of Mars' history, the tidal parameter $k_2/Q$ of the oceanless Mars assumed low values, wherefore the evolution of Nerio's orbit towards the marginal orbit was slow.  It however became rapid as soon as Mars acquired a palaeo-ocean, which boosted the value of Mars' $k_2/Q$.

On reaching the marginal orbit, Nerio left synchronism and spiralled down, further accelerating Mars' spin. Remarkably, Mars' rotation rate at the desynchronisation moment matches its present-day value to the first decimal place. This coincidence should not be overinterpreted, because the post-desynchronisation evolution comprised Mars' continued spin-up during Nerio's fall (until Nerio's demise amid the LHB) and the subsequent despinning of Mars by solar tides.

 A calculation based on a conservative estimate for $k_2/Q$ of a Mars with an ocean shows that, after Nerio's demise, Mars' despinning by solar tides was a weak effect. This indicates that the LHB destroyed Nerio shortly after its orbit reached the marginal point. (We reiterate that Mars' rotation rate at that moment virtually coincided with its present-day value.)

 Based on our current knowledge of dissipation in ocean-bearing planets, Nerio's reaching the Roche limit intact is questionable. To descend all the way to the Roche radius, Nerio would have to accelerate Mars' rotation to a higher value. To subsequently reduce that value by solar tides, the Mars with an ocean would have to be more dissipative that suggested by the currently available models. Unless future models prove Mars' ocean to have been much more dissipative, we should regard Nerio's fall into the Roche limit as unrealistic~---~though this viewpoint may be reconsidered should new data on Mars' palaeo-ocean show up.

 %  Given that the massive primordial moon was eliminated before or during the LHB, it could not serve as a common progenitor of Phobos and Deimos.  That putative progenitor may, however, have been a piece of the primordial moon, had that moon been eliminated by the LHB and not tides.

 \subsection{Relation to tidal rhythmites in Vastitas Borealis and Gale Crater}

{
Aside from the remarkably precise prediction of Mars' rotation rate, our theory gets a less direct confirmation from geological features on Mars.}

{Planar and cross-laminations were discovered by the Zhurong Rover on sedimentary rocks in the Vastitas Borealis Formation. These structures indicate deposition by an alternating aqueous current and are typical of
tidal environments \citep{Xiao}.}

 {High-resolution images from Curiosity's Mast Camera (Mastcam) and Mars Hand Lens Imager (MAHLI) reveal quasi-periodic stratification comprising sub-millimetre to millimetre-scale laminae in Gale Crater. This motivated \citet{Sarkar} to enquire if such laminae are traces of a lost moon. In our view, this question should be answered in the affirmative.}

{In the course of its post-desynchronisation spiralling down, Nerio acquired the ability to generate tidalites within the Martian palaeo-ocean. After Nerio's demise, this generation may have been continued by Nerio's massive remnant(s). The existence of tidal rhythmites found in Vastitas Borealis and Gale Crater therefore serves as an indirect argument in favour of the past existence and subsequent demise of Nerio.
}

\section*{Acknowledgements}

The author extends special thanks to Valeri V. Makarov for valuable advice and numerous stimulating discussions, without which this work would have never been completed. The author is very grateful to Gwena\"{e}l Bou\'{e} for providing Figure \ref{Boue} and for the helpful consultation that accompanied it. The author is also thankful to Erik Asphaug, Pierre Auclair-Desrotour, Rory Barnes, Mohammad Farhat, Julien Frouard, Xuan Ji, and J\'{e}r\'{e}my Leconte for consultations and references. None of these colleagues bears responsibility for the author's hypotheses regarding the putative existence of a lost moon of Mars.

 % \section*{Data Availability}
 %
 % No data have been used in this paper.

\clearpage

~\\
\begin{appendix}
\noindent
{\Large\bf Appendix}

 \section{Symbol key}

 % \begin{table*}[H]
 % \centering
 %    \begin{flushleft}
 %    \begin{minipage}{146mm}
 % \centering
 % \caption{~~~~Symbol key}
 \label{description}
 %  \vspace{3mm}
   %  \hskip-3.2cm
 \begin{tabular}{@{}llll@{}}
 \hline\\
   \vspace{2mm}
 Variable & ~~~~~Value & Explanation & Reference \\
 \hline \\

 \vspace{1.4mm}

 $G$ & $6.67430 \times 10^{-11}$ $\frac{\mbox{m}^{\s 3}}{\mbox{kg~\s s}^{\s 2}}$ & gravitational constant & \citet{gravitational_constant} \\[2pt]

 \vspace{1.4mm}

 $  M                 $   &                $ 6.31\times 10^{23}$ kg                   &  early Mars' mass          &   \citet[Eqn 15]{Mars}                       \\[2pt]

 \vspace{1.4mm}

 $  R                $   &               $3.37\times 10^6$ m                    &  early Mars' mean radius   &   \citet[Eqn 15]{Mars}                   \\[2pt]

 \vspace{1.4mm}

 $  M^{\,\rm (present)}  $   &  $ 6.4169  \times 10^{ 23} $ kg  &  present Mars' mass        &  \citet{Konopliv2016}  \\[2pt]

 \vspace{1.4mm}

 $  R^{\,\rm (present)}  $   &  $ 3.3895 \times 10^{  6} $  m    &  present  Mars'  mean radius & \citet{Seidelmann}   \\[2pt]

 \vspace{1.4mm}

 $ k_2
  % , \s h_2,\s k_3,\s h_3
 $  &                            & early Mars' Love number          &     \\[2pt]

 \vspace{1.4mm}

 $  Q                       $   &                            & early Mars'  quality factor &          \\[2pt]

 \vspace{1.4mm}

 $  K_2\s \equiv\;\frac{\textstyle k_2}{\textstyle Q} $  &    & early Mars'  quality function &                    \\[2pt]

\vspace{1.4mm}

 $ r $   &     & early Mars' synchronous radius     &     Equation (\ref{r})           \\[2pt]

 \vspace{1.4mm}

 $  \stackrel{\bf\centerdot}{\theta\,} $  &    & Mars' rotation rate &                    \\[2pt]

 \vspace{1.4mm}

  $ \Omega $  &    & early synchronous value of $\stackrel{\bf\centerdot}{\theta\,}$ &   Equation (\ref{bills})                  \\[2pt]

 \vspace{1.4mm}

 $ a $   &     & Nerio's semimajor axis     &                  \\[2pt]

 \vspace{1.4mm}

 $ a_{\rm_M} $   &  $2.2794 \times 10^{11}$ m   & Mars' semimajor axis     &                  \\[2pt]

 \vspace{1.4mm}

 $ n $   &     & Nerio's mean motion     &                  \\[2pt]

 \vspace{1.4mm}

 $ n_{\rm{_M}} $   & $1.085\times 10^{-7}$    & early Mars' mean motion     &                  \\[2pt]

 \vspace{1.4mm}

 $ A<B<C $   &     & early Mars' moments of inertia     &                  \\[2pt]

 \vspace{1.4mm}

 $  \xi\,\equiv\,\frac{\textstyle C}{\textstyle M\s R^{\s 2}}
  $      &                                   &  early  Mars' MOI factor              &         \\[2pt]

 \vspace{1.4mm}

  $  \xi^{\rm(present)}
  $      &          0.36379                         &  present  Mars' MOI factor              &  \citet{Khan2017}       \\[2pt]

 \vspace{1.4mm}

 $ J_2 $   &     & early Mars'  oblateness     &      $J_2\equiv\frac{\textstyle C - (A+B)/2}{\textstyle M\s R^2}$            \\[2pt]

  \vspace{1.4mm}

 $ J_{22} $   &     & early Mars'  triaxiality     &      $J_{22}\equiv\frac{\textstyle B\s-\s A}{\textstyle 4\s M\s R^2}
 =\sqrt{C_{22}^2+S_{22}^2}$            \\[2pt]

  \vspace{1.4mm}

 $ J_2^{\rm (present)}  $   &  $1.9566 \times 10^{-3}$   & present  Mars'   oblateness     &    \citet{Konopliv_2011}             \\[2pt]

 \vspace{1.4mm}

 $ J_{22}^{\rm (present)} $   &  $6.3106 \times 10^{-5} $   &  present Mars'  triaxiality     &     \citet[Table S4]{Konopliv2020}
 \\[2pt]

 \vspace{1.4mm}

 $  M_m
  % ^{\,\prime}
  $      &                                   &  Nerio's mass              &         \\[2pt]

 \vspace{1.4mm}

 $  M_{\odot}
  % ^{\,\prime}
  $      &    $2.09\times 10^{30}~\mbox{kg}$                               &  early Sun's mass              &   Appendix \ref{AppendixA}      \\[2pt]

 \hline
 \end{tabular}
  %    \end{minipage}
  %    \end{flushleft}
 ~\\ \vspace{2mm}
 %  \end{table*}

\clearpage

 \section{The mass of the young Sun}\label{AppendixA}

 The combined mass loss to the solar wind and the fusion of hydrogen to helium in the solar core make the young Sun's mass $M_\odot$ higher than the present mass $M_\odot^{\rm(present)}$. \,Currently, the combined effect of the solar wind and hydrogen-into-helium conversion
 is negligible,\,\footnote{~Analysis of the NASA Messenger data establishes a relative change of $GM_\odot$ of $(\,-\s 6.13 \pm 1.47)\times 10^{-14}$ year$^{-1}$ \citep{Genova}.} which was not the case of the young Sun. While evolutionary models suggest a broad range of values $M_\odot\s=\s (1.01 - 1.07)\s M_\odot^{\rm(present)}$, radio observations of young solar analogues indicate that the total amount of mass lost by the Sun in the early stages of its main-sequence evolution was at most 0.4\% \citep{Fichtinger}. A safer estimate probably comes from a suggestion to resolve the Faint Young Sun Paradox by considering a more luminous early Sun than stellar evolution simulations indicate. To that end, the Sun should be endowed at its origin with approximately 5\% more mass than it has now \citep{YoungSun}:
 \ba
 M_\odot\s=\s 1.05\times M_\odot^{\rm(present)}\s=\s 2.09\times 10^{30}~\mbox{kg}\;\,.
 \label{}
 \ea

 \section{
 Derivation of formulae (\ref{refinedk2} - \ref{refinedM})
 }
 %  {\bf Derivation\, of\, formulae~ (\ref{refinedk2} - \ref{refinedM})}
 \label{Appendix1}

 \subsection{First iteration}

 A magma-ocean stage, which Mars underwent within $10 - 20$ million years post-accretion \citep{Magma}, ensured rapid early differentiation.
 Hence a motivation to endow the MOI factor $\s \xi\s$ with its present-day value $\s\xi^{\rm(present)}= 0.36379\s$. Using the other values from the table in Appendix \ref{description}, and setting $\left(1\,+\,{M_m}/{M}  \right)^{1/3}\approx 1$ in inequality (\ref{doubleineq}), we approximate that equality with
 \ba
  1.0914\s\left(\frac{3\s J_2}{k_2}\right)^{2/3}<\,
 \frac{M_m}{M}\,<\,\left(\frac{J_{22}}{J_2}\right)^{\rm (present)}\;\,.
  \label{}
 \ea
 % \ba
 %  1.0914\s\left(2\s J_2\s\left(1\s+\s{\cal B}_2\s\mu\right)\right)^{2/3}<\,
 %  \frac{M_m}{M}\,<\,\left(\frac{J_{22}}{J_2}\right)^{\rm (present)}.\qquad
 %  \label{}
 %  \ea
For values from said table, this becomes
\ba
3.5514\times 10^{-2}\,k_2^{-2/3}
\,<\,
 \frac{M_m}{M}\,<\,3.2253\times 10^{-2}\;\,.
\label{M}
\ea
 %  \ba
 %  2.7102\times 10^{-2}
 %  \left(1\s+\s{\cal B}_2\s\mu\right)^{2/3}<\,
 %  \frac{M_m}{M}\,<\,3.2253\times 10^{-2}\;\,.
 %  \label{ddouble}
 %  \ea
 The ensuing inequality
 $\s 3.5514\times 10^{-2}\,k_2^{-2/3} < \s 3.2253\times 10^{-2}\s$ entails a limitation on the Love number value: $k_2 >1.155\s.\s$
 %  and on the shear rigidity: $\mu < 1.545$ GPa.
 Rounding this, we arrive at the following constraints:
 \ba
 1.16\,\lesssim \,k_2\,\lesssim\,1.44\,\;,
 \label{36}
 \ea
  \ba
 1.54~\mbox{GPa}\,\gtrsim\,\mu\,\gtrsim\,0.20~\mbox{GPa}\,\;,
 \label{newmu}
 \ea
 \ba
 5.84  \,\lesssim\,  \frac{r}{R}  \,\lesssim\, 6.26 \;\,.
 \label{newdouble}
 \ea
 The left bound in inequality (\ref{36}) originates from formula (\ref{M}). The left bound in inequality (\ref{newmu}) was obtained from equations (\ref{36}), (\ref{kl}) and (\ref{Bl}). The left bound in inequality (\ref{newdouble}) was then deduced by combining equations (\ref{36}) and (\ref{inta}), with $\left(1\,+\,{M_m}/{M}  \right)^{1/3}\approx 1$ set in the latter.

 The right-hand bounds in these inequalities originate from geophysical considerations provided in Section~\ref{geoph}, e.g. equations (\ref{mu}), (\ref{kl}), and (\ref{double}). Given the low value of $\mu$, these bounds correspond to a situation where the moon synchronised Mars already at the magma ocean stage.

  We would reiterate that the left-hand bounds stem from the spin-orbit state stability condition, equation (\ref{23}). Given the higher value of $\mu$, these bounds correspond, arguably, to synchronisation at the beginning of solidification.

 Inserting into expression (\ref{M}) the largest value of $k_2$ suggested by geophysical reasoning, $k_2= 1.44$,  we obtain the minimum value for $M_m/M$. After rounding, we have:
 \ba
 2.78\times 10^{-2}
 \lesssim\,
 \frac{M_m}{M}\,\lesssim\,3.23\times 10^{-2}\;\,,
 \label{doubledouble}
 \ea
 the left and right bounds having the same meaning as those in equations (\ref{36} - \ref{newdouble}).

 \subsection{Second iteration}

 The double inequality (\ref{doubledouble}) demonstrates that $M_m/M$ is close to $3\times 10^{-2}$, and therefore $\left(1\,+\,{M_m}/{M}  \right)^{1/3}\approx 1.0099$. Insertion thereof into expression (\ref{int}) renders us, instead of approximation (\ref{inta}), a more accurate expression
 \ba
 \frac{r}{R}\;=\;5.54\,\left(1\,+\;\frac{M_m}{M}  \right)^{1/3}\!k_2^{\s 1/3}
 \;=\;5.60\;\sqrt[3]{k_2}\,\;.
 \label{refine}
 \ea
 Insertion of $\left(1\,+\,{M_m}/{M}  \right)^{1/3}\approx 1.0099$ into
 inequality (\ref{inequality}) gives
  \ba
 1.1022\s\left(\frac{3\s J_2}{k_2}\right)^{2/3}<\s\frac{M_m}{M}\;\,,
 \label{du}
 \ea
 %  \ba
 %  1.1022\s\left(2\s J_2\s\left(1\s+\s{\cal B}_2\s\mu\right)\right)^{2/3}<\,\frac{M_m}{M}\;\,,
 %  \label{}
 %  \ea
 whence $\s k_2>1.1726\s$ and, by equation (\ref{kl}), $\s\mu<\s 1.4467~\mbox{GPa}\s$. Rounding, we write:
 \ba
 1.17\,< \,k_2\,<\,1.44\,\;,
 \label{aa}
 \ea
 \ba
 1.45~\mbox{GPa}\,>\,\mu\,>\,0.20~\mbox{GPa}\,\;.
 \label{bb}
 \ea
  The right-hand bounds in equations (\ref{a}) and (\ref{bb}) are the same as in equations (\ref{36}) and (\ref{newmu}), correspondingly, because these bounds come from geophysical reasoning explained in Section~\ref{geoph} and are unaltered by our iterations.

 To obtain the maximum value of $r/R$, we insert
 %  $\left(1\,+\,{M_m}/{M}  \right)^{1/3}\approx 1.0099$ and
 the maximum $k_2$ from (\ref{aa}) into expression (\ref{refine}) and arrive at $r/R < 6.32\s$.
 To get the minimum value of $r/R$, we {\underline{do \it not}} insert the minimum $k_2$ into (\ref{refine}), because a stronger constraint $r/R > 5.91$ follows from the insertion of  $\left(1\,+\,{M_m}/{M}  \right)^{1/3}\approx 1.0099$ into the dynamical inequality (\ref{23}). Summing up, we have
 \ba
 5.91  \,<\,  \frac{r}{R}  \,<\, 6.32 \;\,.
 \label{c}
 \ea
 Finally, inserting into equation (\ref{du}) the maximum Love number value permissible by geophysical considerations, $k_2 = 1.44~\mbox{GPa}$, we get
  \ba
 2.81\times 10^{-2}
 \lesssim\,
 \frac{M_m}{M}\,\lesssim\,3.23\times 10^{-2}\;\,,
 \label{d}
 \ea
As in equation (\ref{doubledouble}), in equation (\ref{d}) the right-hand bound is but a rounded right-hand bound from equation (\ref{M}).

We may stop here, because the third iteration will change the bounds on the above quantities by about $1\m\%$.

 \section{Tides on Mars before the ocean formation}\label{quality}

  We model the young Mars' tidal response with that of a homogeneous sphere having a complex compliance
 \ba
 \bar{J}(\chi)\,=\,{\cal R}{\it e}\left[\bar{J}(\chi)\right]\,+\,i\;{\cal I}{\it m}\left[\bar{J}(\chi)\right]\,\;,
 \label{}
 \ea
 where $\chi$ is a short notation for the physical frequency exerted in the body by tides. Such frequencies are equal to the absolute values of the corresponding Fourier tidal modes. Using $\omega$ as a concise notation for a tidal mode,
  \ba
 \omega\,\equiv\,\omega_{lmpq}\,\;,
 \label{}
 \ea
 we write the corresponding frequency as
 \ba
 \chi\,\equiv\,|\omega|\,=\,|\omega_{lmpq}|\;\,.
 \label{}
 \ea
 A degree-$l$ quality function of a tidally perturbed body is defined as
 \ba
 K_l(\omega)\,\equiv\,k_l(\omega)\s\sin\epsilon_l(\omega)\,=\,\frac{k_l(\omega)}{Q_l(\omega)}\,\operatorname{Sign}\,\omega\;\,,
 \label{}
 \ea
 where $k_l(\omega)$, $\epsilon_l(\omega)$, $Q_l(\omega)$ are the degree-$l$ Love number, phase lag, and tidal quality factor.

 For a homogeneous near-spherical body of rheology $\bar{J}(\chi)$, the quality function reads (see, e.g. \citeauthor{Efroimsky2015} \citeyear{Efroimsky2015}, Eqn 40):
 \ba
 \nonumber
 K_{l}(\omega)  \equiv  {k}_l(\omega)\,\sin\epsilon_{l}(\omega)\s=\s
 {k}_l(\chi)\,\sin\epsilon_{l}(\chi)\,\operatorname{Sign}\,\omega\s=\s
 \\
 \nonumber\\
 - \;\frac{\textstyle 3}{\textstyle 2(l-1)}~\frac{\textstyle {\cal{B}}_{\textstyle{_l}}\;\;{\cal I}{\it m}\left[\bar{J}(\chi)\right] \;\,\operatorname{Sign}\,\omega
 }{\textstyle \left(\,{\cal R}{\it e}\left[\bar{J}(\chi)\right]+{\cal{B}}_{\textstyle{_l}\s}\right)^2+\left(\,{\cal I}{\it m}
 \left[\bar{J}(\chi)\right]\,\right)^2}\;\,,
 \label{63a}
 \ea
  where
  \ba
 {\cal{B}}_{l}\,\equiv~\frac{\textstyle{(2\,l^{\,2}\,+\,4\,l\,+\,3)}}{\textstyle{l\,\mbox{g}\,
 \rho\,R}}~=\;\frac{\textstyle{3\;(2\,l^{\,2}\,+\,4\,l\,+\,3)}}{\textstyle{4\;l\,\pi\,
 G\,\rho^2\,R^2}}~\;,
 \label{B}
 \ea
 $G$ being Newton's gravity constant, and $g$, $\rho$, $R$ being respectively the surface gravity, density, and radius of the body. Below we shall also need a convenient quantity ${\cal{A}}_{l}$ often used in the literature:
 \ba
 {\cal{A}}_{l}\s\equiv\s{{\cal{B}}_{l}}\,{J^{-1}}=\s{\cal{B}}_{l}\,\mu\,\;.
 \label{A}
 \ea

\begin{figure}
  %     \begin{minipage}{\textwidth}  \centering
  \includegraphics[width=0.44\textwidth]{
    %  K2(omega).pdf
   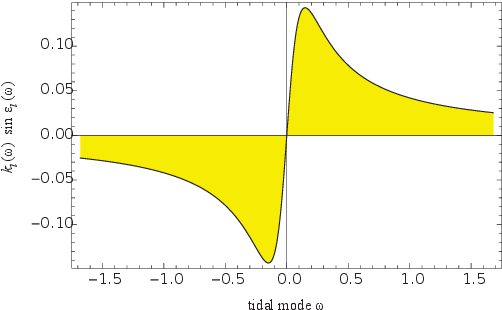
  }
             \caption{. \small A typical shape of the quality function $\,k_l(\omega)\,\sin\epsilon_l(\omega)\,$, ~where $\,\omega\,$ is a shortened notation for the tidal Fourier mode $\,\omega_{\textstyle{_{lmpq}}}\,$. \,(From Noyelles et al. 2014.)}
  %     \end{minipage}
         \label{figure1}
     \end{figure}

 For simple rheologies, like Maxwell or Andrade, the quality function has the form of a kink with one negative and one positive peak, as in Figure \ref{figure1}.

 \section{A Maxwell Mars}

 \subsection{Tidal response of a Maxwell planet}
 \label{TidalMaxwell}

For a Maxwell medium, the complex compliance is
  \ba
 ^{\rm{(Maxwell)}}\bar{J}(\chi)\s=\s J\s+\s i \s\left(\s-\s\frac{1}{\eta\s\chi}  \right)\,=\s J\s\left(1\s-\s\frac{i}{\tau_{\rm_M}\s\chi}  \right)\,\;,
 \label{Maxwell}
 \ea
 $\eta$ being the shear viscosity, $J$ being the unrelaxed sheer compliance (the inverse of the unrelaxed shear rigidity $\mu$), and $\tau_{\rm_M}=\eta\s J=\eta/\mu$ being the Maxwell time.

 Its insertion in equation (\ref{63a}) yields:
 \ba
 K_l(\omega)\,=\,\frac{3}{2(l-1)}~\frac{ \tau_{\rm_M}\s\omega\;{\cal A}_l }{1\,+\, \left(\tau_{\rm_M}\s\omega\right)^2\,\left(1\s+\s{\cal A}_l  \right)^2    }\;\;.
 \label{function}
 \ea
  [\,Mind a misprint in \citet[Eqn B.7]{Pathways}.\s]

 As demonstrated in \citet[Section 4.4]{walterova}, among others, the peak of this kink is located at
 \ba
 {\omega_{\rm peak}}_{\textstyle{_l}}\,=\;\pm\;
 \;\frac{\tau_M^{-1}}{1\,+\,{\cal{A}}_l}
 %  \;\approx \;\pm\;\frac{\tau_M^{-1}}{{\cal{A}}_l}
 \,\;,
 \label{wh}
 \ea
 and has a viscosity-independent magnitude
 \ba
 K_l^{\rm{(peak)}}\s=\;\pm\;\frac{3}{4\s(l-1)}\;\frac{ {\cal A}_l }{ 1 + {\cal A}_l }\,\;.
 \label{see}
 \ea
 [\,Mind a misprint in \citet[Eqn 62]{2022AdGeo..63..231B}.\s]

 The magnitude remains independent of viscosity also when the mantle cools down and begins to obey the Andrade rheology, which is caused by the emergence of transient processes. These dissipation mechanisms usually show up at frequencies much higher than the peak frequency (\ref{wh}), see \citet{walterova}.

 From equation (\ref{function}), we deduce that for a Maxwell body a quality function $\s K_l(\omega)\s$ is nearly linear in the interval between peaks:  \footnote{~In the product $\tau_{\rm_M}\s\omega\s{\cal A}_l\s=\s{\cal B}_l\s\eta\s\omega
 \s$, the dependence on $\mu$ % actually
 cancels, and the resulting quality function in equation (\ref{444}) depends on the product of viscosity and frequency: $\;K_l(\omega)\s=\s\frac{\textstyle 3}{\textstyle 2\s(l-1)}\,{\cal{B}}_l\s\eta\s\omega\,$.}
 \ba
 \nonumber
 |\s\omega\s|\;<\;|\s{\omega_{\rm peak}}_l\s|\quad\Longrightarrow\qquad K_l(\omega)\simeq\qquad\qquad\qquad
 ~\\
 \label{444}\\
 \frac{3}{2\s(l-1)}\s\tau_{\rm_M}\s\omega\s{\cal A}_l=\s
 \frac{3}{2\s(l-1)}\frac{ {\cal A}_l }{ 1 + {\cal A}_l }\frac{\omega}{|\s{\omega_{\rm peak}}_l\s|}
 \,.
 \nonumber
 \ea
 Outside the peaks, the function falls off as the inverse $\omega\,$:
 \ba
 \nonumber
 |\s\omega\s|\;>\;|\s{\omega_{\rm peak}}_l\s|\quad\Longrightarrow\qquad K_l(\omega)
 \simeq\qquad\qquad\qquad
 ~\\
 \label{do}
 \label{555}\\
 \frac{3}{2(l-1)}\frac{ {\cal A}_l (\tau_{\rm_M} \omega)^{-1}}{ (1 + {\cal A}_l)^2 }
 =
 \frac{3}{2 (l-1)}\frac{ {\cal A}_l }{ 1 + {\cal A}_l }\frac{|{\omega_{\rm peak}}_l|}{\omega}
 \,\;.\;
 \nonumber
 \ea

 As explained in Section \ref{action}, in the setting of our concern $\omega$ is the semidiurnal frequency of the solar tides on Mars,
 and is equal, to a good precision, to $\;-\,2\s\Omega$, where $\Omega$ is Mars' rotation rate. Consequently, a quality function $K_{l}(\omega)$ becomes $\;-\s K_{l}(2\s\Omega)\s$.

 We now need to know if the tidal frequency $2\s\Omega$ is located between the peaks in Figure \ref{figure1} or outside the inter-peak interval. To this end, we recall that $\Omega$ is related to the semimajor axis $a$ by equation (\ref{OmegaS}), where the semimajor axis is identified with the synchronous radius: $a=r$.
 Equation (\ref{refinedr}) sets an upper bound for the synchronous radius: $r<6.32\s R\s$. Therefore,
 \ba
 \nonumber
 2\s\Omega=2\s\sqrt{\frac{G\s(M+M_m)}{r^3}} \! &>& \! 2\s\sqrt{\frac{G\s(M+M_m)}{R^3}}\,6.32^{-3/2}\;\;\\
 \label{semidiurnal}\\
                                               \! &=& \! 1.34\times 10^{-4}\;\mbox{s}^{-1}\;\,,
 \nonumber
 \ea
 a value to be compared with the positive peak frequency rendered by expression (\ref{wh}).

 During the magma-ocean stage, $\mu$ was at its minimum, so $1+{\cal{A}}_l\approx1$, and the positive peak frequency was close to the inverse Maxwell time: $\omega_{\rm peak} \approx \tau_{\rm_M}^{-1}=\mu/\eta\s$.  Whether $\tau_{\rm_M}^{-1}$ was higher or lower than $\s 10^{-4}\;\mbox{s}^{-1}$ at the magna-ocean stage is model-dependent. It is therefore unclear whether the tidal response of the planet during that stage should be described by equation (\ref{444}) or (\ref{555}).  On the other hand, the duration of that stage was shorter than 20 million years \citep{Magma}, so the system evolution over that period was insignificant relative to the subsequent changes of $\Omega$ and $r$.

 After the onset of crystallisation, the mantle viscosity could have resided, depending on the moisture content, between $10^{19}$ and $10^{21}~\mbox{Pa~s}$, at a reference temperature of 1600 K \citep{Breuer}. For a shear rigidity of $\mu\simeq 0.2\times 10^9$ Pa, the peak frequency was as low as $\simeq 10^{-11} - 10^{-9}~\mbox{s}^{-1}$. Even for a rigidity lower by four orders of magnitude, $|{\omega_{\rm peak}}_{\textstyle{_l}}|$ would still be lower than the semidiurnal frequency given by equation (\ref{semidiurnal}).

 We thus conclude that over Mars' early evolution, except for a very brief magma-ocean period, the frequency $2\s\Omega$ was residing outside the inter-peak interval in Figure \ref{figure1}. Consequently, except for that short period, the planet's tidal response was governed by expression (\ref{555}).  The quadrupole quality function was
 \ba
 \phantom{12}K_2(\omega)=\frac{3}{2}\frac{{\cal A}_2}{(1+{\cal{A}}_2)^2}\frac{1}{\tau_{\rm_M}\omega}=\frac{3}{2}
 \left( \frac{{\cal B}_2\mu}{ 1+{\cal{B}}_2\mu } \right)^2\frac{1}{{\cal B}_2\s\eta\s\omega}\,\;.
 \label{ref}
 \ea
 We reiterate that this expression is valid only for $\s|\omega|\s>\s|\omega_{\rm peak}|\s$.

 According to equation (\ref{minus}),
 $\s K(\omega)\approx\s-\s K(2\Omega)\,,\,$
 where
  \ba
 K_2(2\Omega)\,=\;\frac{D}{\Omega}
 \label{C.16}
 \ea
 and
 \ba D\,=\,\frac{3}{4}
 \left( \frac{{\cal B}_2\;\mu}{ 1\s +\s{\cal{B}}_2\mu } \right)^2\!\frac{1}{\s{\cal B}_2\s\eta\s}\;\;.
 \label{d}
 \ea
{As the mantle solidifies, $\mu$ attains the value of $18$~MPa or higher, and the expression for $D$ simplifies to}
\ba
D\s\simeq\s\frac{1}{2}\,\frac{1}{{\cal B}_2\,\eta}\;\,.
\label{C.18}
\ea

\subsection{Action of solar tides on a Maxwell Mars synchronised by a moon}
\label{SolarMaxwell}

     Because of the extreme shortness of the magma-ocean stage,  all or most of the evolution from point 2 to point 3 in Figure \ref{Figure_2} was taking place after that stage, i.e., when the mantle was already viscoelastic, not liquid, and the viscosity was high enough to ensure that the forcing frequency was located to the right of the peak in Figure \ref{figure1}. This ensures that the quality function assumed the form given by equation (\ref{C.16}).\,\footnote{~As explained in Appendix \ref{TidalMaxwell}, equation (\ref{555}), these conditions are satisfied for $\s|\omega |\s >\s |{\omega_{\rm peak}}_2 |\s=\s\tau^{-1}_{_M}/(1+\mu{\cal B}_2)\s\approx\s\tau^{-1}_{_M}\s=\s\mu/\eta\s$, i.e., for
 \ba
 \nonumber
 \tau_{_M} \s >\,|\omega |^{-1}\,=\,(2\,\Omega)^{-1}\,\simeq\,0.83\times 10^{4}~\mbox{s}
 %  \eta\,>\,\frac{\mu}{\omega}\,=\,\frac{\mu}{2\,\Omega}
 %  \,\simeq\,\frac{0.2\times 10^9~\mbox{Pa}}{2\times 10^{-4}~\mbox{rad/s}}\,\simeq\,10^{13}~\mbox{Pa~s}
 \;\,,
 \label{}
\ea
where the estimate $\Omega\simeq 0.6\times 10^{-4}$ rad/s is borrowed from Figure~\ref{Figure_2}.\label{estimate}
}
 With equation (\ref{a}) taken into account, expression (\ref{C.16}) becomes
 \ba
 K_2(2\Omega)\,=\;D\;\frac{a^{3/2}}{\sqrt{G(M+M_m)}}\,\;.
 \label{K}
 \ea
 Combining this  with formula (\ref{hyper.eq}), we arrive at the equation
 \ba
 \left(\frac{M_m}{R^2} a^{-2}\s-\s{3}\s\xi(M+M_m)a^{-4}\right)da =\s { -}\s3\s D\,\frac{M_\odot^2}{M}\frac{R^3}{a_{\rm_M}^6}\,dt\;\,,
 \label{capt.eq}
 \ea
% Inserting this expression into equation (\ref{hyper.eq}) furnishes us with
% \ba
% \left(X\s a^{-2}\s-\s Y\s a^{-4}\right)\s da\s=\,Z\s dt\;\,,
% \label{equation}
% \ea
% where
%  \ba
% X\s=\s\frac{G\s M_m}{\xi}\,\frac{1}{\sqrt{\s G(M\s +\s M_m)\s}\s}\s R^{\s -2}
% \;\,,
% \ea
% \ba
% Y\,=\,3\s\sqrt{G\s(M\s+\s M_m)}
% \label{}
% \ea
% and
% \ba
% Z\,=\,\frac{3}{\xi}\,
% \frac{M_\odot}{M\,}\,\frac{GM_\odot}{a_{\rm_M}^3}\,\left( \frac{R}{a_{\rm_M}}  \right)^3\frac{D}{\sqrt{G(M+M_m)}}
% \,\;.
% \label{}
% \ea
%This equation describes the slow evolution of the transient synchronous state of the system.
 integrating
  %  equation (\ref{capt.eq})
 whereof gives us:
 %  \ba
 %  t_3\, -\, t_2\, =\;-\;\frac{X}{Z}\,\left(a_3^{-1} - a_2^{-1}\right)\,+\,\frac{Y}{3\,Z}\,\left(a_3^{-3}\s-\,a_2^{-3}\right) \;\,,
 %  \label{}
 %  \ea
 %  This can be explicitly rewritten as:
 %  \ba
 %  \nonumber
 %  t_3\, -\, t_2\,=\!\!\!& \textcolor{red}{\bf +}& \!\!\! \frac{M_m \,M}{M_{\sun}^2} \,\frac{a_{\rm_M}}{3\s D}\,\left(\frac{a_{\rm_M}}{R}  \right)^5\left(a_3^{-1} - a_2^{-1}\right)\,\\
 %  \label{equation_a}\\
 %  &\textcolor{red}{\bf -}& \!\!\! \frac{
 %  \textcolor{red}{\bf 1}
 %  \s\xi\s (M + M_m)\s M\,a_{\rm_M}^6}{3 M_{\sun}^2 D R^3}\,
 %  (a_3^{-3}\s-\,a_2^{-3})\;\,\qquad~\qquad
 %  \nonumber
 %  \ea
 \ba
   \nonumber
 t_3\, -\, t_2\,=
  %    &\s& \!\!\!\!\!\!\!\!
 \frac{1}{3\s D}\,\left[\,\frac{M_m \,M}{M_{\sun}^2} \,\frac{a_{\rm_M}}{a_3}\,\left(\frac{a_{\rm_M}}{R}  \right)^5\left(1 - \frac{a_3}{a_2}\right)\,
 \right. \qquad\qquad\\
 \label{equation_b}\\
  %    &\s& ~~\,
  \left.
 -\,\xi\left(1 + \frac{\textstyle M_m}{\textstyle M}\right)\s
 \left(\frac{M}{M_{\sun}}\right)^2
 \left(\frac{ a_{\rm_M}}{R}\right)^3\left(\frac{ a_{\rm_M}}{a_3}\right)^3\left(1\s-\s\left(\frac{a_3}{a_2}\right)^{3}\right)\,\right]\;\,,
   \nonumber
 \ea
 where 2 and 3 are the starting and finishing points in Figure~\ref{Figure_2}.

 Using the value of $a_3$ given by expression (\ref{a3}), and borrowing the other parameters values from the table in Appendix \ref{description}, we arrive at
 \ba
   \nonumber
 t_3\, -\, t_2\,=\,\frac{1}{D}\,\left[\,
  1.4257\times 10^{13}\left(1 - \frac{a_3}{a_2}\right)\,
 \right.\qquad\quad\\
 \label{equation_c}\\
  \phantom{t_3\, -\, t_2\,=}{ -}
 \left. 0.47525\times 10^{13}\,
 \left( 1\s-\s\left( \frac{a_3}{a_2}\right)^{3} \right)
 \,\right]\;\,.
   \nonumber
 \ea
  This time interval is highly sensitive to the spatial excursion $\,a_2\s -\s a_3\,$ experienced by the moon between its initialisation of synchronism and its subsequent departure from the synchronous orbit at point 3.

 {For $\Omega\simeq 0.6\times 10^{-4}$~rad/s and realistic values of the viscosity, between $10^{19}$ and $10^{21}~\mbox{Pa~s}$, the insertion of expression (\ref{C.18}) into (\ref{C.16}) yields for $K_2$ unphysically small values $10^{-7} - 10^{-5}$ and, consequently, absurdly large values of $t_3-t_2$. For example, for $a_2-a_3=10,000$ km, the resulting time interval is in the hundreds of Gyr.}

{
Suppose we ignore expression (\ref{C.18}) and, for the sake of argument, set by hand a physically reasonable value  $|K_2|\simeq 0.7\times 10^{-2}$, to which the following value of $D$ corresponds:
 \ba
 D=0.42\times 10^{-6}\;\mbox{s}^{-1}\;.
\label{}
\ea
 For an $\,a_2\s -\s a_3\s =\s 10,000\;\mbox{km}\s$ descent, we then obtain a still prohibitively long value of $\,t_3\s -\s t_2\s =\s 102.37\s$ Gyr much exceeding the universe age.  For a descent as short as $\s a_2 - a_3 = 1000\;\mbox{km}\,$, Nerio's stay in the evolving transient synchronism lasts for $t_3-t_2 = 2.30$~Gyr,
    which is still too long.}

{
    Recall that
    given the current absence of any significant latitudinal distribution in the early Martian cratering record, Nerio's demise should be assigned to the {\ae}ons predating the termination of the LHB, that is, to a time earlier than $\simeq 3.4$ Gyr ago. For the distance  change $\,a_2\s -\s a_3\s$ of several thousand kilometres, this would require the value {$K_2\simeq 0.7\times 10^{-1}$}, which is too high for a Maxwell planet.}

    {
Thus, in the case of a Maxwell Mars,
% , even with frivolous tweaking of the parameters,
Nerio cannot reach the marginal point 3 in Figure \ref{Figure_2} before the formation of a palaeo-ocean on Mars, unless we engage in fine tuning by assuming that Nerio formed within less than a thousand km from the marginal point.}

{
We should, however, bear in mind that the Maxwell model describes mantle rheology only at extremely low frequencies. In practical situations, transient processes also contribute to the viscoelastic response and must therefore be accounted for. This necessitates the use of the Andrade model, which we introduce in Appendix \ref{TidalAndrade}.
}

\section{An Andrade Mars}

 \subsection{Tidal response of an Andrade planet}
 \label{TidalAndrade}

{
The Andrade compliance is a conventional short term for the combined Maxwell-Andrade compliance comprising viscoelastic response with transient processes. Actual minerals and mantles satisfy this rheological law to a reasonable precision. In the old literature
(see
\citet{Castillo} and references therein), this rheological law was expressed as
}
 \begin{subequations}
 \begin{eqnarray}
 {\bar{\mathit{J\,}}}(\chi)&=&J\,+\,\beta\,(i\chi)^{-\alpha}\;\Gamma\,(1+\alpha)\,-\,\frac{i}{\eta\chi}
  \label{112_1}
  \label{E3a}\\
 \nonumber\\
 &=& J\,+\,\beta\,(i\chi)^{-\alpha}\;\Gamma\,(1+\alpha)\,-\,i\,J\,(\chi\,\tau_{_M})^{-1}
 ~~,
 \label{112_2}
 \label{E3b}
 \end{eqnarray}
 {
  where $\s\Gamma\s$ is the Gamma function, while $\s\alpha\s$ and $\s\beta\s$ are the dimensionless and dimensional Andrade parameters.}

{To avoid using the fractional-dimensional quantity $\beta$, the compliance was reparameterised in  \citet{Efroimsky2012} as}
  \begin{eqnarray}
 %  ^{\textstyle{^{(A)}}}
 {\bar{\mathit{J\,}}}(\chi)
&=&J\left[1
  +(i\chi\tau_{_A})^{-\alpha}\s\Gamma(1+\alpha)-i(\chi\tau_{_M})^{-1}\right]\;\qquad
  \label{15a}
  \ea
  or, equivalently,
  \ba
  &
  %  ^{\textstyle{^{(A)}}}
 {\bar{\mathit{J\,}}}(\chi) =
  \nonumber
 J\s \bigg [ 1 \,+\,\Gamma(1+\alpha)\;\zeta^{-\alpha}\,(\chi\tau_M)^{-\alpha}\,\cos \left(\frac{\textstyle\alpha\pi}{\textstyle 2} \right) \bigg]
   ~\\   \label{15b}\\  \nonumber
   &
   \phantom{BB} -
 i\,J\s \bigg [ \Gamma(1+\alpha)\;\zeta^{-\alpha}\, (\chi\tau_M)^{-\alpha}\,\sin \left(\frac{\textstyle \alpha\pi}{\textstyle 2} \right) \,+\, (\chi \tau_M)^{-1} \bigg]\,~.~~~
 \ea
 \label{LL44}
 \end{subequations}
 {\!\!The parameter $\s\tau_{A}\s$
 entering equation (\ref{15a}) is called {\it{the Andrade time}} and is linked to $\beta$ through}
 \begin{eqnarray}
 \beta\,=\,J~\tau_{A}^{-\alpha}~~.
 \label{beta}
 \label{E4}
 \end{eqnarray}
 {The parameter $\zeta$ used in equation (\ref{15b}) and defined as}
  \ba
\zeta\s\equiv\s\frac{\tau_{A}}{\tau_{M}}
 \label{}
 \ea  {is sometimes referred to as
 {\it the dimensionless Andrade time}.
 }

 {While solid-state theory  \citep{Miguel} provides for the Andrade parameter $\alpha$ the value $1/3$, laboratory measurements on actual minerals \citep{Gribb, Fontaine, jackson2010} render values residing in the broad interval of $\s 0.15 - 0.4\s$. On the other hand, geodetic measurements performed on the Earth favour the narrower interval of $0.2 - 0.3$ \citep{Benjamin}.
 As a compromise, we converge on
 \ba
 \alpha\s =\s 0.25\;\,.
 \label{}
 \ea
  Insertion of equation (\ref{15b}) into the expression (\ref{63a}) for $K_l(\omega)$ renders a long and inconvenient formula. Fortunately, to the right of the peak in Figure \ref{figure1}, that formula can be safely approximated with \citep[Appendix C.2.2]{Efroimsky2012}:} {
 \begin{flalign}
 \nonumber
 &~~ K_l(\omega)= \\
 \label{U}\\
 \nonumber
 &\frac{3}{2\,(l-1)}\;\frac{{\cal A}_{\textstyle{_l}}}{({\cal A}_{\textstyle{_l}}+1)^2}\,\sin
 \left(\frac{\alpha\,\pi}{2}\right)\,\frac{\Gamma(\alpha+1)}{\zeta^{\alpha}\left(\tau_{M}\,\chi\right)^{\alpha}}~\operatorname{Sign}\s\omega\;,
 \label{}
 \end{flalign}
 where $\chi=|\omega|$.  Specifically, for $\alpha = 1/4$ and $l=2$,
  \begin{flalign}
 \nonumber
 &~~ K_2(\omega)= \\
 \label{U}\\
 \nonumber
 &\frac{3}{2}\;\left(\frac{{\cal B}_{\textstyle{_2}}\,\mu}{1+{\cal B}_{\textstyle{_2}}\,\mu}\right)^2\sin\left(
 \frac{\pi}{8}\right)\,\Gamma\left(\frac{5}{4}\right)
 \frac{\zeta^{\,-1/4}}{{\cal{B}}_2\s\mu}
 \;\frac{1}{\left(\tau_{_M}\,\chi\right)^{1/4}}\,\operatorname{Sign}\s\omega\;\,,
 \label{a}
 \end{flalign}
 }

 As we saw in Section \ref{action}, of interest to us is the case of $\omega\approx \s-\s 2\s\Omega$, where $\Omega$ is the rate of rotation of Mars. Equation~(\ref{minus}) then entails
 $\s K_2(\omega)\approx\s-\s K_2(2\Omega)\,,\,$
 where
   \ba
 K_2(2\Omega)\,=\;\frac{H\;}{\Omega^{1/4}}
 \label{em}
 \ea
 and
 \ba
 H=\s \frac{3}{2}\s\left(\frac{{\cal B}_{\textstyle{_2}}\,\mu}{1+{\cal B}_{\textstyle{_2}}\,\mu}\right)^2\sin\left(
 \frac{\pi}{8}\right)\,\Gamma\left(\frac{5}{4}\right)
 \frac{\zeta^{\,- 1/4}}{{\cal{B}}_2\s\mu}
 \,\frac{1}{\left(2\s\tau_{_M}\right)^{1/4}}
 \;\,.\;\;
 \label{HH}
 \ea

  On the bad side, there is no agreement in the literature on the span of the values of $\zeta$.\,\footnote{~{While some laboratory datasets indicated that $\zeta$ may be of order unity \citep{Castillo}, other laboratory and geodetic measurements suggest that its value may be orders of magnitude smaller or larger \citep{Barnhoorn, Jackson, Qu, Tan, Amorim_Gudkova_2024}.  Recent analysis of geodetic data by \citet{Amorim_Gudkova_2025} has demonstrated that for Earth's mantle the value of $\zeta$ is likely to lie in the interval $[0.2, \,0.485]$.
  }}

  On the good side, in the above expression this parameter is taken to the power of $-1/4$. So we assume that the values of $\zeta^{-1/4}$ are residing in the interval from $10^{-1}$ to $10^1$.

\subsection{Action of solar tides on an Andrade Mars synchronised by a moon}
\label{SolarAndrade}

 {As the mantle solidifies, its rigidity assumes values up to $10^{11}$~Pa. The viscosity resides in the interval $10^{19}$ to $10^{21}~\mbox{Pa~s}$ \citep{Breuer}.  For the reference values  $\mu=10^{10}$~Pa and $\eta=10^{20}$~Pa~s, we obtain
  \ba
  \label{H}
  H\s\simeq\s 3 \times 10^{-4\s\pm\s 1} ~\mbox{s}^{-1/4}
 \ea
 where $\pm 1$ originates from the uncertainty in our knowledge of the dimensionless Andrade time $\zeta$.
 }

{
Combining equations (\ref{a}) and (\ref{em}), we write
 \ba
 K_2(2\s\Omega)\s=\s\frac{H\, a^{3/8}}{[G\,(M+M_m)]^{1/8}}\;\,.
 \label{}
 \ea
 Plugging this into equation (\ref{hyper.eq}), we obtain the equation of spin-orbit evolution of a system consisting of an Andrade planet and a synchronous moon:
 \ba
 \left(\frac{M_m}{M+M_m} x^{-7/8}\s-\s{3}\s\xi \s x^{-23/8}\right)\s dx\, =\,-\,F\,dt\;\,,
 \label{F}
 \ea
 where the constant $F$
  is expressed through Mars' mean motion $n_{\rm{_M}}= 1.085\times 10^{-7}$ rad/s
 %  \ba
 %  n_{\rm{_M}}\s=\s\sqrt{\frac{G\s M_\odot}{a_{\rm {_M}}^3}}\;\,.
 %  \label{}
 %  \ea
 and the auxiliary quantity
 \ba
 b\s=\s\sqrt{\frac{G\s (M+M_m)}{R^3}} = 1.065\times 10^{-3}\;\mbox{s}^{-1}
 \label{}
 \ea
as
\ba
 F\,=\,3\,\frac{M_\odot}{M}\,\left(\frac{n_{\rm{_M}}}{b}\right)^2\s\left(\frac{R}{a_{\rm {_M}}} \right)^3 b^{3/4}
 H\,=\,5.9\times 10^{-22\pm 1}\,\mbox{s}^{-1}\;\,.
 \label{F}
 \ea
 Integration of equation (\ref{F}) yields:
 \ba
 \label{prohor}
 t_3\s-\s t_2\s=\qquad\qquad\qquad\qquad\qquad\qquad\quad\qquad\qquad\qquad\\
 \nonumber\\
 \nonumber
 \frac{8}{F}\s\left[
 \frac{\xi}{5}\s x_3^{-15/8}\left(\s\left(\frac{a_3}{a_2}  \right)^{15/8} - 1\right)
 +
 \frac{M_m}{M+M_m}\s x_3^{1/8}\left(\s\left(\frac{a_2}{a_3}  \right)^{1/8} - 1\right)\right]\;,
 \ea
 where $x_3\equiv a_3/R=6.12\s$.
 }

{Setting $a_2-a_3=10,000$ km, we arrive at a time interval, which is not as large as in the Maxwell Mars case, but still exceeds the age of the Universe.
For $a_2-a_3=1,000$ km, the time interval $t_3-t_2$ is still too large.\,\footnote{~{For $a_2-a_3=10,000$ km, we arrive at $t_3-t_2\approx 2.47\times 10^{2\pm 1}$ Gyr, which even for $\zeta^{-1/4} = 10$ exceeds the age of the Universe. For $a_2-a_3=1,000$ km, we obtain $t_3-t_2\approx 0.843\times 10^{2\pm 1}$ Gyr, which for $\zeta^{-1/4} = 10$  is longer than the age of the Solar system.}}
While $t_3-t_2$ becomes shorter for $\zeta \ll 1$, it is still unlikely that an Andrade Mars was capable of reaching the marginal point 3 before the ocean formation.
}

\section{A Mars with an ocean}\label{ocean}

\subsection{Tidal response of an ocean-bearing Mars}

For bodily tides, a degree-$l$ component of the decelerating tidal torque scales as  $a_{\rm_M}^{-2(l+1)} {\cal I}\!{\it m}\{\bar{k}_l\}\s$, where $a_{\rm_M}$ is the semimajor axis of Mars' orbit about the Sun \citep[Eqn 106]{Efroimsky2012}. As customary, $\bar{k}_l$ signifies a complex Love number, so that ${\cal I}\!{\it m}\left\{ {\bar{k}}_l\right\}\s=\,-\,k_l\s\sin\epsilon_l\s$, where $k_l$ and $\epsilon_l$ are the frequency-dependent real Love number and phase lag.  As explained by \citet[Appendix K]{Farhat} and \citet[Section 2.8]{Auclair}, the oceanic tidal torque is expressed similarly, though contains different, oceanic Love numbers. Thus, bodily and oceanic tides can be pictured as a single process~---~an overall tidal response of the planet, governed by the total Love numbers
\footnote{~The fact that the tidal torque is proportional to $a_{\rm_M}^{-2(l+1)} {\cal I}\!{\it m}\{\bar{k}_l\}\s$ may imply a complicated overall $a_{\rm_M}$-dependence. Both the solid-body and oceanic Love numbers
depend, in different manners, upon the tidal frequencies $\omega$ involved. There frequencies, in their turn, depend on the mean motion and therefore on the semimajor axis $a_{\rm_M}$. This imposes $a_{\rm_M}$-dependencies on both ${\cal I}\!{\it m}\left\{ {\bar{k}}_l \right\}$ and ${\cal I}\!{\it m}\left\{ {\bar{k}}^{\rm{(ocean)}}_{lm} \right\}$. Such dependencies may be especially pronounced for the oceanic Love numbers, owing to the basin mode resonances.
}
\ba
\bar{k}^{\rm{(total)}}_{lm}\s=\s\bar{k}_{l}\,+\,\bar{k}^{\rm{(ocean)}}_{lm}\;\,,
\label{sum}
\ea
where the first term is the body-tide Love number, while the second is the oceanic Love number. By distinction from the first term, the second one depends on both the
degree $l$ and order $m$ due to the Coriolis force influencing the ocean's response.

The total semidiurnal solar torque (\ref{torque.eq}) then becomes
\ba
 \nonumber
 {\cal T}^{\s\rm(total)}_{22} &=& \frac{3}{2}\,{M_\odot}\,\frac{GM_\odot}{a_{\rm_M}^3}\,\left( \frac{R}{a_{\rm_M}}  \right)^3 \! R^{\s 2} K^{\rm(total)}_{22}(\omega)\\
 \label{torque.total}\\
&=&-\;\frac{3}{2}\,{M_\odot}\,\frac{GM_\odot}{a_{\rm_M}^3}\,\left( \frac{R}{a_{\rm_M}}  \right)^3 \! R^{\s 2}\,{\cal I}\!{\it m}\left\{ \bar{k}^{\rm(total)}_{22}(\omega) \right\}\;\,.
 \nonumber
 \ea

The overall tidal dissipation budget of  Earth is dominated by tides in the ocean. While establishing an exact fluid/solid partition is difficult, it is believed that friction in the solid Earth consumes   % between $3\%$  \citep{Ray1996} and
about $5\%$ \citep{Ray2001} of the tidal power damped. So the ocean dissipates  $\sim\! 20$ times more energy than solid Earth.

For other terrestrial planets, however, the tidal response of the ocean can vary greatly, depending on the form of basin, the range of depth, seafloor topography, and the planet's rotation rate. \citet{Barnes} argue that, depending on the continental shapes, the oceanic dissipation could vary by five  orders of magnitude.

Two thirds of Earth's oceanic dissipation occur in narrow straits and shallow seas. By distinction from Earth, the entire ocean on Mars was mainly shallow~---~which served as a booster of dissipation. On the other hand, the shape of the ocean also played a critical role in tidal dynamics. A near-circular shape
 %  with few straits
would result in a lower dissipation rate. In contrast, a non-symmetric basin with a jagged and rugged coastline would have supported a rich spectrum of normal modes defined by its geometry. Solar forcing could  have excited one or more of these modes into resonance during the ocean's lifetime, potentially increasing oceanic dissipation by orders of magnitude. A further boost could emerge from straights. The exact shape of the Martian palaeo-ocean still remains a subject of exploration.

 \subsection{Estimated magnitude of Mars' tidal response}
\label{estimated}

 To our knowledge, the most accurate investigation of the tidal response of ocean-bearing planets is provided by
 \citet{Auclair2018},
 \citet{Farhat} and \citet{Auclair}.
 Figure 5 in \citet{Auclair2018} illustrates  frequency-dependencies of $|K_2|$ for various models of Earth and TRAPPIST-1{\s}f. The models were built for planets with deep oceans.  While in Earth's case the values of $|K_2|$ at low frequencies are typically below or close to $0.1$, resonant peaks at low frequencies rise close to and even above $10\s$.
 %  Had Mars' ocean demonstrated such peaks, a crossing of those would eject Nerio from its synchronous orbit within several dozens of million years.

 The palaeo-ocean on Mars was shallow and therefore much more dissipative than the oceans modelled in  \citet{Auclair2018, Auclair2019, Auclair} and \citet{Farhat}.  For this reason, we adopt for it the estimate
 \ba
 K_2\,\simeq\;0.7\,\;.
 \label{estim}
 \ea
 Reasonably conservative, this value  ensures  Nerio's arrival at point 3 in Figure~\ref{Figure_2}, see Section \ref{palaeo}. The presence of peaks would boost dissipation and strengthen our conclusions.

      \begin{figure}[h]
   %   \begin{minipage}{\textwidth}  \centering
   \hspace*{-4.0mm}
  \includegraphics[width=0.489\textwidth]{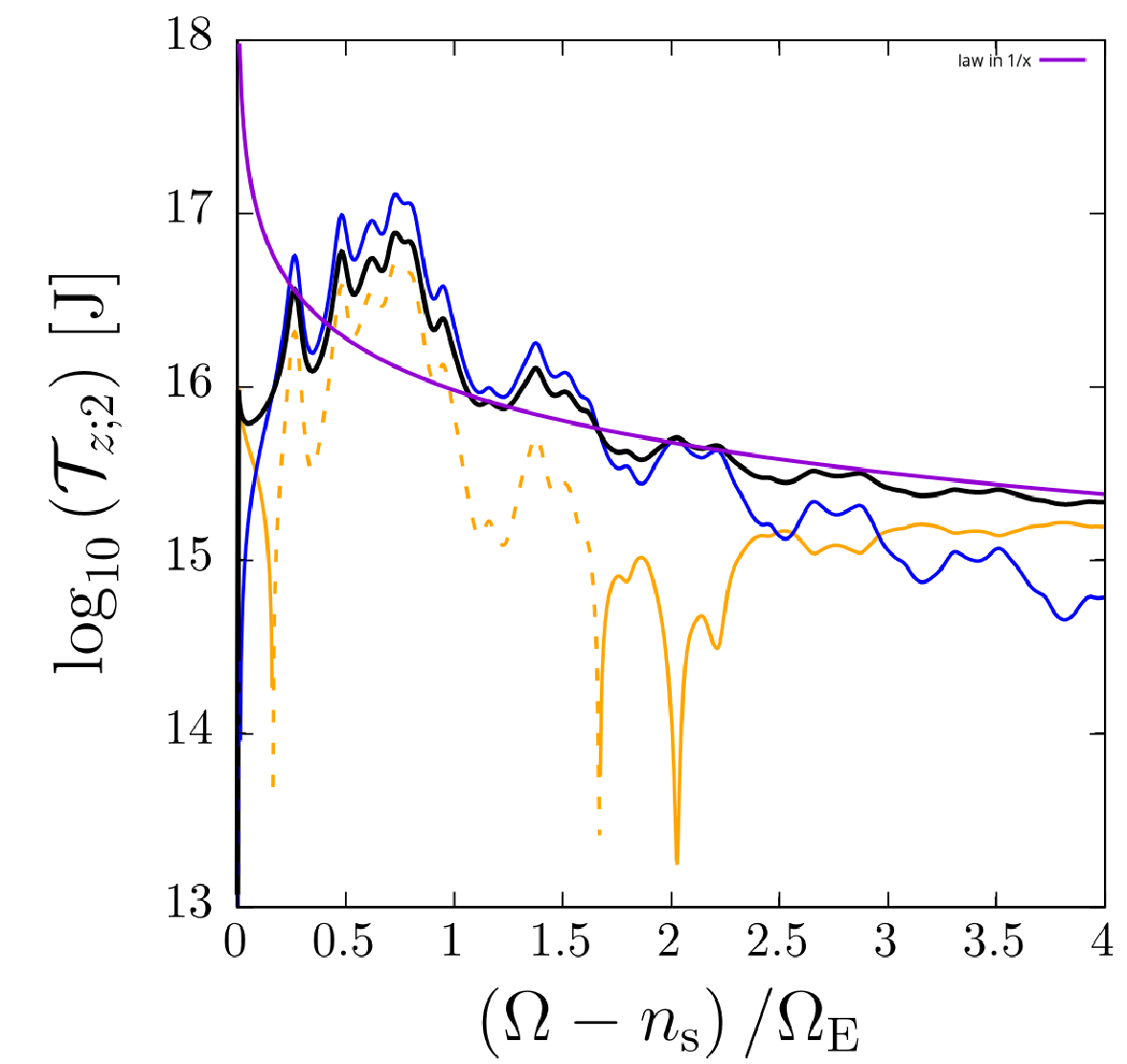}
             \caption{. \small {Courtesy of Gwena\"{e}l Bou\'{e}, this figure is based on the theory developed by \citet{Auclair} for an ocean planet with a continent. Depicted is the quadrupole part of the tidal torque as a function of the semidiurnal tidal frequency $(\Omega - n_s) / \Omega_E$ normalised to
the spin rate $\Omega_E$ of the present-day Earth. The notation $n_s$ stands for the mean motion. In our case, $n_s = n_M\ll\Omega$ and therefore $(\Omega - n_s)/\Omega_E\approx\Omega/\Omega_E$.
The blue and yellow lines show the contributions from the ocean and solid parts, respectively. The total torque is given by the black line.  It is well fit by the  purple curve, which is the inverse tidal frequency. Therefore, except in the low-frequency limit, the average form of the tidal response of an ocean planet with a continent is similar to that of a Maxwell planet. One nonetheless should use this approximation with caution, because at some frequencies it is acting as a lower bound.}
}
  %    \end{minipage}
         \label{Boue}
     \end{figure}
 \subsection{Average form of the frequency dependence}\label{Appendix_Boue}

{
 Kindly produced for us by Gwena\"{e}l Bou\'{e}, Figure \ref{Boue} is based on a model describing tides in an ocean planet with a continent of a spherical cap form \citep{Auclair}.  The plots show the quadrupole tidal torque as a function of the semidiurnal tidal frequency $(\Omega - n_s) / \Omega_E$ normalised to the rotation rate $\Omega_E$ of the present-day Earth. The notation $n_s$ stands for the mean motion. (In our case, $n_s = n_M\ll\Omega$, for which reason $(\Omega - n_s)/\Omega_E\approx\Omega/\Omega_E$.)
The blue and yellow lines show the inputs from the ocean and solid parts, appropriately. The combined torque is given by the black line, and is well fit by the purple curve, which to a factor is the inverse tidal frequency. Therefore, except in the low-frequency limit, the average tidal response of an ocean-bearing planet is similar to that of a Maxwell planet. This being said, one still needs to be cautious with this inverse-
frequency scaling, because at some frequencies the discrepancy
between the black and purple curves becomes appreciable, and
the purple curve serves more like a lower bound.}

\section{\mbox{Did
 Nerio reach the Roche limit?}}
\label{calcula}

On arrival at the marginal point 3 in Figure \ref{Figure_2}, Nerio left synchronism and began to follow the track aimed towards the Roche limit (the straight line which is tangent to the cubic curve at point 3).
Since by the time of Nerio's arrival at  point 3 the LHB was already going on (see Section \ref{palaeo}), Nerio could well have perished soon after its arrival at that point. It would nonetheless be worth exploring if it had a chance to make it all way to the Roche radius. As we shall see shortly, this scenario would be conceivable only for an ocean much more dissipative than proposed by the currently available studies.

\subsection{Mars' rotation rate at the moment of Nerio's hypothetical crossing of the Roche limit}\label{E1}

For $M_m = 0.03 M$, the values of Nerio's semimajor axis and Mars' spin rate at the marginal point were derived in Section~\ref{questions}:
\ba
a_3\s=\s 2.0629\times 10^7\s\mbox{m}\,,~~~~
\Omega_3\s=\,7.0294\times 10^{-5}~\mbox{rad/s}\;\,.
\label{}
\ea
The slope given by expression (\ref{X}) assumes the value of
\ba
X\,=\,4.4643\times 10^{-8}\;\mbox{m}^{-1/2}\,\mbox{s}^{-1}\,\;,
\label{}
\ea
wherefrom the integration constant (\ref{C}) is
\ba
C\,=\,X\sqrt{a_3}\,+\,\Omega_3\,=\,2.7306\times 10^{-4}\,\;.
\label{}
\ea

Approximating Nerio's density with that of our Moon, we obtain for the Roche radius \citep[Section 7.1]{Mars}:
\ba
r_{\rm R}\s=\,2.32\,R\,=\,7.8184\times 10^6\;\mbox{m}\,\;.
\label{}
\ea
 Consequently, by equation (\ref{Omega}), the Martian spin rate at the moment of Nerio's crossing the Roche limit is
 \ba
 \Omega_{\rm R}\s=\s-\s X\sqrt{r_{\rm R}}\s+\s C
 \s=\,1.4823\times 10^{-4}\;\mbox{rad~s}^{-1}\,\;,
 \label{r_R}
 \ea
 as illustrated by Figure \ref{Figure_3}.

\begin{figure}
  \hspace*{-6.0mm}\includegraphics[width=0.52\textwidth]{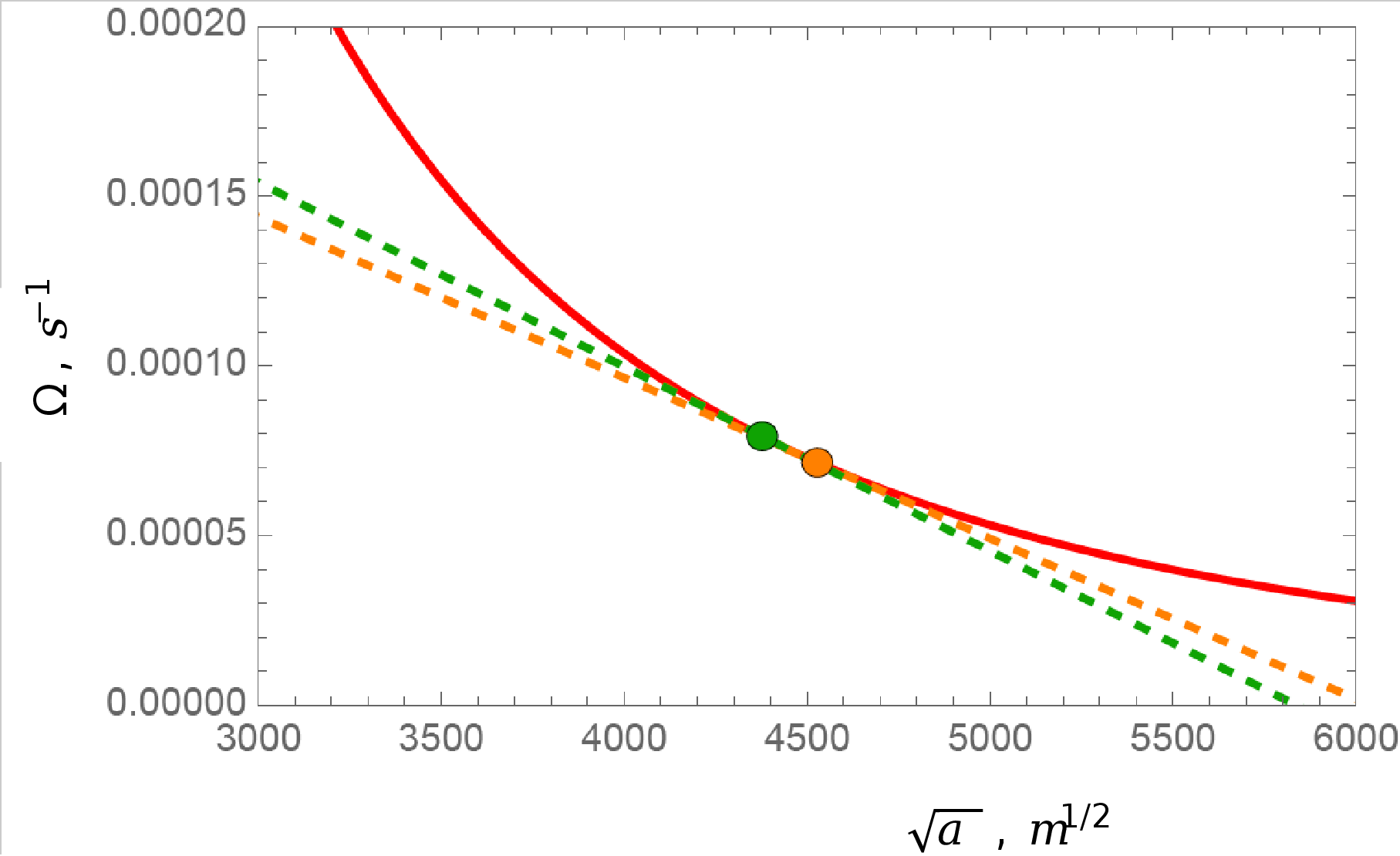}
             \caption{. \small This figure illustrates the sensitivity of $a_3$, $\Omega_3$, and $\Omega_{\rm R}$ to Nerio's mass $M_m$.
             The green tangent like, one with a steeper slope, is the evolution track corresponding to $M_m = 0.0323 \s M$ and rendering $\s a_3 = 1.9903\times 10^7\s\mbox{m}\s$,
 $\s\Omega_3 = 7.4258\times 10^{-5}$ rad/s, and
             $\Omega_{\rm R} = 1.5741\times 10^{-4}$ rad/s.
             The yellow line, one with a shallower slope, is the evolution track corresponding to $M_m = 0.0281 \s M$ and producing
             $\s a_3 = 2.1295\times 10^7\s\mbox{m}\s$,
 $\s\Omega_3 = 6.6960\times 10^{-5}$ rad/s, and
             $\s\Omega_{\rm R}= 1.4612\times 10^{-4}$ rad/s.
             }
          \label{Figure_5}
     \end{figure}

 The fourth and third decimal points in the above values are unreliable because of the uncertainties in Nerio's mass and in Mars' radius prior to the LHB.  Figure \ref{Figure_5} shows the sensitivity of
  $\Omega_{\rm R}$ to Nerio's mass $M_m$. The range of this mass is established by the double inequality (\ref{refinedM}).

   The maximum possible value $M_m = 0.0323 \s M$ produces, via formulae (\ref{marginal1}), (\ref{OmegaS}), (\ref{X}), (\ref{C}), and (\ref{Omega}), correspondingly:
 $\s a_3 = 1.9903\times 10^7\s\mbox{m}\s$,
 $\s\Omega_3 = 7.4258\times 10^{-5}\s\mbox{rad~s}^{-1}\s$,
 $\s X\s= 4.9933\times 10^{-8}\s\mbox{m}^{-1/2}\s\mbox{s}^{-1}$,
 $\s C = 2.9703\times 10^{-4}\s\mbox{rad~s}^{-1}\s$, and
 $\s\Omega_{\rm R}= 1.5741\times 10^{-4}\s\mbox{rad~s}^{-1}$. In Figure \ref{Figure_5},
 this situation is illustrated by the green tangent track, one with a steeper slope.

 Similarly, the yellow track, one with a shallower slope, corresponds to $M_m = 0.0281 \s M$, $\s a_3 = 2.1295\times 10^7\s\mbox{m}\s$,
 $\s\Omega_3 = 6.6960\times 10^{-5}\s\mbox{rad~s}^{-1}\s$,
 $\s X\s= 4.3530\times 10^{-8}\s\mbox{m}^{-1/2}\s\mbox{s}^{-1}$,
 $\s C = 2.6784\times 10^{-4}\s\mbox{rad~s}^{-1}\s$, and
 $\s\Omega_{\rm R}= 1.4612\times 10^{-4}\s\mbox{rad~s}^{-1}$.

 \subsection{Subsequent despinning of Mars by solar tides}
 \label{AppendixD}\label{E2}

The question now becomes whether solar tides in a Mars with an ocean were capable of slowing down Mars's rotation to its present rate. From equation (\ref{deceleration}), we observe that to overcome the difference $\Delta\Omega =  \Omega^{\rm\s(present)} -\s \Omega_{\rm R} = -\s 0.7735\times 10^{-4}$ rad/s over a time span of $\Delta t \simeq 0.4$~Gyr, the ocean-bearing Mars' quality function should have assumed the value
  \ba
K_2\,\lesssim\,\frac{2\s\xi}{3}\;\frac{M}{M_\odot}\left(\frac{a_{\rm_M}}{R}\right)^3\,\frac{a_{\rm_M}^{3}}{G\s M_\odot}\,\frac{|\Delta\Omega |}{\Delta t}\,\approx\, 12\;\,,
 \label{}
  \ea
which may be too high in the light of the currently available models of oceans' tidal response.

{Things will get even worse if we assume that the entire material of Nerio and its entire orbital momentum were deposited on Mars after destruction. In that case, the final angular velocity of Mars will be given not by equation (\ref{r_R}) but by the expression $\;-\s X\sqrt{R}\s+\s C
 \s=\,1.9110\times 10^{-4}\;\mbox{rad~s}^{-1}$, with Mars' radius $R$ standing under the root instead of $r_R$. This will render $\Delta\Omega=\s-\s 1.2020\times 10^{-4}\;\mbox{rad~s}^{-1}$
 and will result in a larger estimate for the quality function value, $K_2\,\lesssim\, 18\;$.}

This prevents us from concluding that Nerio reached the Roche limit, unless future studies indicate that the palaeo-ocean on Mars was more dissipative
than presumed presently.

\end{appendix}

\bibliography{references}

@ARTICLE{2006Natur.441..192A,
       author = {{Agnor}, C. B. and {Hamilton}, D. P.},
        title = "{Neptune's capture of its moon Triton in a binary-planet gravitational encounter}",
      journal = {Nature},
         year = 2006,
        month = may,
       volume = {441},
       number = {7090},
        pages = {192-194},
          doi = {10.1038/nature04792},
       adsurl = {https://ui.adsabs.harvard.edu/abs/2006Natur.441..192A}
}

@ARTICLE{Amorim_Gudkova_2024,
       author = {{Amorim}, D. O. and {Gudkova}, T.},
        title = "{Constraining Earth's mantle rheology with Love and Shida numbers at the M$_{2}$ tidal frequency}",
      journal = {Physics of the Earth and Planetary Interiors},
         year = 2024,
       volume = {347},
          eid = {107144},
        pages = {107144},
          doi = {https://10.1016/j.pepi.2024.107144},
          url = {https://10.1016/j.pepi.2024.107144},
          eprint = {https://10.1016/j.pepi.2024.107144},
       adsurl = {https://ui.adsabs.harvard.edu/abs/2024PEPI..34707144A}
}

@ARTICLE{Amorim_Gudkova_2025,
       author = {{Amorim}, D. O. and {Gudkova}, T.},
        title = "{Constraining the parameters of the Andrade rheology in Earth's mantle with Love numbers of 12 tidal constituents}",
      journal = {Physics of the Earth and Planetary Interiors},
         year = 2025,
       volume = {359},
          eid = {107304},
        pages = {107304},
          doi = {10.1016/j.pepi.2024.107304},
       adsurl = {https://ui.adsabs.harvard.edu/abs/2025PEPI..35907304A}
}

@INPROCEEDINGS{Andrews,
       author = {{Andrews-Hanna}, J.~C. and {Bottke}, W.~B.},
        title = "{Mars During the Pre-Noachian}",
    booktitle = {Fourth International Conference on Early Mars: Geologic, Hydrologic, and Climatic Evolution and the Implications for Life},
         year = 2017,
       editor = {{LPI Editorial Board}},
       series = {LPI Contributions},
       volume = {2014},
        month = oct,
          eid = {3078},
        pages = {3078},
       adsurl = {https://ui.adsabs.harvard.edu/abs/2017LPICo2014.3078A}
}

@article{Arnold,
doi = {10.1070/RM1998v053n01ABEH000005},
year = {1998},
month = {feb},
publisher = {},
volume = {53},
number = {1},
pages = {229},
author = {Vladimir I. Arnold},
title = {On teaching mathematics},
journal = {Russian Mathematical Surveys}
}

@ARTICLE{Auclair2018,
       author = {{Auclair-Desrotour}, P. and {Mathis}, S. and {Laskar}, J. and {Leconte}, J.},
        title = "{Oceanic tides from Earth-like to ocean planets}",
      journal = {Astronomy and Astrophysics},
         year = 2018,
       volume = {615},
          eid = {A23},
        pages = {A23},
          doi = {10.1051/0004-6361/201732249},
archivePrefix = {arXiv},
       eprint = {1801.08742},
 primaryClass = {astro-ph.EP},
       adsurl = {https://ui.adsabs.harvard.edu/abs/2018A&A...615A..23A}
}

@ARTICLE{Auclair2019,
       author = {{Auclair-Desrotour}, P. and {Leconte}, J. and {Bolmont}, E. and {Mathis}, S.},
        title = "{Final spin states of eccentric ocean planets}",
      journal = {Astronomy and Astrophysics},
         year = 2019,
        month = sep,
       volume = {629},
          eid = {A132},
        pages = {A132},
          doi = {10.1051/0004-6361/201935905},
archivePrefix = {arXiv},
       eprint = {1907.06451},
 primaryClass = {astro-ph.EP},
       adsurl = {https://ui.adsabs.harvard.edu/abs/2019A&A...629A.132A}
}

@ARTICLE{Auclair,
       author = {{Auclair-Desrotour}, P. and {Farhat}, M. and {Bou{\'e}}, G. and {Gastineau}, M. and {Laskar}, J.},
        title = "{Can one hear supercontinents in the tides of ocean planets?}",
      journal = {Astronomy and Astrophysics},
         year = 2023,
       volume = {680},
          eid = {A13},
        pages = {A13},
          doi = {10.1051/0004-6361/202347301},
archivePrefix = {arXiv},
       eprint = {2310.06635},
 primaryClass = {astro-ph.EP},
       adsurl = {https://ui.adsabs.harvard.edu/abs/2023A&A...680A..13A}
}

@ARTICLE{2022AdGeo..63..231B,
       author = {{Bagheri}, Amirhossein and {Efroimsky}, Michael and {Castillo-Rogez}, Julie and {Goossens}, Sander and {Plesa}, Ana-Catalina and {Rambaux}, Nicolas and {Rhoden}, Alyssa and {Walterov{\'a}}, Michaela and {Khan}, Amir and {Giardini}, Domenico},
        title = "{Tidal insights into rocky and icy bodies: an introduction and overview}",
      journal = {Advances in Geophysics},
         year = 2022,
        month = aug,
       volume = {63},
        pages = {231-320},
          doi = {10.1016/bs.agph.2022.07.004},
archivePrefix = {arXiv},
       eprint = {2206.04370},
 primaryClass = {astro-ph.EP},
       adsurl = {https://ui.adsabs.harvard.edu/abs/2022AdGeo..63..231B}
}

@ARTICLE{Barnes,
       author = {{Blackledge}, B.~W. and {Green}, J.~A.~M. and {Barnes}, R. and {Way}, M.~J.},
        title = "{Tides on Other Earths: Implications for Exoplanet and Palaeo-Tidal Simulations}",
      journal = {Geophysical Research Letters},
         year = 2020,
       volume = {47},
       number = {12},
          eid = {e85746},
        pages = {e85746},
          doi = {10.1029/2019GL085746},
       adsurl = {https://ui.adsabs.harvard.edu/abs/2020GeoRL..4785746B}
}

@ARTICLE{Barnhoorn,
       author = {{Barnhoorn}, A. and {Jackson}, I. and {Fitz Gerald}, J.~D. and {Kishimoto}, A. and {Itatani}, K.},
        title = "{Grain size-sensitive viscoelastic relaxation and seismic properties of polycrystalline MgO}",
      journal = {Journal of Geophysical Research (Solid Earth)},
         year = 2016,
        month = jul,
       volume = {121},
       number = {7},
        pages = {4955-4976},
          doi = {10.1002/2016JB013126},
       adsurl = {https://ui.adsabs.harvard.edu/abs/2016JGRB..121.4955B}
}

@ARTICLE{Benjamin,
       author = {{Benjamin}, David and {Wahr}, John and {Ray}, Richard D. and {Egbert}, Gary D. and {Desai}, Shailen D.},
        title = "{Constraints on mantle anelasticity from geodetic observations, and implications for the J$_{2}$ anomaly}",
      journal = {Geophysical Journal International},
         year = 2006,
       volume = {165},
       number = {1},
        pages = {3-16},
          doi = {10.1111/j.1365-246X.2006.02915.x},
       adsurl = {https://ui.adsabs.harvard.edu/abs/2006GeoJI.165....3B}
}

@article{Bills,
author = {Bills, Bruce G. and Neumann, Gregory A. and Smith, David E. and Zuber, Maria T.},
title = {Improved estimate of tidal dissipation within Mars from MOLA observations of the shadow of Phobos},
journal = {Journal of Geophysical Research: Planets},
volume = {110},
number = {E7},
pages = {E07004},
doi = {https://doi.org/10.1029/2004JE002376},
eprint = {https://doi.org/10.1029/2004JE002376},
year = {2005}
}

@ARTICLE{Boue,
       author = {{Bou{\'e}}, Gwena{\"e}l and {Efroimsky}, Michael},
        title = "{Tidal evolution of the Keplerian elements}",
      journal = {Celestial Mechanics and Dynamical Astronomy},
         year = 2019,
        month = jul,
       volume = {131},
       number = {7},
        pages = {30},
          doi = {10.1007/s10569-019-9908-2},
archivePrefix = {arXiv},
       eprint = {1904.02253},
 primaryClass = {astro-ph.EP},
       adsurl = {https://ui.adsabs.harvard.edu/abs/2019CeMDA.131...30B}
}

@ARTICLE{Broquet,
       author = {{Broquet}, A. and {Plesa}, A.-C. and {Klemann}, V. and {Root}, B.~C. and {Genova}, A. and {Wieczorek}, M.~A. and {Knapmeyer}, M. and {Andrews-Hanna}, J.~C. and {Breuer}, D.},
        title = "{Glacial isostatic adjustment reveals Mars's interior viscosity structure}",
      journal = {Nature},
         year = 2025,
        month = mar,
       volume = {639},
       number = {8053},
        pages = {109-113},
          doi = {10.1038/s41586-024-08565-9},
       adsurl = {https://ui.adsabs.harvard.edu/abs/2025Natur.639..109B}
}

@ARTICLE{Borg,
       author = {{Borg}, Lars and {Drake}, Michael J.},
        title = "{A review of meteorite evidence for the timing of magmatism and of surface or near-surface liquid water on Mars}",
      journal = {Journal of Geophysical Research (Planets)},
         year = 2005,
        month = sep,
       volume = {110},
       number = {E12},
          eid = {E12S03},
        pages = {E12S03},
          doi = {10.1029/2005JE002402},
       adsurl = {https://ui.adsabs.harvard.edu/abs/2005JGRE..11012S03B}
}

@ARTICLE{Bottke2012,
       author = {{Bottke}, William F. and {Vokrouhlick{\'y}}, David and {Minton}, David and {Nesvorn{\'y}}, David and {Morbidelli}, Alessandro and {Brasser}, Ramon and {Simonson}, Bruce and {Levison}, Harold F.},
        title = "{An Archaean heavy bombardment from a destabilized extension of the asteroid belt}",
      journal = {Nature},
         year = 2012,
        month = may,
       volume = {485},
       number = {7396},
        pages = {78-81},
          doi = {10.1038/nature10967},
       adsurl = {https://ui.adsabs.harvard.edu/abs/2012Natur.485...78B}
}

@ARTICLE{Magma,
       author = {{Bouvier}, Laura C. and {Costa}, Maria M. and {Connelly}, James N. and {Jensen}, Ninna K. and {Wielandt}, Daniel and {Storey}, Michael and {Nemchin}, Alexander A. and {Whitehouse}, Martin J. and {Snape}, Joshua F. and Bellucci, Jeremy J. and {Moynier}, Fr{\'e}d{\'e}ric and {Agranier}, Arnaud and {Gueguen}, Bleuenn and {Sch{\"o}nb{\"a}chler}, Maria and {Bizzarro}, Martin},
        title = "{Evidence for extremely rapid magma ocean crystallization and crust formation on {Mars}}",
      journal = {Nature},
         year = 2018,
        month = jun,
       volume = {558},
       number = {7711},
        pages = {568-589},
          doi = {10.1038/s41586-018-0222-z},
       adsurl = {https://ui.adsabs.harvard.edu/abs/2018Natur.558..586B}
}

@article{Breuer,
title = {Viscosity of the Martian mantle and its initial temperature: Constraints from crust formation history and the evolution of the magnetic field},
journal = {Planetary and Space Science},
volume = {54},
number = {2},
pages = {153-169},
year = {2006},
issn = {0032-0633},
doi = {https://doi.org/10.1016/j.pss.2005.08.008},
author = {Doris Breuer and Tilman Spohn}
}

@article{Carr,
title = {Geologic history of Mars},
journal = {Earth and Planetary Science Letters},
volume = {294},
number = {3},
pages = {185-203},
year = {2010},
issn = {0012-821X},
doi = {https://doi.org/10.1016/j.epsl.2009.06.042},
author = {Michael H. Carr and James W. Head}
}

@article{Castillo,
  title={The tidal history of Iapetus: Spin dynamics in the light of a refined dissipation model},
  author={Castillo-Rogez, Julie C and Efroimsky, Michael and Lainey, Val{\'e}ry},
  journal={Journal of Geophysical Research: Planets},
  volume={116},
  number={E9},
  year={2011}
}

@article{Clifford,
author = {{Clifford}, Stephen M. and {Parker}, Timothy J.},
title = "{The Evolution of the Martian Hydrosphere: Implications for the Fate of a Primordial Ocean and the Current State of the Northern Plains}",
journal = {Icarus},
volume = {154},
number = {1},
pages = {40-79},
year = {2001},
issn = {0019-1035},
doi = {https://doi.org/10.1006/icar.2001.6671}
}

@book{atlas,
  title={The Atlas of Mars: Mapping its Geography and Geology},
  author={Coles, Kenneth S and Tanaka, Kenneth L and Christensen, Philip R},
  year={2019},
  publisher={Cambridge University Press},
  address={Cambridge},
  isbn={9781107036291}
}

@ARTICLE{Counselman,
       author = {{Counselman}, III, Charles C.},
        title = "{Outcomes of Tidal Evolution}",
      journal = {The Astrophysical Journal},
         year = 1973,
       volume = {180},
        pages = {307-316},
          doi = {10.1086/151964},
       adsurl = {https://ui.adsabs.harvard.edu/abs/1973ApJ...180..307C}
}

@ARTICLE{Darwin1879,
       author = {{Darwin}, G.~H.},
        title = "{The Determination of the Secular Effects of Tidal Friction by a Graphical Method}",
      journal = {Proceedings of the Royal Society of London Series I},
         year = 1879,
        month = jan,
       volume = {29},
        pages = {168-181},
       adsurl = {https://ui.adsabs.harvard.edu/abs/1879RSPS...29..168D}
}

@INPROCEEDINGS{Das,
       author = {{Das}, P. and {Basu Sarbadhikari}, A. and {Sarkar}, R. and {Karunatillake}, S. and {Edgett}, K.~S. and {Paul}, P.~P. and {Brothers}, J.~P. and {Armstrong}, S.~A.},
        title = "{Fine-Scale Rhythmic Sedimentary Layering at Vera Rubin Ridge, Gale Crater, Mars: Possible Tidal Forcing on Ancient Mars}",
    booktitle = {51st Annual Lunar and Planetary Science Conference},
         year = 2020,
       series = {Lunar and Planetary Science Conference},
          eid = {1916},
        pages = {1916},
       adsurl = {https://ui.adsabs.harvard.edu/abs/2020LPI....51.1916D}
}

@ARTICLE{Denton2025,
       author = {{Denton}, C. Adenee and {Asphaug}, Erik and {Emsenhuber}, Alexandre and {Melikyan}, Robert},
        title = "{Capture of an ancient Charon around Pluto}",
      journal = {Nature Geoscience},
         year = 2025,
       volume = {18},
        issue = {1},
          doi = {10.1038/s41561-024-01612-0},
       adsurl = {https://doi.org/10.1038/s41561-024-01612-0}
}

@ARTICLE{Efroimsky2012,
       author = {{Efroimsky}, Michael},
        title = "{Bodily tides near spin-orbit resonances}",
      journal = {Celestial Mechanics and Dynamical Astronomy},
         year = 2012,
       volume = {112},
       number = {3},
        pages = {283-330},
          doi = {10.1007/s10569-011-9397-4},
archivePrefix = {arXiv},
       eprint = {1105.6086},
 primaryClass = {astro-ph.EP},
       adsurl = {https://ui.adsabs.harvard.edu/abs/2012CeMDA.112..283E}
}

@ARTICLE{Mars,
       author = {{Efroimsky}, Michael},
        title = "{A Synchronous Moon as a Possible Cause of Mars' Initial Triaxiality}",
      journal = {Journal of Geophysical Research (Planets)},
         year = 2024,
       volume = {129},
       number = {10},
          eid = {E2023JE008277},
        pages = {E2023JE008277},
          doi = {10.1029/2023JE008277},
archivePrefix = {arXiv},
       eprint = {2408.14725},
 primaryClass = {astro-ph.EP},
       adsurl = {https://ui.adsabs.harvard.edu/abs/2024JGRE..12908277E}
}

@ARTICLE{Efroimsky2005,
       author = {{Efroimsky}, Michael},
        title = "{Long-Term Evolution of Orbits About A Precessing Oblate Planet: 1. The Case of Uniform Precession}",
      journal = {Celestial Mechanics and Dynamical Astronomy},
         year = 2005,
       volume = {91},
       number = {1-2},
        pages = {75-108},
          doi = {10.1007/s10569-004-2415-z},
archivePrefix = {arXiv},
       eprint = {astro-ph/0408168},
 primaryClass = {astro-ph},
       adsurl = {https://ui.adsabs.harvard.edu/abs/2005CeMDA..91...75E}
}

@ARTICLE{Efroimsky2015,
       author = {{Efroimsky}, Michael},
        title = "{Tidal Evolution of Asteroidal Binaries. Ruled by Viscosity. Ignorant of Rigidity.}",
      journal = {The Astronomical Journal},
         year = 2015,
       volume = {150},
       number = {4},
          eid = {98},
        pages = {98},
          doi = {10.1088/0004-6256/150/4/98},
archivePrefix = {arXiv},
       eprint = {1506.09157},
 primaryClass = {astro-ph.EP},
       adsurl = {https://ui.adsabs.harvard.edu/abs/2015AJ....150...98E}
 }

@ARTICLE{Elkins,
       author = {{Elkins-Tanton}, Linda T.},
        title = "{Formation of early water oceans on rocky planets}",
      journal = {Astrophysics and Space Science},
         year = 2011,
        month = apr,
       volume = {332},
       number = {2},
        pages = {359-364},
          doi = {10.1007/s10509-010-0535-3},
archivePrefix = {arXiv},
       eprint = {1011.2710},
 primaryClass = {astro-ph.EP},
       adsurl = {https://ui.adsabs.harvard.edu/abs/2011Ap&SS.332..359E}
}

@ARTICLE{Ray2001,
       author = {{Egbert}, Gary D. and {Ray}, Richard D.},
        title = "{Estimates of M$_{2}$ tidal energy dissipation from TOPEX/Poseidon altimeter data}",
      journal = {Journal of Geophysical Research},
         year = 2001,
       volume = {106},
       number = {C10},
        pages = {22,475-22,502},
          doi = {10.1029/2000JC000699},
       adsurl = {https://ui.adsabs.harvard.edu/abs/2001JGR...10622475E}
}

@article{Farhat,
	author = {{Farhat}, Mohammad and {Auclair-Desrotour}, Pierre and {Bou{\'e}}, Gwena{\"e}l and {Laskar}, Jacques},
	title = {The resonant tidal evolution of the {Earth-Moon} distance},
	DOI= "10.1051/0004-6361/202243445",
	journal = {Astronomy and Astrophysics},
	year = 2022,
	volume = 665,
	pages = "L1",
}

@article{Fichtinger,
	author = {{Fichtinger}, Bibiana and {G\"{u}del}, Manuel and {Mutel}, Robert L. and {Hallinan}, Gregg and {Gaidos}, Eric and
              {Skinner}, Stephen L. and {Lynch}, Christene and {Gayley}, Kenneth G.},
	title = {Radio emission and mass loss rate limits   of four young solar-type stars},
	DOI= "10.1051/0004-6361/201629886",
	journal = {Astronomy and Astrophysics},
	year = 2017,
	volume = 599,
	pages = "A127",
}

@ARTICLE{Fontaine,
       author = {{Fontaine}, Fabrice R. and {Ildefonse}, Benoit and {Bagdassarov}, Nickolai S.},
        title = "{Temperature dependence of shear wave attenuation in partially molten gabbronorite at seismic frequencies}",
      journal = {Geophysical Journal International},
         year = 2005,
        month = dec,
       volume = {163},
       number = {3},
        pages = {1025-1038},
          doi = {10.1111/j.1365-246X.2005.02767.x},
       adsurl = {https://ui.adsabs.harvard.edu/abs/2005GeoJI.163.1025F}
}

@ARTICLE{Genova,
       author = {{Genova}, Antonio and {Mazarico}, Erwan and {Goossens}, Sander and {Lemoine}, Frank G. and {Neumann}, Gregory A. and {Smith}, David E. and {Zuber}, Maria T.},
        title = "{Solar system expansion and strong equivalence principle as seen by the NASA MESSENGER mission}",
      journal = {Nature Communications},
         year = 2018,
        month = jan,
       volume = {9},
          eid = {289},
        pages = {289},
          doi = {10.1038/s41467-017-02558-1},
       adsurl = {https://ui.adsabs.harvard.edu/abs/2018NatCo...9..289G}
}

@ARTICLE{Zircon,
       author = {{Gillespie}, Jack and {Cavosie}, Aaron J. and {Fougerouse}, Denis and {Ciobanu}, Cristiana L. and {Rickard}, William D.~A. and {Saxey}, David W. and {Benedix}, Gretchen K. and {Bland}, Phil A.},
        title = "{Zircon trace element evidence for early hydrothermal activity on Mars}",
      journal = {Science Advances},
         year = 2024,
        month = nov,
       volume = {10},
       number = {47},
          eid = {EADQ3694},
        pages = {EADQ3694},
          doi = {10.1126/sciadv.adq3694},
       adsurl = {https://ui.adsabs.harvard.edu/abs/2024SciA...10.3694G}
}

@Article{Gomes,
  author  = {Gomes, R. and Levison, H. F. and Tsiganis, K. and Morbidelli, A.},
  journal = {Nature},
  title   = {Origin of the cataclysmic Late Heavy Bombardment period of the terrestrial planets},
  year    = {2005},
  month   = {may},
  number  = {7041},
  pages   = {466--469},
  volume  = {435},
  doi     = {10.1038/nature03676},
}

@ARTICLE{Goldreich,
       author = {{Goldreich}, Peter},
        title = "{Inclination of satellite orbits about an oblate precessing planet}",
      journal = {The Astronomical Journal},
         year = 1965,
       volume = {70},
        pages = {5},
          doi = {10.1086/109673},
       adsurl = {https://ui.adsabs.harvard.edu/abs/1965AJ.....70....5G}
}

@ARTICLE{Gribb,
       author = {{Gribb}, Tye T. and {Cooper}, Reid F.},
        title = "{Low-frequency shear attenuation in polycrystalline olivine: Grain boundary diffusion and the physical significance of the Andrade model for viscoelastic rheology}",
      journal = {Journal of Geophysical Research},
         year = 1998,
       volume = {103},
       number = {B11},
        pages = {27,267-27,279},
          doi = {10.1029/98JB02786},
       adsurl = {https://ui.adsabs.harvard.edu/abs/1998JGR...10327267G}
}

@article{gravitational_constant,
  title = "{CODATA} recommended values of the fundamental physical constants: 2018",
  author = {Tiesinga, Eite and Mohr, Peter J. and Newell, David B. and Taylor, Barry N.},
  journal = {Reviews of Modern Physics},
  volume = {93},
  issue = {2},
  pages = {025010},
  numpages = {63},
  year = {2021},
  month = {Jun},
  publisher = {American Physical Society},
  doi = {10.1103/RevModPhys.93.025010}
}

@article{Guinard,
title = {Coupled tidal tomography and thermal constraints for probing Mars viscosity profile},
journal = {Icarus},
volume = {425},
pages = {116318},
year = {2025},
issn = {0019-1035},
doi = {https://doi.org/10.1016/j.icarus.2024.116318},
author = {Alex Guinard and Agn\`es Fienga and Anthony M\'emin and Cl\'ement Ganino}
}

@ARTICLE{Gurfil,
       author = {{Gurfil}, Pini and {Lainey}, Val{\'e}ry and {Efroimsky}, Michael},
        title = "{Long-term evolution of orbits about a precessing oblate planet: 3. A semianalytical and a purely numerical approach}",
      journal = {Celestial Mechanics and Dynamical Astronomy},
         year = 2007,
       volume = {99},
       number = {4},
        pages = {261-292},
          doi = {10.1007/s10569-007-9099-0},
archivePrefix = {arXiv},
       eprint = {astro-ph/0607530},
 primaryClass = {astro-ph},
       adsurl = {https://ui.adsabs.harvard.edu/abs/2007CeMDA..99..261G}
}

@article{Hunten,
title = {Capture of Phobos and Deimos by photoatmospheric drag},
journal = {Icarus},
volume = {37},
number = {1},
pages = {113-123},
year = {1979},
issn = {0019-1035},
doi = {https://doi.org/10.1016/0019-1035(79)90119-2},
author = {Donald M. Hunten}
}

@ARTICLE{Hut,
       author = {{Hut}, P.},
        title = "{Tidal evolution in close binary systems.}",
      journal = {Astronomy and Astrophysics},
         year = 1981,
       volume = {99},
        pages = {126-140},
       adsurl = {https://ui.adsabs.harvard.edu/abs/1981A&A....99..126H}
}

@article{Hartmann,
title = {Possible long-term decline in impact rates: 2. Lunar impact-melt data regarding impact history},
journal = {Icarus},
volume = {186},
number = {1},
pages = {11-23},
year = {2007},
issn = {0019-1035},
doi = {https://doi.org/10.1016/j.icarus.2006.09.009},
author = {William K. Hartmann and Cathy Quantin and Nicolas Mangold}
}

@ARTICLE{Jackson,
       author = {{Jackson}, Ian and {Fitz Gerald}, John D. and {Faul}, Ulrich H. and {Tan}, Ben H.},
        title = "{Grain-size-sensitive seismic wave attenuation in polycrystalline olivine}",
      journal = {Journal of Geophysical Research (Solid Earth)},
         year = 2002,
        month = dec,
       volume = {107},
       number = {B12},
          eid = {2360},
        pages = {2360},
          doi = {10.1029/2001JB001225},
       adsurl = {https://ui.adsabs.harvard.edu/abs/2002JGRB..107.2360J}
}

@ARTICLE{jackson2010,
       author = {{Jackson}, Ian and {Faul}, Ulrich H. and {Suetsugu}, Daisuke and {Bina}, Craig and {Inoue}, Toru and {Jellinek}, Mark},
        title = "{Grainsize-sensitive viscoelastic relaxation in olivine: Towards a robust laboratory-based model for seismological application}",
      journal = {Physics of the Earth and Planetary Interiors},
         year = 2010,
        month = nov,
       volume = {183},
       number = {1-2},
        pages = {151-163},
          doi = {10.1016/j.pepi.2010.09.005},
       adsurl = {https://ui.adsabs.harvard.edu/abs/2010PEPI..183..151J}
}

@ARTICLE{JohnsonMelosh,
       author = {{Johnson}, B.~C. and {Melosh}, H.~J.},
        title = "{Impact spherules as a record of an ancient heavy bombardment of Earth}",
      journal = {Nature},
         year = 2012,
        month = may,
       volume = {485},
       number = {7396},
        pages = {75-77},
          doi = {10.1038/nature10982},
       adsurl = {https://ui.adsabs.harvard.edu/abs/2012Natur.485...75J}
}

@ARTICLE{Khan2017,
       author = {{Khan}, A. and {Liebske}, C. and {Rozel}, A. and {Rivoldini}, A. and {Nimmo}, F. and {Connolly}, J.~A.~D. and {Plesa}, A. -C. and {Giardini}, D.},
        title = "{A Geophysical Perspective on the Bulk Composition of Mars}",
      journal = {Journal of Geophysical Research (Planets)},
         year = 2018,
       volume = {123},
       number = {2},
        pages = {575-611},
          doi = {10.1002/2017JE005371},
       adsurl = {https://ui.adsabs.harvard.edu/abs/2018JGRE..123..575K}
}

@ARTICLE{Konopliv2016,
       author = {{Konopliv}, Alex S. and {Park}, Ryan S. and {Folkner}, William M.},
        title = "{An improved JPL Mars gravity field and orientation from Mars orbiter and lander tracking data}",
      journal = {Icarus},
         year = 2016,
       volume = {274},
        pages = {253-260},
          doi = {10.1016/j.icarus.2016.02.052},
       adsurl = {https://ui.adsabs.harvard.edu/abs/2016Icar..274..253K}
}

@ARTICLE{Konopliv_2011,
       author = {{Konopliv}, Alex S. and {Asmar}, Sami W. and {Folkner}, William M. and {Karatekin}, {\"O}zg{\"u}r and {Nunes}, Daniel C. and {Smrekar}, Suzanne E. and {Yoder}, Charles F. and {Zuber}, Maria T.},
        title = "{Mars high resolution gravity fields from MRO, Mars seasonal gravity, and other dynamical parameters}",
      journal = {Icarus},
         year = 2011,
        month = jan,
       volume = {211},
       number = {1},
        pages = {401-428},
          doi = {10.1016/j.icarus.2010.10.004},
       adsurl = {https://ui.adsabs.harvard.edu/abs/2011Icar..211..401K}
}

@ARTICLE{Konopliv2020,
       author = {{Konopliv}, Alex S. and {Park}, Ryan S. and {Rivoldini}, Attilio and {Baland}, Rose-Marie and {Le Maistre}, Sebastien and {Van Hoolst}, Tim and {Yseboodt}, Marie and {Dehant}, Veronique},
        title = "{Detection of the Chandler Wobble of Mars From Orbiting Spacecraft}",
      journal = {Geophysical Research Letters},
         year = 2020,
       volume = {47},
       number = {21},
          eid = {e90568},
        pages = {e90568},
          doi = {10.1029/2020GL090568},
       adsurl = {https://ui.adsabs.harvard.edu/abs/2020GeoRL..4790568K}
}

@ARTICLE{Laskar,
       author = {{Laskar}, J. and {Robutel}, P.},
        title = "{The chaotic obliquity of the planets}",
      journal = {Nature},
         year = 1993,
        month = feb,
       volume = {361},
       number = {6413},
        pages = {608-612},
          doi = {10.1038/361608a0},
       adsurl = {https://ui.adsabs.harvard.edu/abs/1993Natur.361..608L}
}

@ARTICLE{Laskar2,
       author = {{Laskar}, J. and {Joutel}, F. and {Robutel}, P.},
        title = "{Stabilization of the Earth's obliquity by the Moon}",
      journal = {Nature},
         year = 1993,
        month = feb,
       volume = {361},
       number = {6413},
        pages = {615-617},
          doi = {10.1038/361615a0},
       adsurl = {https://ui.adsabs.harvard.edu/abs/1993Natur.361..615L}
}

@article{LeMaistre,
 year = {2023},
 title = {{Spin state and deep interior structure of Mars from InSight radio tracking}},
 author = {Le Maistre, S. and Rivoldini, A. and Caldiero, A. and Yseboodt, M. and Baland, R.-M. and Beuthe, M. and Van Hoolst, T. and Dehant, V.
 and Folkner, W. M. and Buccino, D. and Kahan, D. and Marty, J.-C. and Antonangeli, D. and Badro, J. and Drilleau, M. and Konopliv, A. and
 P{\`e}ters, M.-J. and Plesa, A.-C. and Samuel, H. and Tosi, N. and Wieczorek, M.
 and Lognonn{\`e}, P. and Panning, M. and Smrekar, S. and Banerdt, B. W.},
 journal = {Nature},
 doi = {10.1038/s41586-023-06150-0},
 pages = {733--737},
 volume = {619}
 }

@ARTICLE{Lissauer,
       author = {{Lissauer}, Jack J. and {Barnes}, Jason W. and {Chambers}, John E.},
        title = "{Obliquity variations of a moonless Earth}",
      journal = {Icarus},
         year = 2012,
        month = jan,
       volume = {217},
       number = {1},
        pages = {77-87},
          doi = {10.1016/j.icarus.2011.10.013},
       adsurl = {https://ui.adsabs.harvard.edu/abs/2012Icar..217...77L}
}

@inproceedings{LoweByerly2015,
  author    = {Lowe, Donald R. and Byerly, Gary R.},
  title     = {The Terrestrial Record of an Extended Late Heavy Bombardment},
  booktitle = {Early Solar System Bombardment III},
  year      = {2015},
       series = {LPI Contributions},
  volume    = {1826},
  adsurl = {https://ui.adsabs.harvard.edu/abs/2015LPICo1826.3015L},
  pages     = {3015}
}

@article{LOWE201839,
title = {The terrestrial record of Late Heavy Bombardment},
journal = {New Astronomy Reviews},
volume = {81},
pages = {39-61},
year = {2018},
issn = {1387-6473},
doi = {https://doi.org/10.1016/j.newar.2018.03.002},
author = {Donald R. Lowe and Gary R. Byerly}
}

@ARTICLE{Pathways,
       author = {{Makarov}, Valeri V. and {Efroimsky}, Michael},
        title = "{Pathways of survival for exomoons and inner exoplanets}",
      journal = {Astronomy and Astrophysics},
         year = 2023,
        month = apr,
       volume = {672},
          eid = {A78},
        pages = {A78},
          doi = {10.1051/0004-6361/202245533},
archivePrefix = {arXiv},
       eprint = {2302.04646},
 primaryClass = {astro-ph.EP},
       adsurl = {https://ui.adsabs.harvard.edu/abs/2023A&A...672A..78M}
}

@Article{synchronisation,
AUTHOR = { {Makarov}, Valeri V. and {Efroimsky}, Michael},
TITLE = {Initial Conditions for Tidal Synchronisation of a Planet by Its Moon},
JOURNAL = {Universe},
VOLUME = {11},
YEAR = {2025},
eid = {309},
pages = {309},
URL = {https://www.mdpi.com/2218-1997/11/9/309},
ISSN = {2218-1997},
DOI = {10.3390/universe11090309}
}

@ARTICLE{2023Univ...10...15M,
       author = {{Makarov}, Valeri V. and {Goldin}, Alexey},
        title = "{Chaotic Capture of a Retrograde Moon by Venus and the Reversal of Its Spin}",
      journal = {Universe},
         year = 2024,
       volume = {10},
       number = {1},
          eid = {15},
        pages = {15},
          doi = {10.3390/universe10010015},
archivePrefix = {arXiv},
       eprint = {2312.17049},
 primaryClass = {astro-ph.EP},
       adsurl = {https://ui.adsabs.harvard.edu/abs/2023Univ...10...15M},
url = {https://www.mdpi.com/2218-1997/10/1/15}
}

@ARTICLE{Malarewicz,
       author = {{Malarewicz}, V. and {Beyssac}, O. and {Zanda}, B. and {Marin-Carbonne}, J. and {Leroux}, H. and {Rubatto}, D. and {Bouvier}, A.~S. and {Deldicque}, D. and {Pont}, S. and {Bernard}, S. and {Bloch}, E. and {Bouley}, S. and {Humayun}, M. and {Hewins}, R.~H.},
        title = "{Evidence for pre-Noachian granitic rocks on Mars from quartz in meteorite NWA 7533}",
      journal = {Nature Geoscience},
         year = 2025,
        month = mar,
       volume = {18},
       number = {3},
        pages = {207-212},
          doi = {10.1038/s41561-025-01653-z},
       adsurl = {https://ui.adsabs.harvard.edu/abs/2025NatGe..18..207M}
}

@article{Marchietal2014,
  author  = {Marchi, Simone and Bottke, William F. and Elkins-Tanton, Linda T. and Bierhaus, Edward B. and Wuennemann, Kai and Morbidelli, Alessandro and Kring, David A.},
  title   = {Widespread mixing and burial of Earth's pristine crust by Hadean impact bombardment},
  journal = {Nature},
  year    = {2014},
  volume  = {511},
  number  = {7511},
  pages   = {578--582},
  doi     = {10.1038/nature13539}
}

@article{Miguel,
  title = {Dislocation Jamming and Andrade Creep},
  author = {Miguel, M.-Carmen and Vespignani, A. and Zaiser, M. and Zapperi, S.},
  journal = {Phys. Rev. Lett.},
  volume = {89},
  issue = {16},
  pages = {165501},
  numpages = {4},
  year = {2002},
  month = {Sep},
  publisher = {American Physical Society},
  doi = {10.1103/PhysRevLett.89.165501}
}

@ARTICLE{YoungSun,
       author = {{Nandy}, Dibyendu and {Martens}, Petrus C.~H. and {Obridko}, Vladimir and {Dash}, Soumyaranjan and {Georgieva}, Katya},
        title = "{Solar evolution and extrema: current state of understanding of long-term solar variability and its planetary impacts}",
      journal = {Progress in Earth and Planetary Science},
         year = 2021,
       volume = {8},
       number = {1},
          eid = {40},
        pages = {40},
          doi = {10.1186/s40645-021-00430-x},
       adsurl = {https://ui.adsabs.harvard.edu/abs/2021PEPS....8...40N}
}

@ARTICLE{Pou,
       author = {{Pou}, L. and {Nimmo}, F. and {Rivoldini}, A. and {Khan}, A. and {Bagheri}, A. and {Gray}, T. and {Samuel}, H. and {Lognonn{\'e}}, P. and {Plesa}, A. -C. and {Gudkova}, T. and {Giardini}, D.},
        title = "{Tidal Constraints on the Martian Interior}",
      journal = {Journal of Geophysical Research (Planets)},
         year = 2022,
       volume = {127},
       number = {11},
          eid = {e2022JE007291},
        pages = {e2022JE007291},
          doi = {10.1029/2022JE007291},
       adsurl = {https://ui.adsabs.harvard.edu/abs/2022JGRE..12707291P}
}

@ARTICLE{Qu,
       author = {{Qu}, Tongzhang and {Jackson}, Ian and {Faul}, Ulrich H.},
        title = "{Low-Frequency Seismic Properties of Olivine-Orthopyroxene Mixtures}",
      journal = {Journal of Geophysical Research (Solid Earth)},
         year = 2021,
        month = oct,
       volume = {126},
       number = {10},
          eid = {e2021JB022504},
        pages = {e2021JB022504},
          doi = {10.1029/2021JB022504},
       adsurl = {https://ui.adsabs.harvard.edu/abs/2021JGRB..12622504Q}
}

@article{Russel,
    author = {Russell, Sara S. and Ballentine, Chris J. and Grady, Monica M.},
    title = {The origin, history and role of water in the evolution of the inner Solar System},
    journal = {Philosophical Transactions of the Royal Society A: Mathematical, Physical and Engineering Sciences},
    volume = {375},
    number = {2094},
    pages = {20170108},
    year = {2017},
    month = {04},
    issn = {1364-503X},
    doi = {10.1098/rsta.2017.0108},
    url = {https://doi.org/10.1098/rsta.2017.0108},
    eprint = {https://royalsocietypublishing.org/rsta/article-pdf/doi/10.1098/rsta.2017.0108/1385541/rsta.2017.0108.pdf},
}

@article{Sarkar,
  author    = {Sarkar, R. and Das, P. and Basu Sarbadhikari, A. and Karunatillake, S. and {\v{C}}uk, M. and Mishra, S. M.},
  title     = {Possible Tidal Rhythmites in {Gale Crater}, {Mars}: Traces of a Lost Moon?},
  booktitle = {AGU Fall Meeting Abstracts},
  volume    = {2025},
  number    = {1951254},
  year      = {2025},
  month     = {December},
  note      = {Presented at AGU25, Paper No. 1951254}
}

@BOOK{Seidelmann,
       author = {{Seidelmann}, P. K. and {Urban}, S.~E.},
        title = "{Explanatory Supplement to the Astronomical Almanac, Third Edition}",
         year = 2013,
         isbn = {978-1-891389-85-6}
}

@ARTICLE{Schmidt,
       author = {{Schmidt}, Fr{\'e}d{\'e}ric and {Way}, Michael J. and {Costard}, Fran{\c{c}}ois and {Bouley}, Sylvain and {S{\'e}journ{\'e}}, Antoine and {Aleinov}, Igor},
        title = "{Circumpolar ocean stability on Mars 3 Gy ago}",
      journal = {Proceedings of the National Academy of Science},
         year = 2022,
        month = jan,
       volume = {119},
       number = {4},
          eid = {e2112930118},
        pages = {e2112930118},
          doi = {10.1073/pnas.2112930118},
archivePrefix = {arXiv},
       eprint = {2310.00461},
 primaryClass = {astro-ph.EP},
       adsurl = {https://ui.adsabs.harvard.edu/abs/2022PNAS..11912930S}
}

@book{Stigler,
  title={Statistics on the Table: The History of Statistical Concepts and Methods},
  author={Stigler, Stephen M},
  year={1999},
  publisher={Harvard University Press},
  address={Cambridge, MA},
  isbn={9780674009790}
}

@ARTICLE{Tan,
       author = {{Tan}, B.~H. and {Jackson}, I. and {Fitz Gerald}, J.~D.},
        title = "{High-temperature viscoelasticity of fine-grained polycrystalline olivine}",
      journal = {Physics and Chemistry of Minerals},
         year = 2001,
        month = jan,
       volume = {28},
       number = {9},
        pages = {641-664},
          doi = {10.1007/s002690100189},
       adsurl = {https://ui.adsabs.harvard.edu/abs/2001PCM....28..641T}
}

@article{Tera,
  title={Isotopic evidence for a terminal lunar cataclysm},
  author={Tera, F. and Papanastassiou, D. A. and Wasserburg, G. J.},
  journal={Earth and Planetary Science Letters},
  volume={22},
  number={1},
  pages={1--21},
  year={1974},
  publisher={Elsevier},
  doi={10.1016/0012-821X(74)90059-4}
}

@article{Thomson,
 author = {Thomson, William}, title = {Dynamical problems regarding elastic spheroidal shells; and On the rigidity of the Earth},
 journal = {Philosophical Transactions of the Royal Society of London},
 year = {1863},
 volume = {153},
 pages = {573--616},
 doi = {10.1098/rstl.1863.0028},
 url = {royalsocietypublishing.org} }

@ARTICLE{Touma,
       author = {{Touma}, J. and {Wisdom}, J.},
        title = "{The Chaotic Obliquity of Mars}",
      journal = {Science},
         year = 1993,
        month = feb,
       volume = {259},
       number = {5099},
        pages = {1294-1297},
          doi = {10.1126/science.259.5099.1294},
       adsurl = {https://ui.adsabs.harvard.edu/abs/1993Sci...259.1294T}
}

@BOOK{book,
       author = {{Kopeikin}, Sergei and {Efroimsky}, Michael and {Kaplan}, George},
        title = "{Relativistic Celestial Mechanics of the Solar System}",
         year = 2011,
          doi = {10.1002/9783527634569},
       adsurl = {https://ui.adsabs.harvard.edu/abs/2011rcms.book.....K}
}

@ARTICLE{walterova,
       author = {{Walterov{\'a}}, Michaela and {B{\v{e}}hounkov{\'a}}, Marie and {Efroimsky}, Michael},
        title = "{Is There a Semi-Molten Layer at the Base of the Lunar Mantle?}",
      journal = {Journal of Geophysical Research (Planets)},
         year = 2023,
       volume = {128},
       number = {7},
          eid = {e2022JE007652},
        pages = {e2022JE007652},
          doi = {10.1029/2022JE007652},
archivePrefix = {arXiv},
       eprint = {2301.02476},
 primaryClass = {astro-ph.EP},
       adsurl = {https://ui.adsabs.harvard.edu/abs/2023JGRE..12807652W}
}

@ARTICLE{Ward,
       author = {{Ward}, William R. and {Reid}, Mark J.},
        title = "{Solar tidal friction and satellite loss}",
      journal = {Monthly Notes of the Royal Society},
         year = 1973,
       volume = {164},
        pages = {21},
          doi = {10.1093/mnras/164.1.21},
       adsurl = {https://ui.adsabs.harvard.edu/abs/1973MNRAS.164...21W}
}

@article{Williams_and_Zugger,
doi = {10.3847/PSJ/ad5a9a},
url = {https://dx.doi.org/10.3847/PSJ/ad5a9a},
year = {2024},
publisher = {The Planetary Science Journal},
volume = {5},
number = {9},
pages = {208},
author = {Darren M. Williams and Michael E. Zugger},
title = {Forming Massive Terrestrial Satellites through Binary-exchange Capture},
journal = {The Planetary Science Journal}
}

@ARTICLE{Xiao,
       author = {{Xiao}, Long and {Huang}, Jun and {Kusky}, Timothy and {Head}, James W. and {Zhao}, Jiannan and {Wang}, Jiang and {Wang}, Le and {Yu}, Wenchao and {Shi}, Yutong and {Wu}, Bo and {Qian}, Yuqi and {Huang}, Qian and {Xiao}, Xiao},
        title = "{Evidence for marine sedimentary rocks in Utopia Planitia: Zhurong rover observations}",
      journal = {National Science Review},
         year = 2023,
       volume = {10},
       number = {9},
          eid = {nwad137},
        pages = {nwad137},
          doi = {10.1093/nsr/nwad137},
       adsurl = {https://ui.adsabs.harvard.edu/abs/2023NSRev..10D.137X}
}

@ARTICLE{Zellner2017,
       author = {{Zellner}, Nicolle E.~B.},
        title = "{Cataclysm No More: New Views on the Timing and Delivery of Lunar Impactors}",
      journal = {Origins of Life and Evolution of the Biosphere},
         year = 2017,
        month = sep,
       volume = {47},
       number = {3},
        pages = {261-280},
          doi = {10.1007/s11084-017-9536-3},
archivePrefix = {arXiv},
       eprint = {1704.06694},
 primaryClass = {astro-ph.EP},
       adsurl = {https://ui.adsabs.harvard.edu/abs/2017OLEB...47..261Z}
}
\bibliographystyle{plainnat}

 \end{document}